\documentclass{article} 
\usepackage[margin=1in,marginparwidth=0.8in, marginparsep=0.15in]{geometry}

\date{}

\usepackage[square]{natbib}
\usepackage{float}
\usepackage{graphicx}
\usepackage[space]{grffile}
\usepackage[font=small,labelfont=bf]{caption}
\usepackage{amsmath, amssymb, amsthm}
\usepackage{multirow} 
\usepackage{tabu} 
\usepackage{booktabs} 
\usepackage{chngcntr} 
\usepackage[colorlinks=true, urlcolor=blue, citecolor=blue, linkcolor=black, filecolor=blue]{hyperref}
\usepackage[textsize=footnotesize]{todonotes}

\newcommand{\HSIC}{\mbox{HSIC}}
\newcommand{\RDC}{\mbox{RDC}}
\newcommand{\HHG}{\mbox{HHG}}
\newcommand{\DDP}{{S^{DDP}}}
\newcommand{\dCor}{\mbox{dCor}}

\newcommand{\MI}{\mbox{I}}

\newcommand{\onlineSupplementLink}{\url{http://www.exploredata.net/ftp/empirical_supplement.zip}}

\newcommand{\pathToCommon}{.}
\newcommand{\pathToCommonFigs}{.}
\newcommand{\pathToFigures}{.}
\newcommand{\R}{\mathbb{R}}
\newcommand{\Z}{\mathbb{Z}}
\newcommand{\ep}{\varepsilon}
\renewcommand{\Pr}[1]{\mathbf{P} \left( {#1} \right)}

\newcommand{\Q}{\boldsymbol{\mathcal{Q}}}
\newcommand{\F}{\boldsymbol{\mathcal{F}}}
\newcommand{\mcZ}{\mathcal{Z}}
\newcommand{\hvphi}{{\hat{\varphi}}}

\newcommand{\reliablestat}[3]{ {R^{#2}_{#1}\left( #3 \right)} }

\newcommand{\interpretablestat}[3]{ {I^{#2}_{#1}\left( #3 \right)} }

\newcommand{\MIC}{\mbox{MIC}}
\newcommand{\popMIC}{{\mbox{MIC}_*}}
\newcommand{\MICestE}{{\mbox{MIC}_e}}

\newcommand{\TICestE}{{\mbox{TIC}_e}}

\newtheorem{thm}{Theorem}[section]
\newtheorem*{thm*}{Theorem}

\theoremstyle{definition}
\newtheorem{definition}[thm]{Definition}

\newcommand{\figpart}[1]{\textbf{({#1})}}

\makeatletter
\renewcommand*{\@fnsymbol}[1]{\ensuremath{\ifcase#1\or 1\or *\or \dagger\or 2\or 3\or 4\or **\or 5\or \mathsection\or \mathparagraph\or \|\or \ddagger\or \dagger\dagger
   \or \ddagger\ddagger \else\@ctrerr\fi}}
\makeatother

\title{An Empirical Study of Leading Measures of Dependence}
\author{
David N.\ Reshef\footnote{Department of Computer Science, Massachusetts Institute of Technology.} \footnote{Co-first author.} \footnote{To whom correspondence should be addressed. Email: \url{dnreshef@mit.edu}}
\and
Yakir A. Reshef\footnote{School of Engineering and Applied Sciences, Harvard University.} \footnotemark[2]
\and
Pardis C. Sabeti\footnote{Department of Organismic and Evolutionary Biology, Harvard University.} \footnote{Broad Institute of MIT and Harvard.} \footnote{Co-last author.}
\and
Michael M. Mitzenmacher\footnotemark[1] \footnotemark[7]
}

\begin{document}
\maketitle

\begin{abstract}
In exploratory data analysis, we are often interested in identifying promising pairwise associations for further analysis while filtering out weaker, less interesting ones. This can be accomplished by computing a measure of dependence on all possible variable pairs and examining the highest-scoring pairs, provided the measure of dependence used assigns similar scores to equally noisy relationships of different types. This property, called {\em equitability}, is formalized in \citet{reshef2015equitability}. In addition to equitability, measures of dependence can also be assessed by the power of their corresponding independence tests as well as their runtime.

Here we present extensive empirical evaluation of the equitability, power against independence, and runtime of several leading measures of dependence. These include two statistics newly introduced in \citet{reshef2015estimating}: $\MICestE$, which has equitability as its primary goal, and $\TICestE$, which has power against independence as its primary goal.

Regarding equitability, our analysis finds that $\MICestE$ is the most equitable method on functional relationships in most of the settings we considered, although mutual information estimation proves the most equitable at large sample sizes in some specific settings. Regarding power against independence, we find that $\TICestE$, along with Heller and Gorfine's $\DDP$, is the state of the art on the relationships we tested. Our analyses also show evidence for a trade-off between power against independence and equitability consistent with the theory in \citet{reshef2015equitability}. In terms of runtime, $\MICestE$ and $\TICestE$ are significantly faster than many other measures of dependence tested. Moreover, computing either one makes computing the other trivial. This suggests that a fast and useful strategy for achieving a combination of power against independence and equitability may be to filter relationships by $\TICestE$ and then to examine the $\MICestE$ of only the significant ones.

We conclude with a discussion of the settings in which $\MICestE$ and $\TICestE$ are (and are not) appropriate tools. It is our hope that this work provides a practical guide for the use of $\MICestE$, $\TICestE$, and related statistics, and for the role of equitability more generally.
\end{abstract}

\section{Introduction}
Suppose we have a high-dimensional data set with hundreds or thousands of dimensions and we wish to find interesting associations within it to analyze further.  Even if we only search for pairwise associations among the variables, the number of potential relationships to examine is unmanageably large, necessitating automation to assist in the search. In this context, a common, simple approach is to compute some statistic on each combination of variables, rank the variable pairs from highest- to lowest-scoring, and then examine a small number of the top-scoring variable pairs in the resulting list.

The success of this strategy depends heavily on the statistic used. One natural approach is to use a measure of dependence, that is, a statistic whose population value is zero when the variables in question are statistically independent and non-zero otherwise. However, this is not sufficient to guarantee success. To see this, imagine using such a statistic $\hvphi$ on a data set containing many noisy linear relationships as well as a smaller number of strong sinusoidal relationships. The fact that $\hvphi$ is a measure of dependence guarantees us that, given sufficient sample size, all of these relationships will receive non-trivial scores. Unfortunately though, it tells us nothing about how those non-trivial scores will compare to each other. For example, $\hvphi$ could systematically assign higher scores to linear relationships than to sinusoidal relationships. If that is the case, then when we rank relationships by $\hvphi$ the noisy linear relationships may crowd out the sinusoidal relationships from the top of the list. Since we can only manually examine a relatively small number of relationships from the top of the list, we may therefore miss the sinusoidal relationships even though they are strong.

If our goal were simply to detect as many relationships as possible, then the measure of dependence $\hvphi$ would perform well to the extent that its associated independence test has good power. But a high-dimensional data set may contain a very large number of non-trivial relationships, some strong and others weak, and a list of all of them may be too large to allow for manual follow-up of each identified relationship \cite{reshef2015equitability,emilsson2008genetics}. Thus, in the exploration of large data sets, our goal is often not only to detect as many of the non-trivial associations in the data set as possible, but also to rank them by some notion of strength. For this task, deviation from independence can be too weak a search criterion.

One framework to address this challenge utilizes a property called {\em equitability}. Loosely, an equitable measure of dependence is one that gives similar scores to equally noisy relationships of different types \citep{MINE}. This definition is formalized in \citet{reshef2015equitability} and shown there to be equivalent to power against a range of null hypotheses corresponding to different relationship strengths rather than the single null hypothesis of statistical independence (i.e., zero relationship strength). While the general concept of equitability is quite broad, one intuitive and natural instantiation is that, when used on functional relationships, the value of an equitable measure of dependence should reflect the coefficient of determination ($R^2$) with respect to the generating function with as weak a dependence as possible on the particular function in question.

Equitability is a difficult property to achieve, and most measures of dependence do not have high equitability on functional relationships. (This is understandable, as they are not designed with that goal in mind.)  One statistic that has shown good equitability on functional relationships is the maximal information coefficient ($\MIC$) \citep{MINE}. In \citet{reshef2015estimating} a new, efficiently computable, consistent estimator of the population $\MIC$, called $\MICestE$, is introduced, along with a related measure of dependence called the total information coefficient $\TICestE$, which is essentially free to compute when $\MICestE$ is computed.

In this paper, we demonstrate how the theoretical advances of \citet{reshef2015equitability,reshef2015estimating} translate into practical benefits via extensive empirical analyses, under a wide range of settings, of the equitability, power, and runtime of $\MICestE$, $\TICestE$, and several leading measures of dependence: $\MIC$ \citep{MINE}, distance correlation \citep{szekely2009brownian,szekely2007measuring}, mutual information estimation \citep{Kraskov}, maximal correlation \citep{renyi1959measures, breiman1985estimating}, the randomized dependence coefficient ($\RDC$) \citep{lopez2013randomized}, the Heller-Heller-Gorfine distance ($\HHG$) \citep{heller2013consistent}, $\DDP$ \citep{heller2014consistent}, and the Hilbert-Schmidt Independence Criterion ($\HSIC$) \citep{gretton2005measuring,gretton2008kernel,gretton2012kernel}. Throughout our analyses, we show how the theoretical framework of \citet{reshef2015equitability} can be used to rigorously quantify equitability in practice.

Our analyses yield four main conclusions. First, with regard to equitability, they show that estimation of the population $\MIC$ via $\MICestE$ is more equitable than other methods across the majority (32 out of 36) of the settings of noise/marginal distributions and sample size that we tested. (In the remaining four settings, the Kraskov mutual information estimator outperforms $\MICestE$.)

The second conclusion we draw is that the total information coefficient $\TICestE$ achieves overall statistical power against independence that is state-of-the-art. State-of-the-art power against independence is also achieved by Heller and Gorfine's $\DDP$, which outperforms $\TICestE$ by some metrics and is outperformed by $\TICestE$ in others. The power of $\TICestE$ is high not just overall, but also on each individual alternative hypothesis relationship type we examined, meaning that we did not identify any one relationship type that $\TICestE$ is especially poorly suited for detecting.

The third conclusion is that the power against independence of $\MICestE$, the new estimator of the population $\MIC$, is competitive with other state-of-the-art techniques, albeit with a different setting of its parameter $\alpha$ than the one that confers good equitability. This observation leads us to characterize a power-equitability trade-off that is captured by this parameter and appears consistent with the theory of equitability developed in \citet{reshef2015equitability} together with ``no free lunch" considerations.

Our final conclusion concerns runtime. We find that $\MICestE$ and $\TICestE$ are as fast as or faster than most other methods tested. Even at a sample size of $n=5,000$, running $\MICestE$/$\TICestE$ on all variable pairs in a $1,000$-variable data set using a $100$-node cluster, with parameters that yield state-of-the-art power against independence and near-optimal equitability, takes just $8.1$ minutes. Moreover, once either $\MICestE$ or $\TICestE$ is computed, the other can be computed trivially.

Taken together, our results suggest that $\MICestE$ can be efficiently used in conjunction with $\TICestE$ to achieve a useful mix of power against independence (by filtering results using $\TICestE$) and equitability (by using $\MICestE$ on the remaining variable pairs) when exploring a data set.

Together, this paper, \citet{reshef2015equitability}, and \citet{reshef2015estimating} have three primary objectives. The first is to formalize the theory behind both equitability and the maximal information coefficient. The second is to introduce and analyze a new estimator of the population $\MIC$ as well as a new measure of dependence called the total information coefficient. The third is to provide an extensive comparison of the performance of a set of state-of-the-art measures of dependence in a wide range of settings in terms of equitability, power against independence, and runtime.  While this paper is focused primarily on the performance comparison, providing direct and in-depth comparisons to existing methods, we hope these papers together expand the use of both this framework for data analysis and the existing algorithms.

The rest of this paper is organized as follows. In Section~\ref{sec:prelim} we cover preliminaries, in Section~\ref{sec:equitabilityreview} we give a brief review of equitability, in Section~\ref{sec:equitabilityAnalysis} we analyze the equitability of the methods in question, in Section~\ref{sec:power} we analyze their power against independence, in Section~\ref{sec:power_equitability_tradeoff} we characterize the tradeoff between power against independence and equitability, in Section~\ref{sec:runtime} we analyze runtime, and in Section~\ref{sec:discussion} we offer a concluding discussion.

\section{Preliminaries}
\label{sec:prelim}
As we extensively analyze several statistics introduced in \citet{MINE} and \citet{reshef2015estimating}, we start by reviewing the definitions of those statistics and related objects. The informed reader may skip this section and refer to it as needed.

\subsection{Overview and notation}
The statistics we present here are two estimators of the maximal information coefficient, as well as the total information coefficient. For all of these statistics, we have a sample from the distribution of some two-dimensional random variable $(X,Y)$. The goal in estimating the maximal information coefficient is to provide a score in the form of a number between 0 and 1 that quantifies the strength of the relationship between $X$ and $Y$ in an equitable way (see Section~\ref{sec:equitabilityreview} for a review of equitability). The goal in computing the total information coefficient is to obtain a statistic for testing for the presence or absence of statistical independence between $X$ and $Y$.

For all statistics, we use the following notational conventions. Let $G$ be a finite grid drawn on the Euclidean plane. Given a point $(x,y) \in \R^2$, we define the function $\mbox{row}_G(y)$ to be the row of $G$ containing $y$ and we define $\mbox{col}_G(x)$ analogously. For a pair $(X,Y)$ of jointly distributed random variables, we write $(X,Y)|_G$ to denote the discrete random variable $\left( \mbox{col}_G(X), \mbox{row}_G(Y) \right)$. For natural numbers $k$ and $\ell$, we use $G(k, \ell)$ to denote the set of all $k$-by-$\ell$ grids (possibly with empty rows/columns). Given a finite sample $D$ from the distribution of $(X,Y)$, we use $D$ to refer both to the set of points in the sample as well as to a point chosen uniformly at random from $D$. In the latter case, it then makes sense to talk about, e.g., $D|_G$ and $I(D|_G)$.

\subsection{The maximal information coefficient}
The maximal information coefficient (MIC) is a statistic introduced in \citet{MINE} as a way to achieve good equitability on a wide range of relationship types. In \citet{reshef2015estimating}, the population value of this statistic is computed and a new estimator of that population value is given. Here we define all three of these objects.

\subsubsection{The population MIC}
We begin by defining the population value of MIC, which we denote by $\popMIC$. To define this quantity, we must first define an object called the {\em population characteristic matrix}. The population MIC will then be the supremum of this matrix.

\begin{definition}[\citet{reshef2015estimating}]
\label{def:charmatrix}
Let $(X,Y)$ be jointly distributed random variables. Let
\[ I^*((X, Y), k, \ell) = \max_{G \in G(k,\ell)} I((X,Y)|_G) \]
where $I$ represents the mutual information.
The {\em population characteristic matrix} of $(X,Y)$, denoted by $M(X,Y)$, is defined by
\[ M(X,Y)_{k,\ell} = \frac{I^*((X,Y),k,\ell)}{\log \min \{k, \ell\}} \]
for $k, \ell > 1$.
\end{definition}
For more on mutual information see, e.g., \citet{Cover2006} and \citet{csiszar2004information}).

The characteristic matrix is so named because in \citet{MINE} it was hypothesized that this matrix takes on different ``shapes" that are characteristic of different relationship types, so that different properties of the matrix may correspond to different properties of relationships. One such property was the maximal value of the matrix. This is called the maximal information coefficient (MIC), and its corresponding population quantity is defined below.

\begin{definition}[\citet{reshef2015estimating}]
Let $(X,Y)$ be jointly distributed random variables. The {\em population maximal information coefficient} ($\popMIC$) of $(X, Y)$ is defined by
\[ \popMIC(X, Y) = \sup M(X,Y) .\]
\end{definition}

The population MIC has several alternate characterizations, both as a canonical smoothing of mutual information and as the supremum of the boundary of the characteristic matrix. For more, see \citet{reshef2015estimating}.

\subsubsection{Estimators of $\popMIC$}
In this work we study two different estimators of the population MIC.

\paragraph{The first estimator: $\MIC$}
The first statistic we analyze is the original statistic introduced in \citet{MINE}, which estimates $\popMIC$ by first estimating each entry of the characteristic matrix until a sample size-dependent maximal grid resolution. This estimated characteristic matrix is called the {\em sample characteristic matrix} and is defined below.

\begin{definition}[\citet{MINE}]
\label{def:samplecharmat}
Let $D \subset \R^2$ be a set of ordered pairs. The {\em sample characteristic matrix} $\widehat{M}(D)$ of $D$ is defined by
\[
\widehat{M}(D)_{k,\ell} = \frac{I^*(D, k, \ell)}{ \log \min \{k, \ell \}} .
\]
\end{definition}

$\MIC$ is then the maximum of the sample characteristic matrix, subject to a sample size-dependent limit on the maximal allowed grid resolution.

\begin{definition}[\citet{MINE}]
Let $D \subset \R^2$ be a set of $n$ ordered pairs, and let $B : \Z^+ \rightarrow \Z^+$. We define
\[ \MIC_B(D) = \max_{k\ell \leq B(n)} \widehat{M}(D)_{k,\ell} . \]
\end{definition}

The statistic $\MIC$ is proven in \citet{reshef2015estimating} to be a consistent estimator of the population MIC, provided $\omega(1) < B(\cdot) \leq O(n^{1-\ep})$ for $\ep > 0$. However, it is not known how to efficiently compute the exact value of $\MIC$, and so in practice a heuristic dynamic-programming approximation algorithm is used.

\paragraph{The second estimator: $\MICestE$}
The second statistic we analyze is $\MICestE$, a statistic introduced in \citet{reshef2015estimating} and proven there to be a consistent estimator of $\popMIC$. In contrast to $\MIC$, it {\em is} known how to compute $\MICestE$ exactly in polynomial time (although in practice other, still more efficient statistics may nevertheless be used; see below). Rather than attempting to estimate any entries of the characteristic matrix, $\MICestE$ estimates a different matrix, the {\em equicharacteristic matrix}, whose supremum is the same as that of the characteristic matrix. Estimates of entries of this other matrix turn out to be both much easier to compute and sufficient for estimating $\popMIC$.

We first define the sample equicharacteristic matrix, along with a prerequisite definition.

\begin{definition}[\citet{reshef2015estimating}]
Let $(X,Y)$ be a pair of jointly distributed random variables. Define
\[ I^* \left( (X,Y), k, [\ell] \right) = \max_{G \in G(k, [\ell])} I \left( (X,Y)|_G \right) \]
where $G(k,[\ell])$ is the set of $k$-by-$\ell$ grids whose y-axis partition is an equipartition of size $\ell$. Define $I^*((X,Y), [k], \ell)$ analogously.

Define $I^{[*]}((X,Y), k, \ell)$ to equal $I^*((X,Y), k, [\ell])$ if $k \leq \ell$ and $I^*((X,Y), [k], \ell)$ otherwise.
\end{definition}

\begin{definition}[\citet{reshef2015estimating}]
Let $D \subset \R^2$ be a set of ordered pairs. The {\em sample equicharacteristic matrix} $\widehat{[M]}(D)$ of $D$ is defined by
\[
\widehat{[M]}(D)_{k,\ell} = \frac{I^{[*]}(D, k, \ell)}{ \log \min \{k, \ell \}} .
\]
\end{definition}

We can now define the second estimator, $\MICestE$.
\begin{definition}[\citet{reshef2015estimating}]
Let $D \subset \R^2$ be a set of $n$ ordered pairs, and let $B : \Z^+ \rightarrow \Z^+$. We define
\[ \MICestE_{,B}(D) = \max_{k\ell \leq B(n)} \widehat{[M]}(D)_{k,\ell} . \]
\end{definition}

$\MICestE$ can be computed using dynamic programming, resulting in a search procedure that takes time $O(n^2B(n)^2)$, which equals $O(n^{2+2\alpha})$ when $B(n) = n^\alpha$. In practice, however, this algorithm can be modified to include a parameter $c$ that controls the coarseness of the discretization of the grid-maximization search. The modified statistic remains a consistent estimator of $\popMIC$ and runs in time $O(c^2 B(n)^{5/2}) = O(c^2n^{5\alpha/2})$ \citep{reshef2015estimating}. In this work we use $\MICestE$ to refer both to the statistic as defined above and to the result of this modified algorithm. For more, see \citet{reshef2015estimating}.

\subsection{The total information coefficient}
\label{sec:TICdef}
While the maximal information coefficient aims to measure the strength of a relationship equitably, the total information coefficient (TIC), introduced in \citet{reshef2015estimating}, provides a way of testing for the presence or absence of statistical independence with good power and is a trivial side-product of the computation of the maximal information coefficient.

The intuition behind the total information coefficient is that while estimating $\popMIC$ has many advantages, this estimation involves taking a maximum over many estimates of entries of the characteristic or equicharacteristic matrix. Since the maximum of a set of random variables tends to become large as the number of variables grows, one can imagine that this procedure can lead to an unwanted positive bias in the case of statistical independence, when the population characteristic matrix equals 0, and a consequent reduction in power against independence.

To circumvent this problem, the total information coefficient is not the maximum but the {\em sum} of the entries of the matrix. Since this property of the matrix has better statistical properties, we might expect it to have a smaller bias in the case of statistical independence and therefore better power. Stated alternatively, if our only goal is to distinguish any dependence at all from complete noise, then disregarding all of the sample characteristic/equicharacteristic matrix except for its maximal value throws away useful signal, and the total information coefficient avoids this by summing all the entries.

\subsubsection{The statistic $\TICestE$}
The version of the total information coefficient studied in this work is analogous to the statistic $\MICestE$ presented above in that it proceeds via summation not of the sample characteristic matrix $\widehat{M}$ but rather of the sample equicharacteristic matrix $\widehat{[M]}$.

\begin{definition}
Let $D \subset \R^2$ be a set of $n$ ordered pairs. Given a function $B: \Z^+ \rightarrow \Z^+$, we define $\TICestE_{,B}(D)$ to be
\[ \TICestE_{,B}(D) = \sum_{k \ell \leq B(n)} \widehat{[M]}(D)_{k,\ell} \]
where $\widehat{[M]}(D)$ is the sample equicharacteristic matrix.
\end{definition}

In \citet{reshef2015estimating} it is proven that $\TICestE$ yields a consistent right-tailed independence test, provided $\omega(1) < B(n) \leq O(n^{1-\ep})$ for $\ep > 0$. As with $\MICestE$, there is an additional parameter $c$ that controls the coarseness of the discretization of the grid search when $\TICestE$ is computed. However, this does not affect the consistency of the corresponding independence test. See \citet{reshef2015estimating} for more detail.

\subsection{Summary of MIC and TIC-related statistics}
Table~\ref{tab:objectsDefined} lists the objects discussed in this section.

\begin{table}[ht!]
\centering
	\begin{tabular}{p{0.5in}p{3in}p{1.3in}}
	\textbf{Object} &
	\textbf{Description} &
	\textbf{Defined in} \\ \hline \\
	$\MIC$ &
	Statistic for quantifying relationship strength &
	\citet{MINE} \vspace{1\bigskipamount} \\ \hline \\
	 
	$\popMIC$ &
	Population value of $\MIC$ &
	\citet{reshef2015estimating} \vspace{1\bigskipamount} \\ \hline \\
	
	$\MICestE$ &
	Estimator of $\popMIC$ via equicharacteristic matrix &
	\citet{reshef2015estimating} \vspace{1\bigskipamount} \\ \hline \\
	
	$\TICestE$ &
	Statistic for testing for independence &
	\citet{reshef2015estimating} \vspace{1\bigskipamount} \\ \hline \\
	
\end{tabular}
\caption{Statistics and estimands related to the maximal and total information coefficients.} \label{tab:objectsDefined}
\end{table}

\section{A review of equitability}
\label{sec:equitabilityreview}
Equitability is a property of measures of dependence introduced in \citet{MINE} and formalized in \citet{reshef2015equitability} that is particularly useful in the context of data exploration. Because this paper analyzes the equitability of several leading measures of dependence, we first present here a review of the basic definitions of- and results about equitability from \citet{reshef2015equitability}.

There are two different ways to view equitability, each with its corresponding intuition. The first states roughly that an equitable measure of dependence ``give[s] similar scores to equally noisy relationships of different types"~\citep{MINE}. In this viewpoint, a highly equitable measure of dependence allows us notionally to find the ``strongest $K$" relationships in our data set for any $K$.

The second view of equitability is based on statistical power: an equitable measure of dependence provides good tests for distinguishing between relationships with different, potentially non-zero amounts of noise. In other words, instead of yielding tests that only reject a null hypothesis of independence (i.e., ``relationship strength = 0"), an equitable measure of dependence yields tests for rejecting null hypotheses of the form ``relationship strength $\leq x_0$" for all possible $x_0$. That is, a highly equitable measure of dependence allows us to find with high power all the relationships in our data set with ``strength at least $x_0$" for any $x_0$.

These two viewpoints are formalized and shown to be equivalent in \citet{reshef2015equitability}. We now summarize those formalizations as well as their equivalence, together with some examples and intuition.

\subsection{Defining equitability via power}
Let $\hvphi$ be some statistic. To be able to talk rigorously about the equitability of $\hvphi$, we must specify two things: a set $\Q$ of distributions on which we can state what we mean by relationship strength, and a corresponding function $\Phi : \Q \rightarrow [0,1]$ that computes that strength. The set $\Q$ is called the set of {\em standard relationships} and the function $\Phi$ is called the {\em property of interest}.

A natural setting to keep in mind is that $\Q$ is some diverse set of functional relationships with noise added and $\Phi$ is $R^2$, i.e., the coefficient of determination with respect to the generating function. We return to this example often as a way to build intuition.

We can now define equitability in terms of power against a broad class of null hypotheses.\footnote{We deviate here from \citet{reshef2015equitability} in that we use the term ``equitability" for arbitrary properties of interest $\Phi$, rather than using ``interpretability" in general and reserving ``equitability" for cases in which $\Phi$ specifically reflects some notion of relationship strength. We do this because in this paper $\Phi$ always reflects a notion of relationship strength. However, we note that the concepts and tools here can be readily applied even if this is not the case.}
\begin{definition}
Let $\hvphi$ be a statistic, let $\Q$ be a set of standard relationships, let $\Phi : \Q \rightarrow [0,1]$, and fix some $0 \leq \alpha \leq 1/2$. The statistic $\hvphi$ is {\em $1/d$-equitable} with respect to $\Phi$ with confidence $1-2\alpha$ if and only if for every $x_0, x_1 \in [0,1]$ satisfying $x_1 - x_0 > d$, there exists a right-tailed level-$\alpha$ test based on $\hvphi$ that can distinguish between $H_0 : \Phi(\mcZ) \leq x_0$ and $H_1 : \Phi(\mcZ) \geq x_1 $ with power at least $1-\alpha$.
\end{definition}

The smaller $d$ is the better, and consequently the best equitability that can be achieved is when $d=0$, and the statistic in question is $\infty$-equitable. This is called {\em perfect equitability}, and is generally discussed as a property of the population value of a statistic.

\begin{figure}
	\centering
	\includegraphics[clip=true, trim = 0.05in 1.25in 0in 6.7in, width=0.88\textwidth]{\pathToCommonFigs/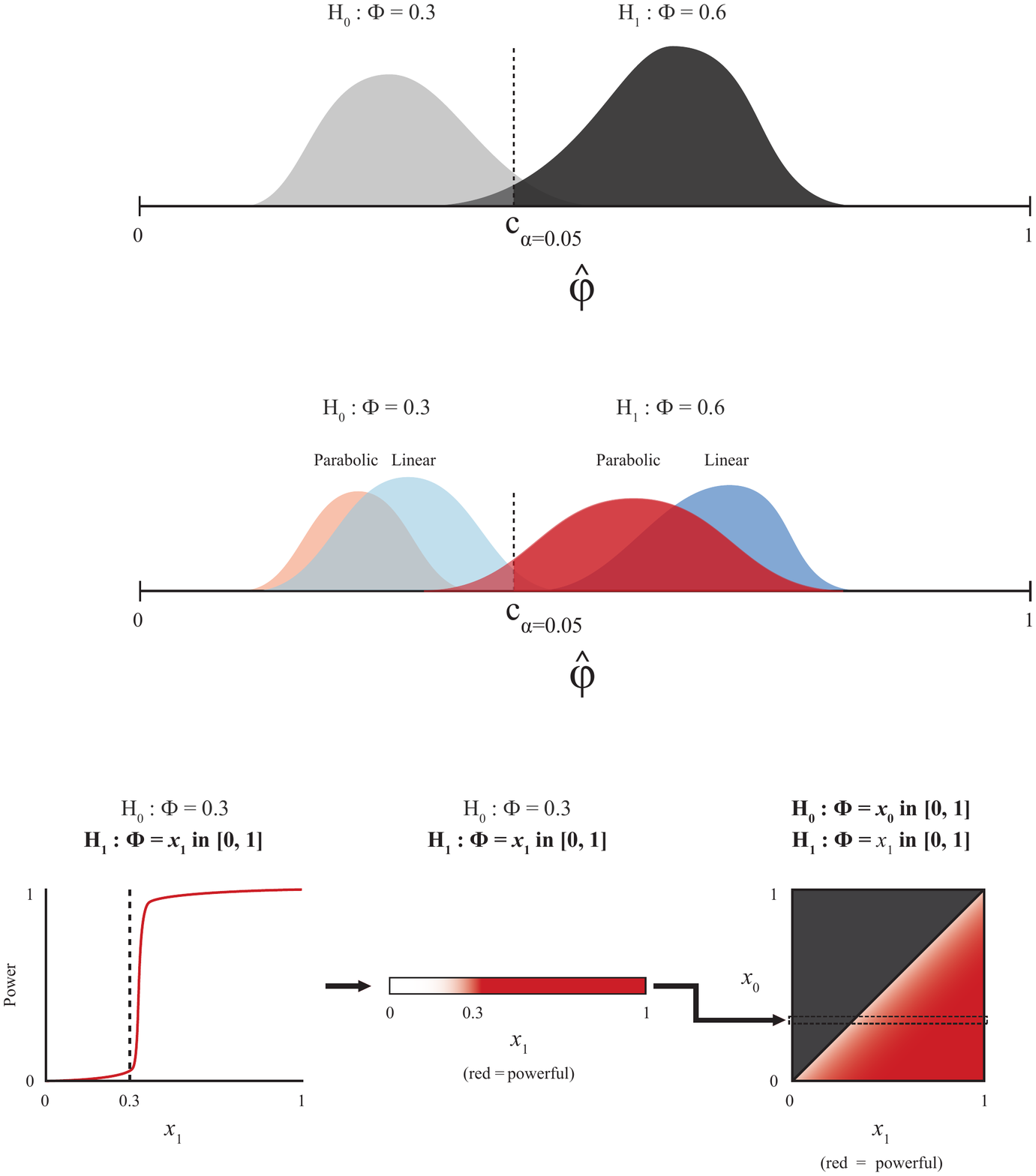}
	\begin{tabu} to 0.88\textwidth {X[c] X[c] X[c]} 
	(a) & (b) & \hspace{0.135in} (c)
	\end{tabu}
	\caption[Equitability as a generalization of power against independence]{
	    Equitability as a generalization of power against independence.
	    \figpart{a} The power function of a size-$\alpha$ right-tailed test based on a statistic $\hvphi$ with null hypothesis $H_0: \Phi = 0.3$. The curve shows the power of the test as a function of $x_1$, the value of $\Phi$ in the alternative hypothesis.
    	\figpart{b} The power function can be depicted instead as a heat map. 
    	\figpart{c} Instead of considering just one null hypothesis/critical value, we can consider a set of null hypotheses (with corresponding critical values) of the form $H_0: \Phi = x_0$ and plot each of the resulting power curves as a heat map. The result is a plot in which the intensity of the color in the coordinate $(x_1, x_0)$ corresponds to the power of a size-$\alpha$ right-tailed test based on $\hvphi$ at distinguishing $H_1 : \Phi = x_1$ from $H_0 : \Phi = x_0$. A $1/d$-equitable statistic is one for which this power surface attains the value $1-\alpha$ within distance $d$ of the diagonal along each row.}
	\label{fig:equitabilityAndPower}
\end{figure}

This definition of equitability is illustrated schematically in Figure~\ref{fig:equitabilityAndPower}. It implies that when $\Phi$ is 0 precisely in cases of statistical independence, equitability can be viewed as a generalization of power against statistical independence on $\Q$. Specifically, when we set $x_0 = 0$, a statistic being $1/d$-equitable means that that statistic yields a test that has good power against independence on any alternative hypothesis as extreme or more extreme than $H_1: \Phi = d$. In general, the definition says that a $1/d$-equitable statistic allows us to, given some threshold $x_0$ of relationship strength as measured by $\Phi$, successfully identify all the relationships in a data set with strength greater than $x_0 + d$. This may be important if our data set has many weak relationships and a smaller number of strong relationships that we would like to find.

As the formalization just presented makes clear, an analysis of equitability must differ from conventional analyses of power against independence in two ways. First, statistical independence represents only one null hypothesis, in contrast to the many null hypotheses against which equitability requires good power. Second, since in the setting of equitability the model $\Q$ will contain multiple distinct classes of relationship types (e.g., linear, exponential, etc.), the null and alternative hypotheses that must be analyzed are composite.

\subsection{Defining equitability via interpretability}
In addition to the view that defines equitability in terms of power, we can take an alternative approach that directly formalizes the intuition that an equitable statistic assigns similar scores to equally noisy relationships of different types. To do so, we must define two concepts, \textit{reliability} and \textit{interpretability}, which invoke acceptance regions and interval estimates, respectively. For clarity of exposition, we avoid using the term ``equitability" in the following, since we have already defined it previously. However, what we describe here as ``worst-case interpretability" will turn out to be equivalent to equitability.

We begin with the definition of reliability.
\begin{definition}[\citet{reshef2015equitability}]
Let $\hvphi : \R^{2n} \rightarrow [0,1]$ be a statistic, let $x, \alpha \in [0,1]$. The $\alpha$-reliable interval of $\hvphi$ at $x$, denoted by $\reliablestat{\alpha}{\hvphi}{x}$, is the smallest closed interval $A$ with the property that, for all $\mcZ \in \Q$ with $\Phi(\mcZ) = x$,
\[
\Pr{\hvphi(D) < \min A} < \alpha/2 \quad \mbox{and} \quad \Pr{\hvphi(D) > \max A} < \alpha/2
\]
where $D$ is a sample of size $n$ from $\mcZ$.

The statistic $\hvphi$ is {\em $1/d$-reliable} with respect to $\Phi$ on $\Q$ at $x$ with probability $1-\alpha$ if and only if the diameter of $\reliablestat{\alpha}{\hvphi}{x}$ is at most $d$.
\end{definition}

\begin{figure}
	\centering
	\begin{tabular}{rccc}
	\includegraphics[clip=true, trim = 0in 9.2in 8.35in 0in, height=0.17\textheight]{\pathToCommonFigs/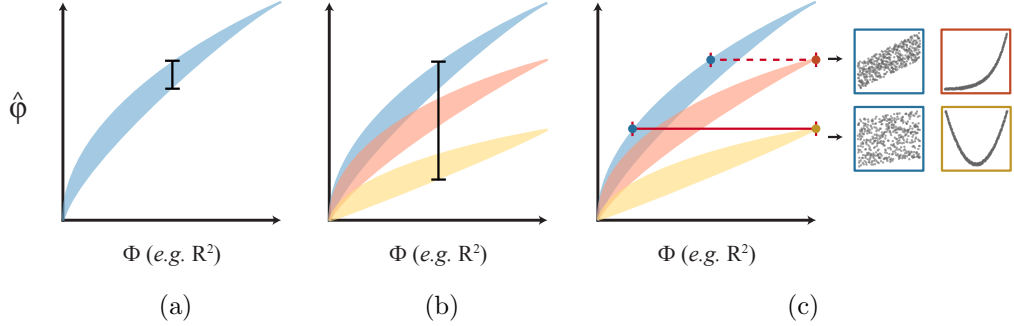}
	    &
	    \includegraphics[clip=true, trim = 2.07in 9.2in 5in 0in, height=0.17\textheight]{\pathToCommonFigs/WhatIsEquitability_v4.pdf}\quad
	    &
	    \includegraphics[clip=true, trim = 3.83in 9.2in 3.21in 0in, height=0.17\textheight]{\pathToCommonFigs/WhatIsEquitability_v4.pdf}\quad
	    &
	    \includegraphics[clip=true, trim = 5.6in 9.2in 0.3in 0in,    height=0.17\textheight]{\pathToCommonFigs/WhatIsEquitability_v4.pdf} \\
	    & (a) & (b) & (c)
	\end{tabular}
	\caption[A schematic illustration of interpretability/equitability]{
	    A schematic illustration of interpretability/equitability with three relationship types: linear (blue), exponential (red), and parabolic (yellow).
    	Here the property of interest ($\Phi$) is $R^2$ and the statistic in question ($\hvphi$) is the sample Pearson correlation coefficient $\hat\rho$.
    	\figpart{a} A plot of central intervals of the sampling distributions of $\hvphi = \hat{\rho}$ against $R^2(\mcZ)$ for $\mcZ \in \Q$, when $\Q$ consists only of linear relationships with varying amounts of added noise; one reliable interval is pictured. Since there is exactly one relationship in $\Q$ corresponding to each $R^2$ value, the reliable interval is simply a central interval of the relevant sampling distribution.
    	\figpart{b} The analogous plot in the case where $\Q$ contains noisy functional relationships ranging over three different functions: linear (blue), exponential (red), and parabolic (yellow). Now the reliable interval interval is the smallest interval containing all three of the relevant central intervals.
    	\figpart{c} The same plot, with interpretable intervals pictured. The interpretable interval at each value of $\hat{\rho}$ is composed of the $R^2$ values whose reliable intervals contain that value of $\hat{\rho}$. The shorter the interpretable intervals, the more interpretable/equitable the statistic. The worst-case interpretable interval is denoted by a solid red line; an additional interpretable interval is shown with a dashed red line. The thumbnails to the right of each interval show representative relationships from the endpoints of that interval, both of which have the same $\hat\rho$ but dramatically different values of $R^2$.}
	\label{fig:whatIsEquitability}
\end{figure}

The reliable interval at $x$ is an acceptance region for a size-$\alpha$ test of the null hypothesis $H_0 : \Phi = x$. This is a convex hull of central intervals of the sampling distributions of $\hvphi$ over all distributions $\mcZ \in \Phi^{-1}(\{x\})$. If there is only one $\mcZ$ such that $\Phi(\mcZ) = x$, then the reliable interval is simply a central interval of the sampling distribution of $\hvphi$ on $\mcZ$.

Figures~\ref{fig:whatIsEquitability}a and~\ref{fig:whatIsEquitability}b show schematic illustrations of reliable intervals in the case where $\Q$ is a set of noisy functional relationships, $\Phi = R^2$, and $\hvphi$ is the sample Pearson correlation coefficient. In Figure~\ref{fig:whatIsEquitability}a, the set $\Q$ contains only one relationship type: linear. Consequently, each possible value of $R^2$ has only one distribution $\mcZ \in \Q$ with that $R^2$. In this case, the reliable interval at that $R^2$ value is simply a central interval of the sampling distribution of the sample correlation. In Figure~\ref{fig:whatIsEquitability}b, the set $\Q$ contains not one but three relationship types: linear, exponential, and parabolic. This means that at every $R^2$ value there are three different distributions in $\mcZ \in \Q$ with that $R^2$, and consequently three different sampling distributions of the sample correlation. In this setting, the reliable interval at that $R^2$ value is the smallest interval that contains the union of the central intervals we constructed of those three sampling distributions.

Having defined the reliable interval as an acceptance region, we can now define the interpretable interval as an interval estimate of $\Phi$.

\begin{definition}[\citet{reshef2015equitability}]
Let $\hvphi : \R^{2n} \rightarrow [0,1]$ be a statistic, and let $y, \alpha \in [0,1]$. The $\alpha$-interpretable interval of $\hvphi$ at $y$, denoted by $\interpretablestat{\alpha}{\hvphi}{y}$, is the smallest closed interval containing the set
\[ \left\{ x \in [0,1] : y \in \reliablestat{\alpha}{\hvphi}{x} \right\} .\]

The statistic $\hvphi$ is {\em $1/d$-interpretable} with respect to $\Phi$ on $\Q$ at $y$ with confidence $1-\alpha$ if and only if the diameter of $\interpretablestat{\alpha}{\hvphi}{y}$ is at most $d$.
\end{definition}

Figure~\ref{fig:whatIsEquitability}c shows schematic illustrations of two different interpretable intervals in the setting discussed above, in which $\Q$ is a set of noisy functional relationships with three different function types (linear, exponential, parabolic), $\Phi = R^2$, and $\hvphi$ is the sample Pearson correlation coefficient.

When we are discussing the interpretability or reliability of a statistic, we need to speak about more than one $x$ or $y$ value at a time. There are many potential ways to do this. Here we limit ourselves to two basic ones.
\begin{definition}[\citet{reshef2015equitability}]
A measure of dependence is {\em worst-case} $1/d$-reliable (resp. interpretable) if it is $1/d$-reliable (resp. interpretable) at all $x$ (resp. $y$) $\in [0,1]$.

A measure of dependence is {\em average-case} $1/d$-reliable (resp. interpretable) if its reliability (resp. interpretability), averaged over all $x$ (resp. $y$) $\in [0,1]$, is at least $1/d$.
\end{definition}

Here and throughout, we use ``worst-case" to refer to the worst-seen performance, as opposed to a proven bound, and we use ``interpretability" with no qualifier to refer to worst-case interpretability.

To gain some intuition for the definition of interpretability, let us consider what values $d$ can take. The lowest possible interpretability happens when one of the interpretable intervals has size 1. In this case, the (worst-case) interpretability of the statistic is 1 as well. In the best case, when all interpretable intervals of a statistic are of size 0, the interpretability is $\infty$, and the statistic is said to be {\em perfectly interpretable}. (As before, the perfect case is only expected to arise, if at all, as a property of the population value of the statistic.)

To complete our example, let us find the worst-case interpretability of the sample correlation coefficient in the example of noisy functional relationships depicted in Figure~\ref{fig:whatIsEquitability}c. To do this, we locate the widest interpretable interval in the figure; this happens to be the lower of the two intervals pictured. If the length of this interval is $d$, the sample Pearson correlation coefficient is worst-case $1/d$-interpretable with respect to $R^2$ on our set $\Q$. Thus, the shorter the interpretable intervals, the more interpretable the statistic.

\subsection{The equivalence of the two formalizations}
\label{sec:equitabilityAsGeneralizationOfPower}
It turns out that equitability and worst-case interpretability as defined above are equivalent under modest assumptions~\citep{reshef2015equitability}. We state this result below.

\begin{thm}[\citet{reshef2015equitability}]
Let $\Q$ be a set of standard relationships, let $\Phi : \Q \rightarrow [0,1]$, and let $0 < \alpha < 1/2$. Let $\hvphi$ be a statistic with the property that $\max \reliablestat{\alpha}{\hvphi}{x}$ is a strictly increasing function of $x$. Then for all $d > 0$, the following are equivalent.
\begin{enumerate}	
\item $\hvphi$ is $1/d$-equitable with respect to $\Phi$ with confidence $1-2\alpha$.
\item $\hvphi$ is worst-case $1/d$-interpretable with respect to $\Phi$ with confidence $1-\alpha$.
\end{enumerate}
\end{thm}

This result can be interpreted in two ways. One interpretation is that a statistic that allows us to approximately rank the relationships in a data set by strength as measured by $\Phi$ will also allow us, for any $x_0$, to find all the relationships in the data set that have strength at least $x_0$ as measured by $\Phi$, and vice versa. Another interpretation arises if $\Phi$ reflects relationship strength, in particular if $\Phi=0$ corresponds to the relationships in $\Q$ exhibiting statistical independence. If this is the case, then the above theorem tells us that equitability is a generalization of power against statistical independence on $\Q$.

This is good news and bad news. On the one hand, it provides a link between equitability and power and clarifies the relationship between the two. On the other hand, it shows that equitability \--- by virtue of being stronger than power against independence \--- will also be more difficult to achieve, as it requires simultaneously attaining power against a much larger set of null hypotheses. This hints at a trade-off between equitability and power against independence for which we provide empirical evidence in Section~\ref{sec:power_equitability_tradeoff}.

\subsection{Equitability on functional relationships}
So far we have discussed equitability in general, conceptual terms, and it has many different concrete interpretations depending on the choice of $\Phi$ and $\Q$. We define here a concrete instantiation of equitability on functional relationships that is used throughout this paper. To do this, we first must state what we mean by ``functional relationship".
\begin{definition}[\citet{reshef2015equitability}]
A random variable distributed over $\R^2$ is called a {\em noisy functional relationship} if and only if it can be written in the form $(X + \ep, f(X) + \ep')$ where $f : [0,1] \rightarrow \R$, $X$ is a random variable distributed over $[0,1]$, and $\ep$ and $\ep'$ are (possibly trivial) random variables. We denote the set of all noisy functional relationships by $\F$.
\end{definition}

Equitability on functional relationships in the sense of \citet{MINE} and \citet{reshef2015equitability} now just amounts to the use of $R^2$ as the property of interest.
\begin{definition}[\citet{reshef2015equitability}]
Let $\Q \subset \F$ be a set of noisy functional relationships. A measure of dependence is worst-case (resp. average-case) {\em $1/d$-equitable} on $\Q$ if it is worst-case (resp. average case) $1/d$-equitable with respect to $R^2$ on $\Q$.
\end{definition}

In this paper we often abuse terminology by simply writing ``equitability" to mean equitability with respect to $R^2$ on various sets of functional relationships as defined above. Alternative definitions of this concept with other sets $\Q$ and functions $\Phi$ have been proposed. These are discussed in detail in \citet{reshef2015equitability}.

\subsection{Equitability: an example}
\label{sec:equitabilityExample}
\begin{figure}
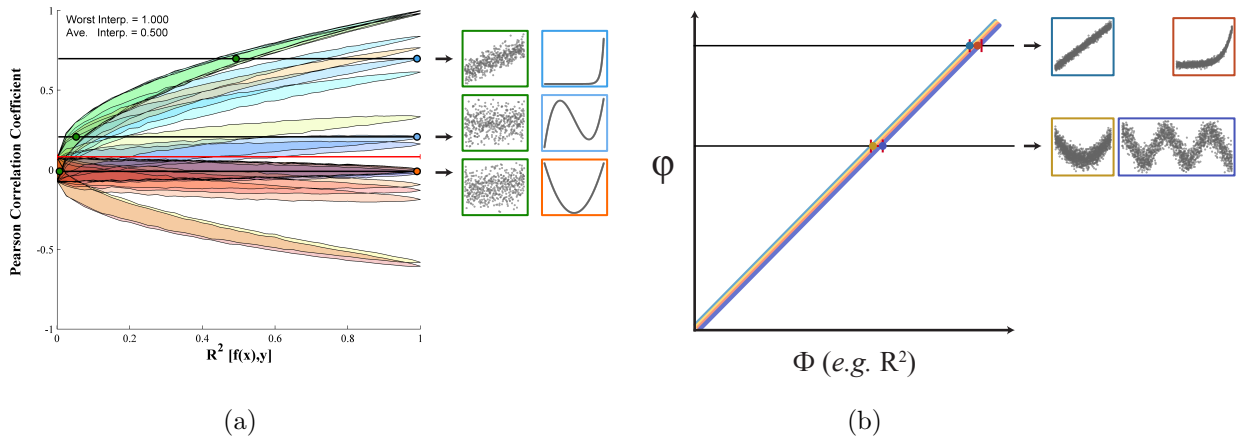

	\centering
	\begin{tabular}{@{}cc@{}} 
	    \includegraphics[clip=true, trim = 0in 6in 4.32in 2in, width=0.49\textwidth]{\pathToCommonFigs/WhatIsEquitability_v4.pdf}
	    &
	    \includegraphics[clip=true, trim = 4.41in 6in 0in 2in, width=0.48\textwidth]{\pathToCommonFigs/WhatIsEquitability_v4.pdf} \\
	    (a)\hspace{0.7in} & (b)\hspace{0.8in}
	\end{tabular}
	\caption[Examples of equitable and non-equitable behavior]{
	    Examples of equitable and non-equitable behavior on a set of noisy functional relationships. (Reproduced from \citet{reshef2015equitability}.)
    	\figpart{a} The equitability with respect to $R^2$ of the sample Pearson correlation coefficient $\hat\rho$ over the set $\Q$ of relationships described in Section~\ref{sec:equitabilityExample}, with $n=500$. Each shaded region is an estimated $90\%$ central interval of the sampling distribution of $\hat\rho$ for a given relationship at a given noise level. The fact that the interpretable intervals of $\hat\rho$ are large indicates that a given $\hat\rho$ value could correspond to relationships with very different $R^2$ values. This is illustrated by the pairs of thumbnails corresponding to relationships with the same $\hat\rho$ but different $R^2$ values. The largest interpretable interval is indicated by a red line. Because it has width 1, the worst-case equitability with respect to $R^2$ in this case is 1, the lowest possible.
    	\figpart{b} An illustration of a hypothetical measure of dependence that achieves \textit{perfect equitability} in the large-sample limit. Here, the population quantity $\varphi$ depends only on the $R^2$ of the relationships and increases monotonically with $R^2$. Thus, $\varphi$ can be used as a proxy for $R^2$ on $\Q$ with no loss. Thumbnails are shown for sample relationships that receive the same $\varphi$ score, which corresponds to the fact that they have equal $R^2$ scores.
	}
	\label{fig:equitabilityExample}
\end{figure}

Using the framework reviewed here, Figure~\ref{fig:equitabilityExample}a demonstrates how one might analyze the equitability of a statistic in practice from the standpoint of interpretable intervals. We take as an example the sample Pearson correlation coefficient ($\hat\rho$). This statistic is not a measure of dependence in the sense that its population value can be zero even in cases of non-trivial dependence. However, we analyze it here due to its widespread familiarity and the intuitiveness of its scores.

In this example, as before, our property of interest will be $\Phi = R^2$. The set of standard relationships $\Q$ will be a set of noisy functional relationships of the form $(X + \ep, f(X) + \ep'_\sigma)$ with $\ep = 0$, $\ep'_\sigma \sim \mathcal{N}(0, \sigma^2)$, and $f$ ranging over the functions in Table~\ref{table:fctSuiteEquitability}.

To analyze the equitability of $\hat\rho$, we generate, for 41 different noise levels $\sigma$ and for every function $f$ in our set, 500 samples from the relationship $Z=(X, f(X) + \ep'_\sigma)$ with a sample size of $n=500$. Using these, we estimate the 5th and 95th percentiles of the sampling distribution of $\hat\rho$ on $Z$. These allow us to estimate the reliable interval at the value of $R^2$ corresponding to each noise level. The reliable intervals then enable us to construct interpretable intervals, and our estimate of the equitability is then the reciprocal of the length of the longest interpretable interval.

The fact that the interpretable intervals at many values of $\hat\rho$ are large indicates that a given value of $\hat\rho$ could correspond to samples from relationships of different types that have very different $R^2$ values. This is illustrated by the pairs of thumbnails corresponding to relationships that received the same $\hat\rho$ but have different amounts of noise. This means that $\hat\rho$ is not very interpretable with respect to $R^2$ on this set $\Q$ and is thus said to have poor equitability with respect to $R^2$ on $\Q$. As a contrast, Figure~\ref{fig:equitabilityExample}b contains a hypothetical illustration of the notion of \textit{perfect equitability}, which would require that all the interpretable intervals be of size $0$.

Of course, equitability is a function not only of the method in question but also of the standard relationships and the property of interest. For instance, while $\hat\rho$ has poor equitability with respect to $R^2$ on the $\Q$ above, it is (trivially) asymptotically {\em perfectly} equitable with respect to the correlation on the set $\Q$ of bivariate normals.

\section{Equitability analysis}
\label{sec:equitabilityAnalysis}
Having reviewed equitability and how to quantify it, we turn to evaluating the equitability of $\MICestE$ and several other leading measure of dependence.  We begin by quantifying the equitability of each measure of dependence using interpretable intervals.  This is followed by an alternate visualization of the equitability of each measure of dependence using conventional power analysis via the connection described in the previous section.

\subsection{Setting up the analysis}
\label{sec:equitability_setting_up}
\subsubsection{Choice of methods to analyze}
The set of existing measures of dependence is too large for us to analyze exhaustively, even in a paper that aims to be comprehensive. We therefore strive to include in our analysis a collection of methods that is representative of the broad approaches prevalent in the field today.

\paragraph{Grid-based methods}
The methods based on the maximal information coefficient and the total information coefficient can be viewed as exploring the space of possible grids that can be drawn on the sampled data, assigning a score to each grid via some metric, and then aggregating the scores. For $\MIC$ \citep{MINE}, the metric is a normalized mutual information score and the aggregation is a supremum. $\MICestE$ \citep{reshef2015estimating} is similar except it explores a more restricted set of grids. $\TICestE$ \citep{reshef2015estimating} is like $\MICestE$ except it aggregates by summation.

We also include other recent grid-based methods introduced since the maximal information coefficient \citep{MINE}. $\HHG$ \citep{heller2013consistent} uses Pearson's $\chi^2$ test statistic as its score, explores a set of two-by-two grids defined by individual data points, and aggregates by summation. Though similar to Hoeffding's $D$ \citep{hoeffding1948non} in that it considers only two-by-two grids, it differs in the use of the $\chi^2$ test statistic.  $\DDP$ \citep{heller2014consistent} explores a larger set of grids defined by subsets of the data points, uses non-normalized mutual information as its score, and also aggregates by summation.\footnote{There are other variations on these statistics presented in \citet{heller2013consistent, heller2014consistent}. However, we omit those results as they were generally similar or worse than the ones we display.} Another notable grid-based method introduced recently is dynamic slicing \citep{jiang2014non}, which like $\MIC$ explores all possible grids and aggregates by maximization, but uses as its score a version of mutual information that is regularized according to a prior on the space of possible grids. We did not include dynamic slicing in our comparison, however, because it is formulated only for performing a $k$-sample test whereas our focus here is on measuring dependence between two continuous random variables.

\paragraph{Mutual information estimation}
Since many of the grid-based methods we consider either use some form of mutual information as their score or have variants that do, we also included a standard mutual information estimator introduced by Kraskov~\citep{Kraskov}. This estimator was compared against $\MIC$ in previous work~\citep{MINE,reshef2013equitability,kinney2014equitability,reshef2014comment}, but those comparisons were more limited in scope and did not include $\MICestE$. (For convenience, in this work we represent the estimated mutual information values in terms of the squared Linfoot correlation~\cite{speed2011correlation, linfoot1957informational}, defined by $L^2(X,Y) = 1-2^{-2I(X,Y)}$, which takes values in $[0,1]$.)

\paragraph{Distance/kernel-based statistics}
We include distance correlation ($\dCor$) \citep{szekely2009brownian}, an analogue of the Pearson correlation coefficient that is defined using a different notion of covariance that uses pairwise distances between points. In addition, we include the Hilbert-Schmidt Information Criterion ($\HSIC$) \citep{gretton2005measuring, gretton2008kernel}, a more general statistic defined on reproducing kernel Hilbert spaces of which $\dCor$ is a special case~\citep{sejdinovic2013equivalence}.

\paragraph{Correlation-based methods}
As an intuitive benchmark for the reader, we include the Pearson correlation coefficient ($\rho$). However, there are many successful tools that use $\rho$ after computing a non-linear transformation of the data. We include perhaps the best-known one, maximal correlation \citep{renyi1959measures}, which given random variables $X$ and $Y$ searches for arbitrary measurable functions $f$ and $g$ such that $\rho(f(X), g(Y))$ is maximized. There is no known algorithm for finding the optimal $f$ and $g$ in general, but the (approximate) method of alternating conditional expectations \citep{breiman1985estimating} is widely used and we use it here as well. We also include a more recent related method, the randomized dependence coefficient \citep{lopez2013randomized}, which applies many random transformations to $X$ and $Y$ and then searches for the linear combinations of the transformed features that maximize the correlation.

\subsubsection{Choice of $\Q$, $\Phi$, and sample sizes}
In an ideal world, when assessing equitability in a specific instance, we would know the true underlying model $\Q$ governing the relationships in our data set.  Knowledge of $\Q$ would, for example, include information about the types of relationships present and the noise distribution (e.g., Gaussian, zero-mean, heteroscedastic, etc.). Of course, in reality we generally do not have this information and, to make matters worse, the results of an equitability analysis may depend strongly on the choice of $\Q$. Thus, in evaluating the equitability of measures of dependence, it is important to aim for robustness: we would like to have a measure of dependence with good equitability over as many different relationship types as possible.

However, there is a central tension between the need to use as large a set $\Q$ as possible in order to assess robustness and the need to use a $\Q$ that is sufficiently small that a reasonable property of interest $\Phi$ can be defined for the relationships in $\Q$. To take an extreme example, setting $\Q$ to be the set of all bivariate relationships would certainly ensure that we do not leave any stone unturned, but at the same time it begs the original question of how one can measure relationship strength in such a general context.

For this reason, following \citet{MINE}, we choose to focus on noisy functional relationships since these represent a broad, easily definable class of relationships commonly found in practical applications that comes with an intuitive and natural measure of relationship strength: $R^2$, the coefficient of determination with respect to the generating function. To ensure robustness, we vary the relationships tested along as many dimensions as possible including relationship type, the type of noise added, marginal distributions, and sample size. 

We note here that our goal in this analysis is not to establish the equitability of any method across the {\em entire} set of noisy functional relationships. In fact, under some of the sampling/noise models we considered, there are functions whose inclusion leads to poor equitability across all methods. We therefore attempted to characterize as broad a set of functions as possible that still allowed for non-trivial equitability.

To that end, our analyses include some 16-21 different functional relationships (depending on noise model; see Appendix~\ref{app:functionsUsed}), each with increasing levels of additive Gaussian noise, considered under twelve different sampling/noise models, at four sample size regimes ($n=250, 500, 5000,$ and the infinite data limit).  Each of the 12 sampling/noise models $\Q$ is defined using a combination of an independent variable marginal distribution from the set
\[
\left\{
\begin{array}{ll}
\mbox{points sampled evenly along the curve described by $f(X)$} & \quad \left(E_{f(X)}\right) \\
\mbox{points sampled evenly along the $X$ range} & \quad \left(E_{X}\right) \\
\mbox{points sampled uniformly along the curve described by $f(X)$} & \quad \left(U_{f(X)}\right) \\
\mbox{points sampled uniformly along the $X$ range} & \quad \left(U_{X}\right)
\end{array}
\right\}
\]
and a noise distribution from the set
\[
\left\{
\begin{array}{l}
\mbox{normally distributed noise added to the dependent variable $\left( \mathcal{N}_y \right)$} \\
\mbox{normally distributed noise added to both variables $\left( \mathcal{N}_x, \mathcal{N}_y \right)$} \\
\mbox{normally distributed noise added to the independent variable $\left( \mathcal{N}_x \right)$}
\end{array}
\right\}.
\]

We refer to these noise models using abbreviations of the form $E_{f(X)}[\mathcal{N}_y]$, which would correspond to a model in which the independent variable is sampled evenly along the curve described by $f(X)$ and Gaussian noise is added only to the dependent coordinate. Appendix~\ref{app:functionsUsed} contains definitions of the functions used.

\subsubsection{Parameters of the analysis}
For each $\Q$, for each sample size $n$, we examine 41 different $R^2$ values evenly spaced in the unit interval. At each of these $R^2$ values, we generate $500$ independent realizations of a sample of size $n$ from each relationship in $\Q$ with the given $R^2$ value. These are used to estimate sampling distributions for $\hvphi$. (See Appendix~\ref{app:dataGen} for details regarding data generation.)

\subsubsection{Parameters of statistics tested}
Several of the methods tested are parametrized, including $\MICestE$, $\HSIC$, the Kraskov mutual information estimator, $\RDC$, and $\DDP$. For each of these methods, we performed a parameter sweep to assess the effect of parameter settings on equitability. In most cases, we found that parameter settings did not significantly affect equitability and so we present here results obtained with default parameters. For $\MICestE$ and the Kraskov mutual information estimator, however, parameter settings did affect equitability. Therefore, for these methods, we present for each sample size the best results across parameter values tested. Results for all parameter values tested can be found in the online supplement at \onlineSupplementLink. In Section~\ref{sec:choosingAlpha} we discuss guidelines for how to set parameters for $\MICestE$ more generally.

\subsubsection{Quantification of equitability}
The equitability of each measure of dependence is quantified using interpretable intervals, as discussed in Section~\ref{sec:equitabilityExample}. In the equitability plots presented, shaded regions denote central intervals containing $90\%$ probability mass of the sampling distribution of each measure of dependence at each $R^2$ value; these reliable intervals correspond to $0.05-$interpretable intervals. In general, we report both average-case and worst-case equitability in our analyses, and the interval plotted in red on each plot represent the worst-case $0.05-$interpretable interval for that plot. (The shorter the interval, the more equitable the statistic.)

\subsection{Results}
\begin{figure}
	\begin{minipage}[b][\textheight][t]{0.62\linewidth}
		\includegraphics[clip=true, trim = 0.15in 1in 0.7in 0.6in, width=\textwidth]{\pathToFigures/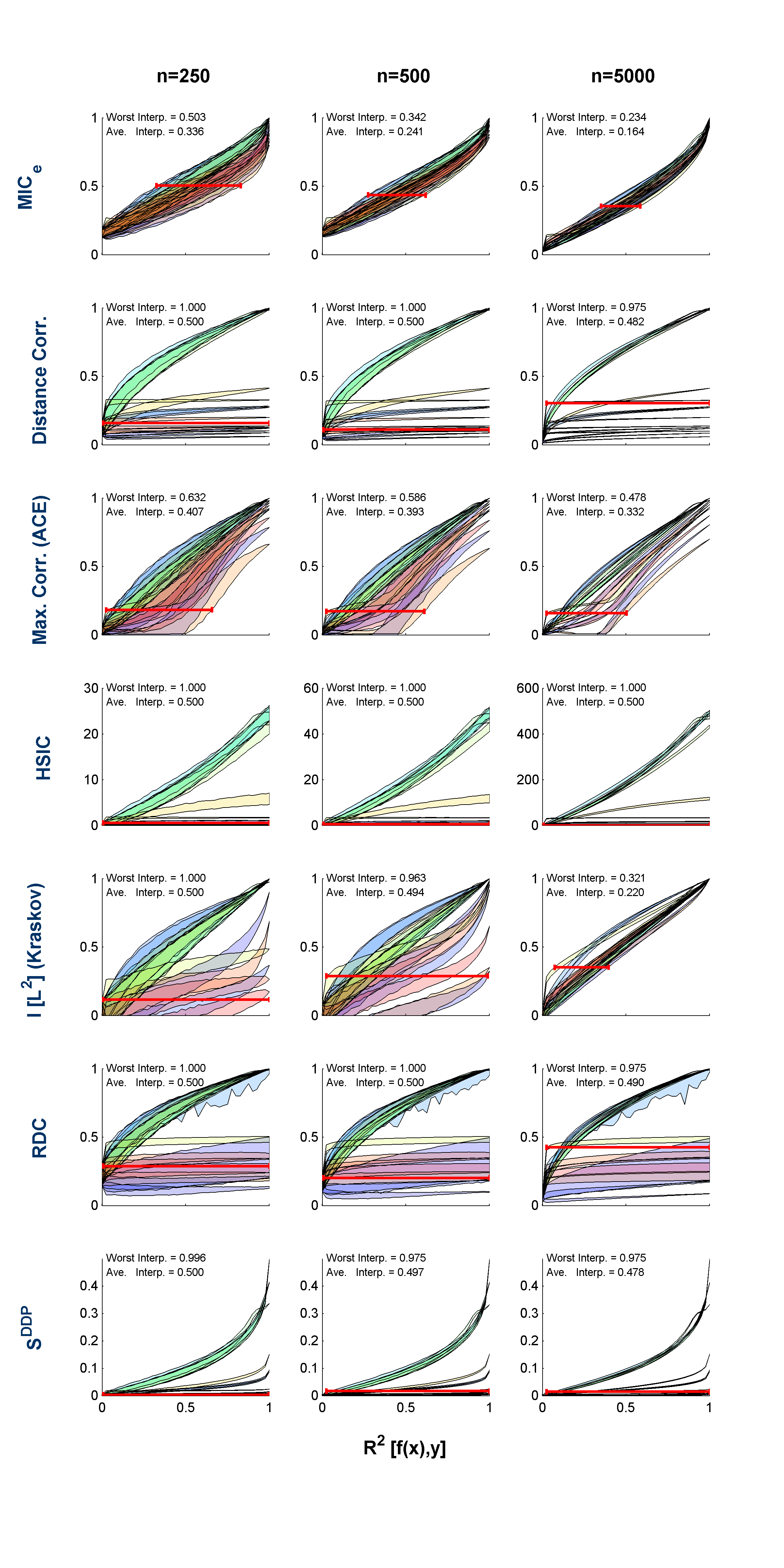}
	\end{minipage}
	\hfill
	\begin{minipage}[b][\textheight][t]{0.34\linewidth}
	    \includegraphics[clip=true, trim = 4.7in 0.25in 2.25in 0.35in, width=\textwidth]{\pathToFigures/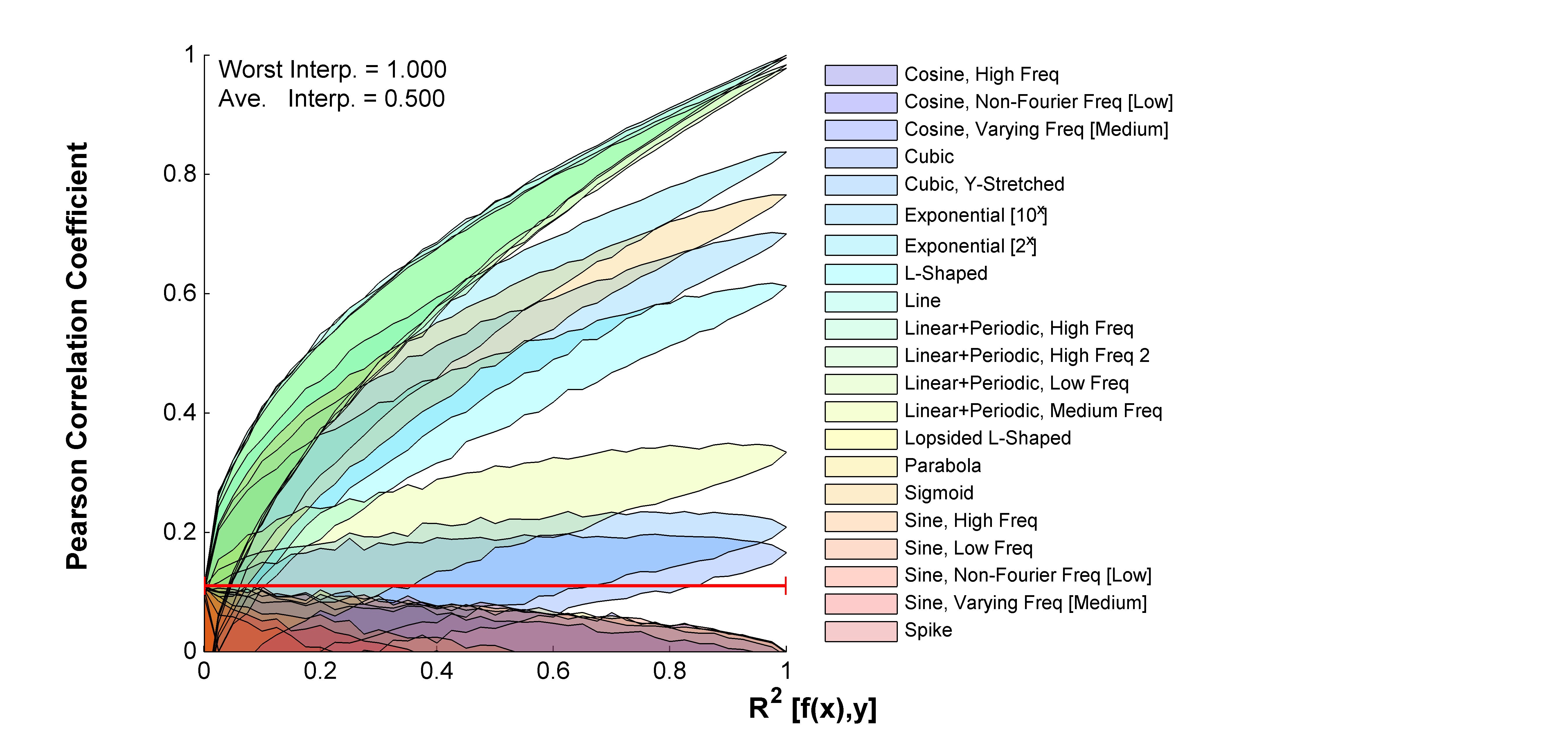}
		\caption[The equitability of measures of dependence on noisy functional relationships]{
		    The equitability of measures of dependence on a set $\Q$ of noisy functional relationships. \textit{[Narrower is more equitable.]}
		    The relationships take the form $(X+\ep, f(X)+\ep')$ where $\ep$ and $\ep'$ are i.i.d. normals of varying amplitude, and relationship strength is quantified by $\Phi = R^2$. The plots were constructed as described in Figure~\ref{fig:whatIsEquitability}. In each plot, the worst-case interpretable interval is indicated by a red line, and both the worst- and average-case equitability are listed. The fact that the worst-case interpretable intervals of $\MICestE$ are small indicates that a given $\MICestE$ score reflects the coefficient of determination ($R^2$) with respect to the generating function $f$ with a relatively weak dependence on the function $f$ in question. That is, $\MICestE$ has high equitability with respect to $\Phi=R^2$ for this choice of $\Q$.
		    Mutual information, estimated using the Kraskov estimator, is represented using the squared Linfoot correlation. For every parametrized statistic whose parameter meaningfully affects equitability, results are presented at each sample size using parameter settings that maximize equitability across all twelve of the noise/marginal distributions tested at that sample size.}
	    \label{fig:equitabilityAnalysis_evenCurve_XYNoise}
	\end{minipage}
\end{figure}

Figures~\ref{fig:equitabilityAnalysis_evenCurve_XYNoise} and~\ref{fig:equitabilityAnalysis_evenCurve_YNoise} demonstrate the equitability of $\MICestE$, distance correlation, maximal correlation, $\HSIC$, the Kraskov mutual information estimator, $\RDC$, and $\DDP$ for noise models $E_{f(X)}[\mathcal{N}_x, \mathcal{N}_y]$ and $E_{f(X)}[\mathcal{N}_y]$ at a range of sample sizes.  Results for all other noise models are presented in the supplemental materials, along with results for $\TICestE$, $\HHG$, and $\rho$.  Tables~\ref{table:equitability_worstCase} and~\ref{table:equitability_aveCase} summarize the worst-case and average-case equitability, respectively, for all measures of dependence across all models and sample sizes, as measured by $0.05-$interpretability intervals.

We offer here some discussion of the salient questions answered by these analyses.

\subsubsection{Comparing the equitability of $\MICestE$ and mutual information}
Given the connections between $\MICestE$ and mutual information, which are discussed in depth in~\citet{reshef2015estimating}, it is natural to ask whether the direct estimation of mutual information achieves a similar level of equitability to that of $\MICestE$. In general, among the variety of models and sample sizes tested, the answer appears to be `no', but we present a more detailed breakdown of the results below.

\begin{figure}[t!]
	\centering
	\includegraphics[clip=true, trim = 0.55in 0.2in 0.9in 0in, width=\textwidth]{\pathToFigures/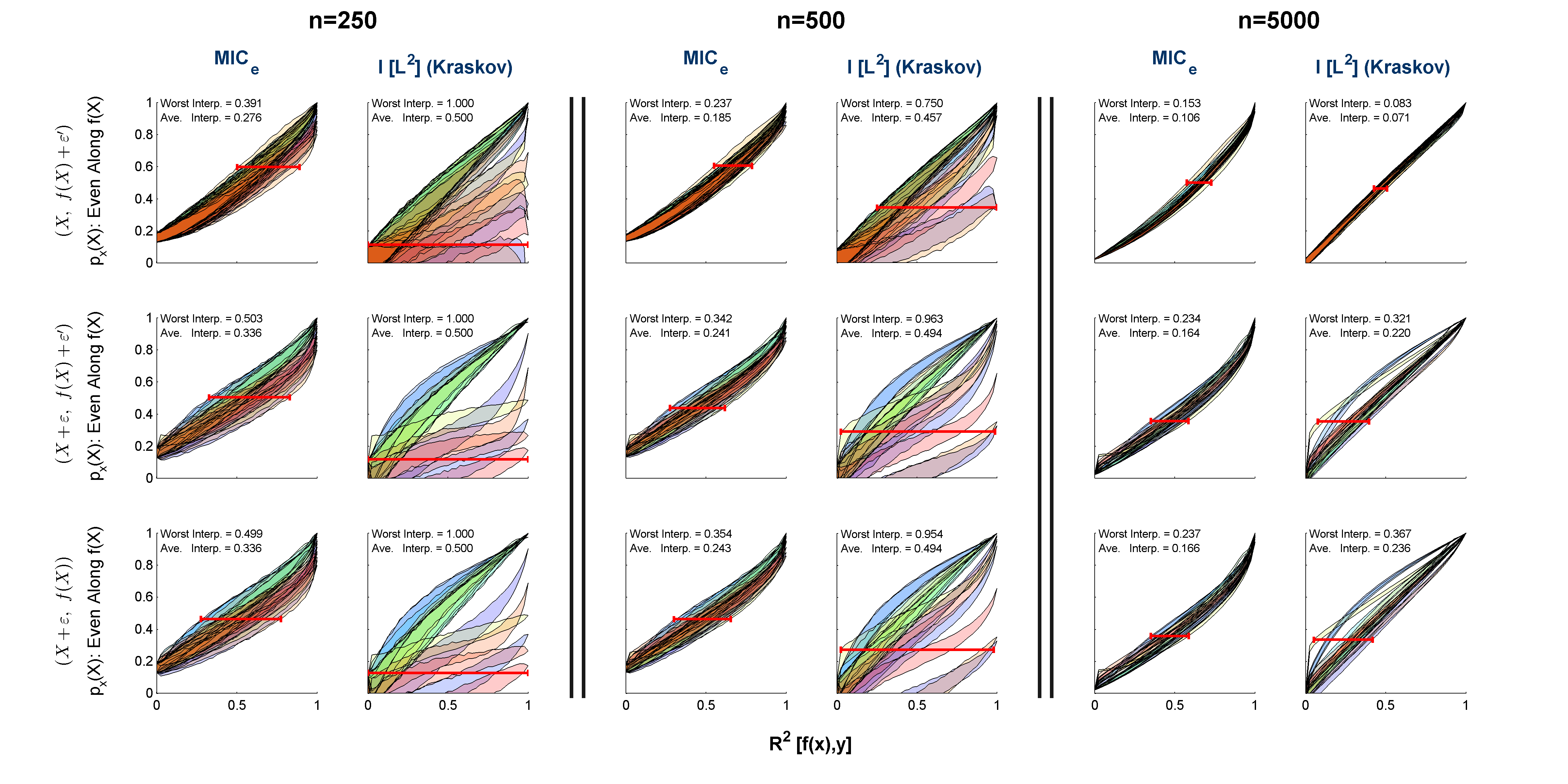}
	\caption[A comparison of the equitability of $\MICestE$ and mutual information estimation]{
	    A comparison of the equitability of $\MICestE$ and mutual information estimation under three noise models including the one in Figure~\ref{fig:equitabilityAnalysis_evenCurve_XYNoise}. \textit{[Narrower is more equitable.]}
	    Plots were constructed as in Figure~\ref{fig:whatIsEquitability}. In each plot, the worst-case interpretable interval is indicated by a red line, and both the worst- and average-case equitability are listed.
	    As in Figure~\ref{fig:equitabilityAnalysis_evenCurve_XYNoise}, results for both statistics are presented for each sample size using parameter settings that maximize equitability across all twelve of all twelve of the noise/marginal distributions tested at that sample size. Mutual information, estimated using the Kraskov estimator, is represented using the squared Linfoot correlation. 
	    While mutual information estimation using the Kraskov estimator is equitable at high sample size on some of the sets $\Q$ that were tested, the equitability of $\MICestE$ is more robust to noise model, independent variable marginal distribution, and limited sample size. For versions of this analysis using additional independent variable marginal distributions, see the supplemental materials.}
	\label{fig:equitability_MICvMI}
\end{figure}

\paragraph{Effect of model choice on equitability}
Figure~\ref{fig:equitability_MICvMI}, as well as Tables~\ref{table:equitability_worstCase} and~\ref{table:equitability_aveCase}, demonstrate the relative robustness of the equitability of $\MICestE$ to the choice of model $\Q$ compared to that of the Kraskov mutual information estimator.  At each sample size, the equitability of $\MICestE$ is fairly stable with respect to the variations in noise models and independent variable marginal distributions tested. On the other hand, while mutual information estimation sometimes has good equitability, it more often has poor equitability under the models tested. More specifically, mutual information estimation can be equitable in models that only contain noise added in the dependent coordinate, while $\MICestE$ performs equitably even outside this domain, such as in the case of models that include noise added to either or both the dependent and independent coordinates. The performance of mutual information estimation is also improved when the independent variable is stochastic rather than fixed, though this distinction never affects whether it outperforms $\MICestE$ or not.

\paragraph{Effect of sample size on equitability} Estimating mutual information from finite samples is a challenging problem that has inspired many non-trivial methods~\citep{paninski2003estimation,moon1995estimation,Kraskov}, and Tables~\ref{table:equitability_worstCase} and~\ref{table:equitability_aveCase}, as well as Figures~\ref{fig:equitabilityAnalysis_evenCurve_XYNoise}, \ref{fig:equitabilityAnalysis_evenCurve_YNoise}, and~\ref{fig:equitability_MICvMI}, demonstrate the strong influence of finite-sample effects on the equitability of mutual information estimation. Consistent with the fact that $\popMIC$ is uniformly continuous while mutual information is not \citep{reshef2015estimating}, estimation of $\popMIC$ suffers less from this problem: for $n=250$ and $n=500$, $\MICestE$ has both superior worst-case and average-case equitability over mutual information estimation (using $k=1$, $6$, $10$, and $20$ in the Kraskov estimator) in every model $\Q$ tested, and in most cases by substantial margins. For $n=5000$, mutual information estimation has better equitability than $\MICestE$ in settings where there is only noise in the dependent variable, while $\MICestE$ has superior equitability in all other models tested. Aspects of this phenomenon have previously been noted in~\citet{reshef2013equitability}, and subsequently in~\citet{kinney2014equitability}, and~\citet{reshef2014comment}.

\paragraph{Equitability in the large-sample limit} Departures from perfect equitability can occur either as a result of finite sample effects, or because of the lack of interpretability of the population value of the statistic. To disentangle these two potential effects, we compare the equitability of $\popMIC$ and the Kraskov mutual information estimator in the large-sample limit (Figure~\ref{fig:equitabilityAnalysis_InfDataLimit}). This analysis yields two important insights. First, it demonstrates that when finite sample effects are minimal, $\popMIC$ has both superior worst-case and average-case equitability in the four models $\Q$ that contain noise added in the independent variable or in both the independent and dependent variables, while mutual information is more equitable than $\popMIC$ in the two remaining settings, where noise is added only in the dependent variable. Second, more generally, it shows that neither $\popMIC$ nor mutual information is worst-case perfectly interpretable with respect to $\Phi=R^2$ over the sets $\Q$ examined.  This is not surprising given the broad range of relationships, noise models, and independent variable marginal distributions tested.

\paragraph{Relationship to equitability analysis from~\citet{kinney2014equitability}}
\label{sec:kinneyAnalysis}
A more limited analysis of the equitability of $\MIC$ and mutual information estimation was presented in~\citet{kinney2014equitability}.  There, the authors examined the equitability of $\MIC$ and mutual information estimation specifically at a large sample size ($n=5000$) and under one choice of $\Q$ ($E_{f(X)}[\mathcal{N}_y]$). From this, they concluded that mutual information estimation was more equitable than $\MIC$. As our analysis here shows, though that is true for this specific choice of $\Q$ and sample size, it is not true in general. To the contrary, the general picture seems to be that the equitability of estimators of $\popMIC$ is more robust than that of estimators of mutual information due to a combination of finite-sample effects and differences between the population values themselves.

For more on this discussion, see the technical comment \citep{reshef2014comment} published by the authors of this paper about~\citet{kinney2014equitability}. For a discussion of the theoretical results of \citet{kinney2014equitability}, see \citet{reshef2015equitability} and \citet{Murrell2014comment}.

\subsubsection{The equitability of $\rho$, $\dCor$, maximal correlation, $\HSIC$, $\RDC$, $\TICestE$, $\HHG$, and $\DDP$}
Figures~\ref{fig:equitabilityAnalysis_evenCurve_XYNoise} and~\ref{fig:equitabilityAnalysis_evenCurve_YNoise}, as well as Tables~\ref{table:equitability_worstCase} and~\ref{table:equitability_aveCase}, demonstrate that $\rho$, distance correlation, maximal correlation, $\HSIC$, $\RDC$, $\TICestE$, $\HHG$, and $\DDP$ all display relatively poor equitability over the models $\Q$ tested. (We note that these methods were not designed with equitability in mind and so do not make claims about equitability.) Of these methods, maximal correlation displays the highest degree of equitability. Additionally, the equitability profiles of both $\dCor$ and $\RDC$ are similar to that of the correlation $\rho$.

\subsection{Alternate equitability analysis via connection with statistical power}
\begin{figure}
	\centering
	\begin{minipage}[t!]{0.66\linewidth}
	\includegraphics[clip=true, trim = 0.125in 1.0in 0.125in 0.8in, width=0.95\linewidth]{\pathToFigures/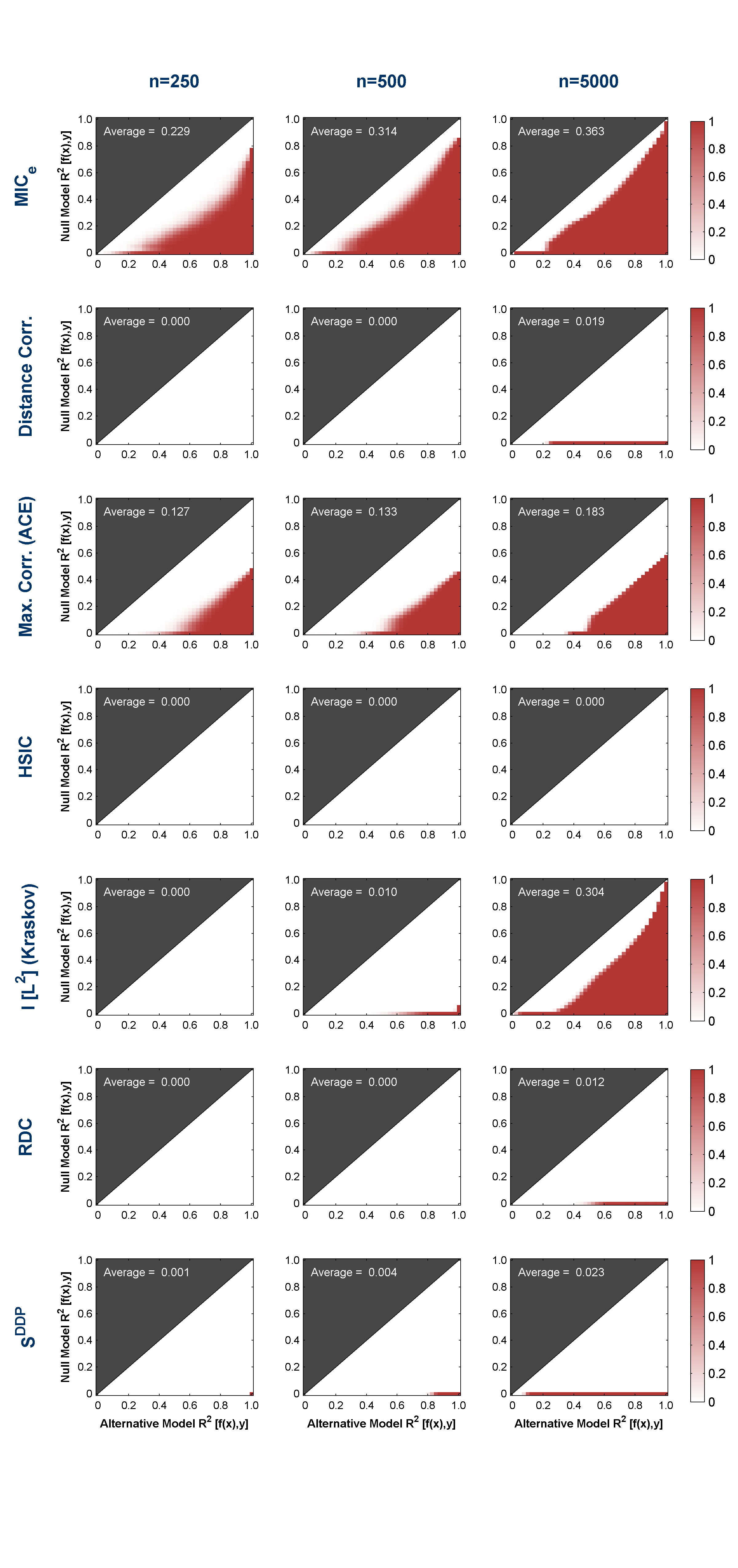}
	\end{minipage}
	\begin{minipage}[t!][0.9\textheight][t]{0.30\linewidth}
	
	\caption[The equitability of measures of dependence on noisy functional relationships, visualized in terms of power]{
    	The equitability of measures of dependence on noisy functional relationships, visualized in terms of power. \textit{[Redder is more equitable.]}
    	The set of noisy functional relationships analyzed is the same as in Figure~\ref{fig:equitabilityAnalysis_evenCurve_XYNoise}, and relationship strength is again quantified by $\Phi = R^2$.  Plots were generated as in Figure~\ref{fig:equitabilityAndPower}. The intensity of the pixel at coordinate $(x_1, x_0)$ in each heat map shows the power of a right-tailed test based on the statistic in question at distinguishing the (composite) alternative hypothesis $H_1 : R^2 = x_1$ from the (composite) null hypothesis $H_0 : \Phi = x_0$ with type I error at most $\alpha=0.05$. An optimal statistic would yield tests with 100\% power for every $x_1 > x_0$. $\MICestE$ comes closest to achieving this ideal, and performs particularly well relative to other methods at lower sample sizes. For each plot, the average area under the power curve across the entire set of null hypotheses is listed. (The maximum achievable such area is 0.5.)
    	Mutual information, estimated using the Kraskov estimator, is represented using the squared Linfoot correlation. For every parametrized statistic whose parameter meaningfully affects equitability, results are presented at each sample size using parameter settings that maximize equitability across all twelve of noise/marginal distributions tested at that sample size.}
	\label{fig:equitabilityAndPower_evenCurve_XYNoise}
	\end{minipage}
\end{figure}

Figures~\ref{fig:equitabilityAndPower_evenCurve_XYNoise} and~\ref{fig:equitabilityAndPower_evenCurve_YNoise} quantify the equitability of the set of measures of dependence examined above via a power analysis.  This is achieved as demonstrated in Figure~\ref{fig:equitabilityAndPower}. Analyses are presented for the same range of models and sample sizes examined in the equitability analysis performed using interpretable intervals, and results for all other models are presented in the supplemental materials.

Assessing equitability using statistical power analysis confirms the conclusions that are reached by the quantification of equitability using interpretable intervals above.  That is, in this analysis, $\MICestE$ is the only measure of dependence that is able to distinguish any null hypothesis of the form $H_0 : R^2 = x_0$ from any alternative hypothesis of the form $H_1 : R^2 = x_1$ with high power across the full range of models $\Q$ and sample sizes examined, even when $x_1 - x_0$ is relatively small. As in the equitability analysis using interpretable intervals, the Kraskov mutual information estimator is not able to achieve this task for sample sizes tested lower than $n=5,000$, and even at $n=5,000$ it is only able to do so for models that contain noise only in the dependent variable.  This is true regardless of the choice of parameter used in the Kraskov estimator. (See Appendix~\ref{app:additionalEquitabilityResults} for results achieved using additional parameters.) Finally, as before we see that $\rho$, distance correlation, $\HSIC$, $\RDC$, $\HHG$, and $\DDP$ are highly non-equitable, with maximal correlation being the only other measure of dependence tested that displayed any degree of equitability.

In this analysis, in which $\Q$ is a set of noisy functional relationships and the property of interest is $R^2$, methods such as distance correlation and $\HSIC$, which are traditionally considered to be well powered for detecting deviations from independence, do not yield tests that achieve high power, even in the case where the null hypothesis is statistical independence. This is due to the fact that even when we consider a null of independence, we have a composite alternative hypothesis due the the multiple different functional forms present in $\Q$. This requires methods to yield tests that are highly powered at simultaneously detecting deviations from independence in {\em all} of the relationship types present in $\Q$.  The poor power displayed by tests based on distance correlation, $\HSIC$, and $\RDC$ is due to the fact that, while they may be highly powered at detecting deviations from independence in, say, linear relationships, they are worse at simultaneously doing so for the more nonlinear relationships.  Of course,  when both the null and alternative hypotheses are allowed to take on non-zero values of $R^2$, the task of differentiating between the null and alternative becomes even harder as both the null and alternative are now composite, and correspondingly the performance of these methods suffers further.

\subsection{Discussion}
In this section we analyzed the equitability with respect to $R^2$ of $\MICestE$ alongside several leading measures of dependence, on many different sets of relationships with varying sample sizes, noise types, and marginal distributions. Our main finding is that in most (32 out of 36) of the settings we considered, $\MICestE$ is substantially more equitable than the other methods. In the remaining four settings, all of which had a sample size of $n=5,000$ and no noise added in the independent variable, mutual information estimation using the Kraskov estimator outperformed $\MICestE$ by a small margin; however, the equitability of the Kraskov estimator at lower sample sizes or on other noise models is otherwise poor.

As we show later, the equitability of $\MICestE$ does seem to come at a price. Specifically, though $\MICestE$ does, with certain parameter settings, yield tests with good power against independence (see Section~\ref{sec:power}), the settings that confer the equitability demonstrated above do not have this property. This suggests that there is an inherent trade-off in the statistic between power against independence and equitability, and in Section~\ref{sec:power_equitability_tradeoff} we establish that this is indeed the case.

Interestingly, besides $\MICestE$ and the Kraskov estimator, the other method with non-trivial equitability with respect to $R^2$ in our experiments is maximal correlation as computed using the method of alternating conditional expectations (ACE). This is interesting because, on the one hand, one can show from its definition that the squared maximal correlation is bounded from below by $R^2$, and on the other hand the lack of equitability of maximal correlation in our experiments seems to stem from the ACE method returning results below this lower bound. We therefore wonder whether maximal correlation --- were it computable exactly --- would be highly equitable with respect to $R^2$.

The analyses presented in this section demonstrate that equitability with respect to $R^2$ is achievable to a significant extent, at least on the relationships tested here. However, while the noise models, marginal distributions, and functions used were chosen to be representative of real-world relationships, they by no means form a large enough set to allow us to make claims about the performance of these methods in general. Given this state of affairs, a better theoretical understanding of $\MICestE$ and also of equitability --- with respect to $R^2$ and otherwise --- is crucial for allowing us to determine when and to what extent equitability can be achieved. Though this is an ambitious goal, we feel it is important for guiding the development of methods for coping with the growing complexity of today's data sets. It is our hope that the empirical insights presented here, together with the theory presented in \citet{reshef2015equitability,reshef2015estimating}, will inform and enable further investigation of both equitability and $\MICestE$.

\section{Statistical power analysis}
\label{sec:power}
There are many settings that call simply for testing for {\em any} deviation from independence rather than relationship ranking, or in which relationship ranking is simply not feasible.  These settings require a measure of dependence that yields tests with high power against a null hypothesis of statistical independence.

Here, we turn to assessing the power against independence of the set of measures of dependence examined. This has been done previously, most notably by Simon and Tibshirani~\citep{simon2012comment}. Our analysis expands upon the power analysis performed by Simon and Tibshirani in three key ways. First, we examine power not as a function of absolute amount of noise in the alternative hypothesis but rather as a function of the $R^2$ of the alternative hypothesis, allowing us to aggregate across relationship types to gain a more global view of the power of each method. Second, for each of the statistics we analyze that has a free parameter, we perform a parameter sweep to understand the power of the corresponding tests as a function of that parameter, and to determine what the optimal value of the parameter is. Last, we analyze a larger set of methods, with a greater variety of sample sizes. The result is an in-depth portrait of statistical power, assembled using the best achievable performance of a large number of leading methods.

\subsection{Setting up the analysis}
\subsubsection{Choice of methods to analyze}
The methods analyzed were the same as those examined in the equitability analysis. See Section~\ref{sec:equitability_setting_up} for more details.

\subsubsection{Choice of relationships and sample size}
For all of the power analyses performed, we use both the set of relationships and noise model ($U_{X}[\mathcal{N}_y]$) chosen by Simon and Tibshirani \citep{simon2012comment}.  For consistency with the sample sizes used throughout this work, we show results for $n=500$, but results for all analyses using $n=100$ are similar and are provided in the supplemental materials.

\subsubsection{Parameters of the analysis}
In order to make power results for different relationships comparable, we sought to compute power as a function of $R^2$, in a manner similar to the equitability analyses above, rather than as a function of absolute magnitude of added noise. To do this, we determined, for each of the eight relationship types chosen by Simon and Tibshirani\footnote{
Note that one of the relationship types chosen by Simon and Tibshirani was a circle. Since this relationship is not a noisy functional relationship, one cannot truly discuss its $R^2$. Therefore, as a heuristic workaround, we defined the $R^2$ of a noisy circle to be the average of the $R^2$ values, computed separately, of the top and bottom halves.},
$100$ noise levels evenly distributed over the range of noise levels yielding $R^2=1.0$ (no noise) and $R^2 = 10^{-2.5}$ (substantial noise). (See Appendix~\ref{app:dataGen}.) We then drew 1000 independent samples, each of size $n=500$, from the corresponding distribution. This was our alternative hypothesis. We also drew 1000 independent samples from a corresponding null hypothesis chosen to have the same marginals. All analyses were performed at a significance level of 0.05.

\subsubsection{Parameters of statistics tested}
\label{subsubsec:power_param_sweep_description}
To understand how choice of parameter affects statistical power in the case of each measure of dependence, we performed a parameter sweep for each method that has a parameter.\footnote{
Some methods, such as $\RDC$, will in the future automatically select optimal parameters in a relationship-type-dependent way \citet{lopez2015personal}.}
To do this, we needed a way of quantitatively summarizing power across eight relationship types, so that we could then graph performance as a function of parameter value and then choose the optimal parameter value. We did this in two different ways. For both ways, having power computed as a function of $R^2$, so that power on different relationships could be directly compared, was crucial.

The first way that we summarized power was by computing the area under the power curve for each relationship type, integrating with respect to absolute noise level. That is, we computed the power curve for a given relationship type (e.g., linear) as a function of amount of noise added, and then computed the area under that curve up to a pre-specified limit on the amount of noise (as measured by $R^2$). The resulting number measures the expected power of tests based on the statistic in question when the amount of noise added in the alternative-hypothesis is chosen uniformly at random.

The second way that we summarized power was by computing the minimum alternative hypothesis $R^2$ necessary to achieve a certain level of power \citep{kinney2014equitability}. Another way of thinking of this is ``what is the maximum amount of noise that can be added to a relationship before power for differentiating that relationship from independence drops below a pre-set threshold?" The results presented here use a threshold of 50\% power; results for other thresholds (95\%, 75\%, 25\%, and 10\%) are similar and can be found in the online supplement.

\begin{figure}
	\centering
	\begin{tabular}{@{} r p{0.9\textwidth} @{}}
	    (a) & \vspace{-7pt} \includegraphics[clip=true, trim = 0.825in 0.125in 1.1in 0.4in, width=0.89\textwidth]{\pathToFigures/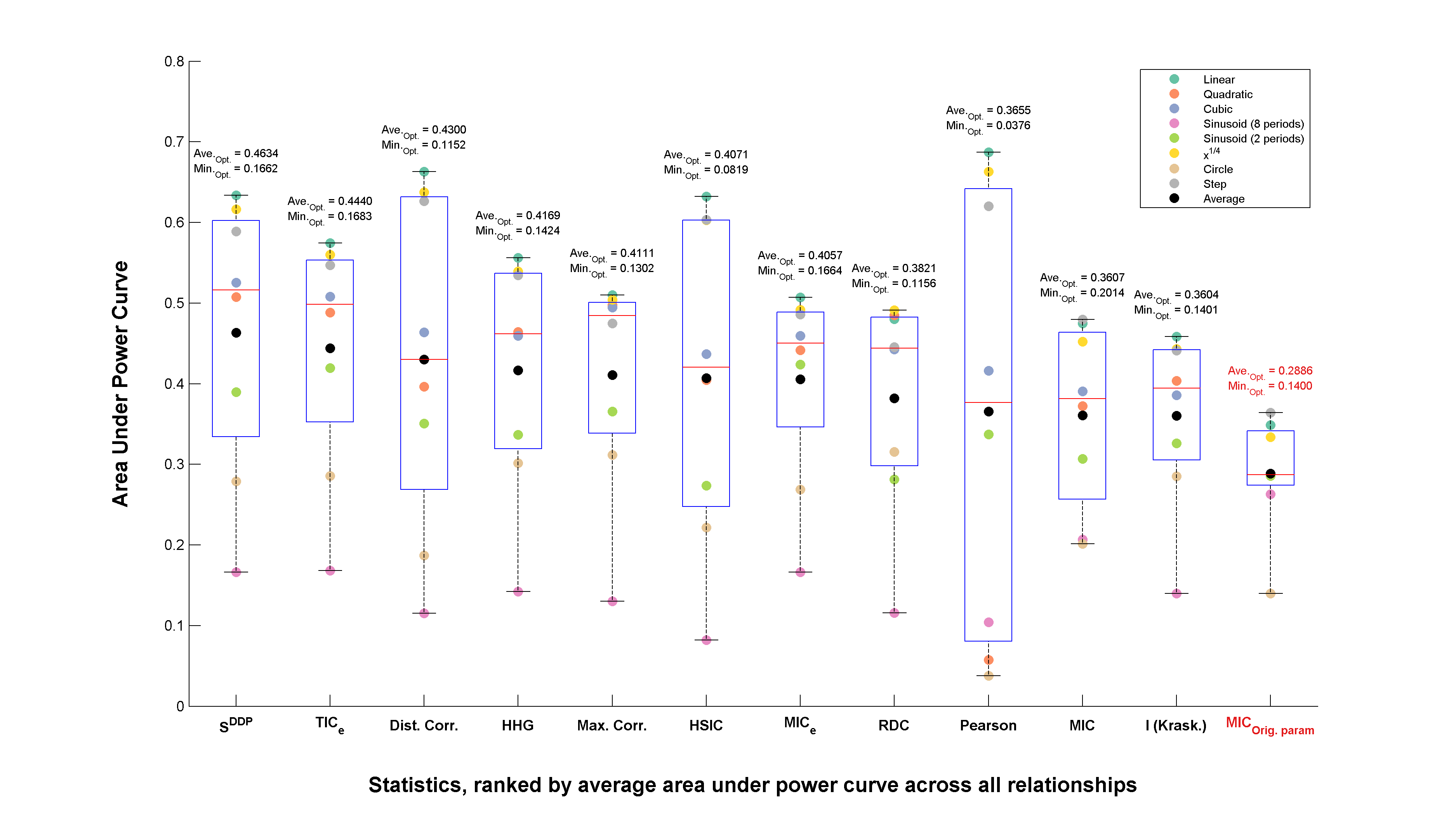} \\
	    (b) & \vspace{-7pt} \includegraphics[clip=true, trim = 0.775in 0.275in 1.1in 0.125in, width=0.89\textwidth]{\pathToFigures/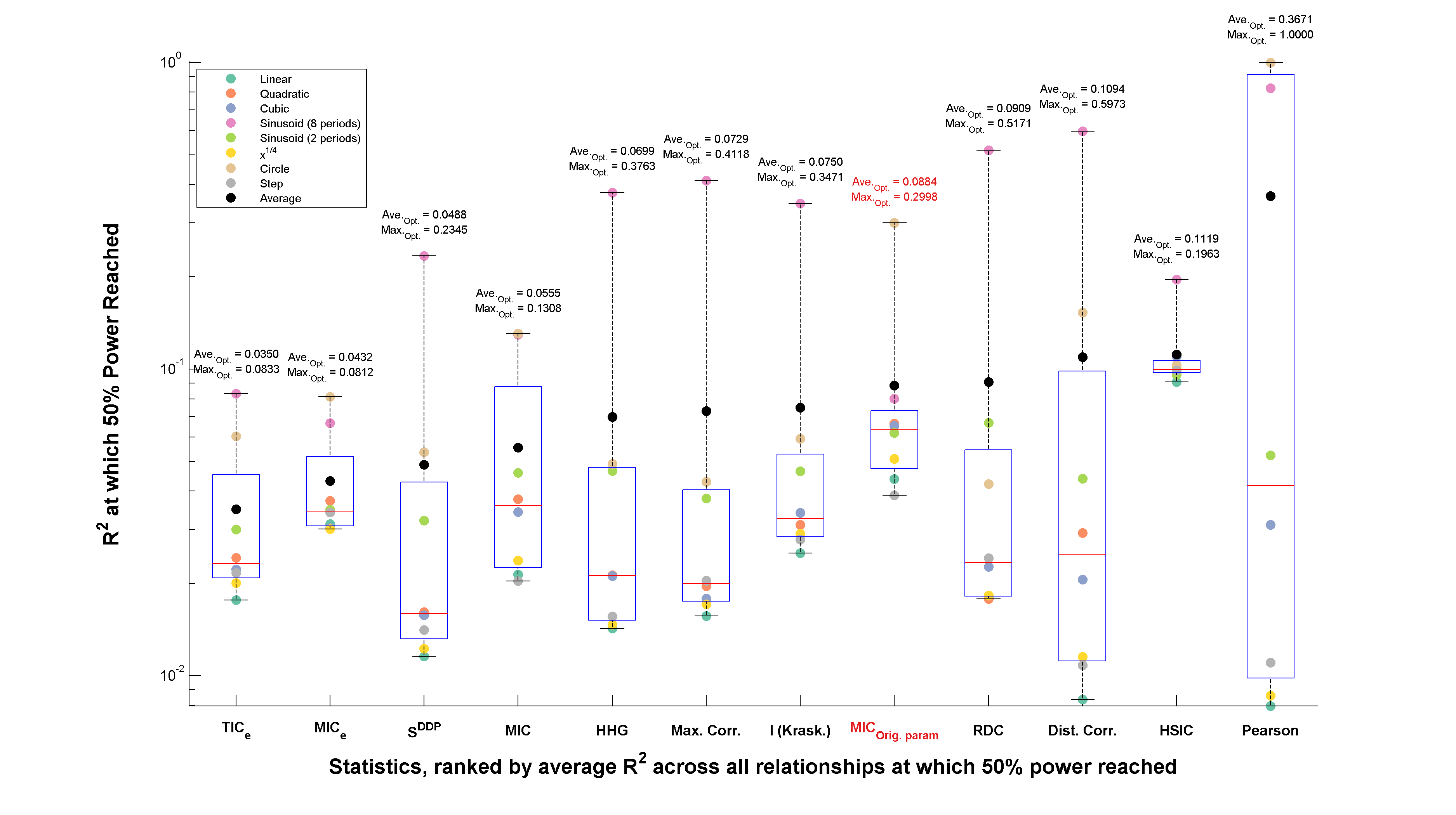}
    \end{tabular}
	
	\caption[Quantitative ranking of measures of dependence by the power of their corresponding independence tests]{
	    Measures of dependence ranked by the power of their corresponding independence tests. For each measure of dependence and each relationship type, power was quantified using \figpart{a} the area under the power curve \textit{[higher is more powerful]}, or \figpart{b} the minimal $R^2$ at which at least $50\%$ power is achieved \textit{[lower is more powerful]}. The collection of these scores across relationship types is then plotted for each method along with quartiles, and both average- and worst-case performance across relationship types are listed. Optimal parameter values for each test statistic were chosen to maximize average-case performance; see (a) Figure~\ref{fig:indivRelPowerParamSweep}, or (b) Figure~\ref{fig:indivRelPowerParamSweepMaxR2}. The $\MIC$ statistic from~\cite{MINE} with the parameters used in~\cite{simon2012comment} is labeled in red; there is a substantial improvement in power when an optimal parameter is chosen. A further improvement in power is attained by $\MICestE$, and the performance of $\TICestE$ is state-of-the-art. The sample size was $n=500$; results are similar with $n=100$ and, for (b), with power thresholds besides $50\%$. (See supplementary materials.)}
	\label{fig:powerRanking}
\end{figure}

\subsection{Results}
Figure~\ref{fig:powerRanking} contains quantitative rankings of the measures of dependence by the power of their corresponding tests for independence, using optimal parameter values determined by each of the two methods described above. The parameter sweeps themselves, which characterize power against independence as a function of statistic parameters, are presented in Figures~\ref{fig:indivRelPowerParamSweep} and~\ref{fig:indivRelPowerParamSweepMaxR2}.

This analysis yields several insights, which we discuss below.

\subsubsection{Average power across relationship types}
Let us first use the average power across relationship types to rank the measures of dependence from most to least powerful over this set of relationships. Doing so using the quantification of power in Figure~\ref{fig:powerRanking}a\footnote{
Though this quantification of power computes the area under the power curve integrating with respect to absolute noise level, one could integrate with respect to $R^2$ instead. Doing so would measure the expected power of each statistic on an alternative hypothesis with a randomly chosen $R^2$. When this is done and optimal parameters are chosen for each method, the resulting ranking is
\[
\TICestE > \MICestE > \DDP > \MIC > \HHG > \textnormal{max. corr.} > \MI > \MIC_{\textnormal{Orig. param}} > \RDC > \dCor > \HSIC > \hat\rho
\]
This ranking makes sense because integrating with respect to $R^2$ rather than absolute noise level emphasizes performance on stronger relationships, which is more similar to the type of performance quantified by equitability. Correspondingly, the optimal parameters determined for the methods in this analysis were more similar to the parameters yielding optimal equitability. For this reason, we did not focus on this method for quantifying power against independence.
}
yields (from most to least powerful):
\[ \DDP > \TICestE > \dCor > \HHG > \textnormal{max. corr.} > \HSIC > \MICestE > \RDC > \hat\rho > \MIC > \MI > \MIC_{\textnormal{Orig. param}} \]
Doing the same using the quantification of power in Figure~\ref{fig:powerRanking}b yields (from most to least powerful):
\[ \TICestE > \MICestE > \DDP > \MIC > \HHG > \textnormal{max. corr.} > \MI > \MIC_{\textnormal{Orig. param}} > \RDC > \dCor > \HSIC > \hat{\rho} \]
When the largest outlier, a high-frequency sinusoid, is removed from the analysis in Figure~\ref{fig:powerRanking}b (and optimal parameters are re-chosen accordingly), the ordering is as follows\footnote{We chose to remove outliers rather than use median power because since a) the power values for different function types often rank in the same order across methods, and b) there are only eight such numbers and they each vary considerably among methods, the median is very sensitive to the performance of each method on only one or two particular function types.}:
\[ \DDP > \TICestE > \textnormal{max. corr.} > \HHG > \MICestE > \RDC > \HSIC > \MI > \dCor > \MIC > \MIC_{\textnormal{Orig. param}} > \hat{\rho} \]
(See online supplement.)  Finally, when the two largest outliers, a high-frequency sinusoid and a circle, are removed from the analysis, the ordering is as follows:
\[ \DDP > \TICestE > \dCor > \textnormal{max. corr.} > \MICestE > \HHG > \HSIC > \RDC > \MIC > \MI > \MIC_{\textnormal{Orig. param}} > \hat{\rho} \]
(See online supplement.) The orderings produced by these analyses are relatively robust to sample size and power threshold used, with $\TICestE$ or $\DDP$ generally performing the best and occasionally swapping with each other as power threshold is varied. Results obtained with $n=100$ and using $95\%$, $75\%$, $25\%$, and $10\%$ power thresholds are provided in the supplemental materials.

Several aspects of these rankings merit mention. First, state-of-the-art performance is shared between $\TICestE$ and Heller and Gorfine's $\DDP$. This is interesting because the latter statistic is in fact closely related to the theory behind the maximal and total information coefficients in that it too is an aggregation via summation of mutual information scores taken over many different grids. Thus, these results provide evidence that the basic approach of aggregating mutual information scores over a large set of grids, whether via the characteristic matrix or other statistics, is a fundamentally promising avenue for thinking about dependence. 

Second, the average power of independence testing using $\MICestE$, when parameters are optimized for the task of relationship detection rather than ranking, is competitive with the state of the art. In particular it is higher than the power of its predecessor $\MIC$ \citep{MINE}, which estimates the same population quantity ($\popMIC$). This demonstrates that the improved bias/variance properties of $\MICestE$ relative to $\MIC$ \citep{reshef2015estimating} indeed translate into an improvement in power.

We note parenthetically that the power of the $\MIC$ statistic from~\citet{MINE} is substantially higher than has been previously reported. This discrepancy is due to the fact that previous analyses that examined the power of $\MIC$ used the default parameter setting ($\alpha = 0.6$), which was intended to maximize equitability rather than power against independence. As this analysis shows, lower values of $\alpha$ should be used for testing for independence. As we show in Section~\ref{sec:power_equitability_tradeoff}, the same statement holds for $\MICestE$, and both statements follow from a more general power-equitability trade-off.

Our final --- and perhaps most important --- observation about our results is that the differences in power between most of the best-performing methods appear rather small. And indeed, an analysis using many of these methods on a real gene expression data set \citep{heller2014consistent} shows that this observation is true in practice. For example, of the 3312 significant relationships found in the data set using a statistic related to $\DDP$, 3199 (97\%) were also detected by $\HHG$, and the latter found only 84 other relationships; 2845 (86\%) were also detected by $\dCor$ on ranks, and the latter found only 44 other relationships; and 2445 (74\%) were also detected even simply by computing the Pearson correlation coefficient on ranks. $\MICestE$ and $\TICestE$ were not run in this analysis, but the simulation results presented above lead us to believe that they would also have recovered a very similar set of relationships had their corresponding independence tests been used on this data set.\footnote{$\MIC$ was run for this analysis, but with the default value of $\alpha=0.6$, which yields very poor power against independence.}

\subsubsection{Worst-case power against independence across relationship types}
In addition to considering average-case power across relationship types, it is also important to examine worst-case performance. To measure this, we consider the lowest relationship strength $x$ at which each independence test is guaranteed to detect, with a given amount of power, all relationships with strength at least $x$ regardless of relationship type. When a small such $x$ exists, the statistic in question is said to have a {\em low detection threshold} \citet{reshef2015equitability}. This implies that the corresponding independence test will not overlook important relationships because of the test statistic's systematically assigning them lower scores. As described in \citet{reshef2015equitability}, low detection threshold is related to equitability: an equitable statistic provably has a low detection threshold on its set of standard relationships, whereas the converse is not true.

The detection threshold of the independence tests we consider can be read from Figure~\ref{fig:powerRanking}b: for if $x$ is the maximum, across relationship types, of the $R^2$ required to achieve $50\%$ power on each relationship type, then $x$ is also the minimal $R^2$ such that we can guarantee at least $50\%$ power on any relationship with an $R^2$ of $x$ regardless of type.

As the figure shows, the detection threshold of $\TICestE$ and $\MICestE$ on the set of relationships examined is an order of magnitude lower than the detection thresholds of the other statistics we evaluated. This phenomenon is robust to power thresholds besides $50\%$; see the online supplementary materials. It implies that $\TICestE$ is a good candidate for a ``first-pass" filtering of the relationships in a data set before other, more fine-grained analyses are conducted. In contrast, the high detection thresholds of the other statistics imply that, for a fixed relationship strength, their power against independence may be more sensitive to relationship type. Using such statistics for pre-processing may therefore result in certain relationship types being missed in downstream analyses.

\subsubsection{Power on specific relationships}
\begin{figure}
	\centering
	\includegraphics[clip=true, trim = 0.85in 0.8in 0.85in 0.6in, width=0.975\textwidth]{\pathToFigures/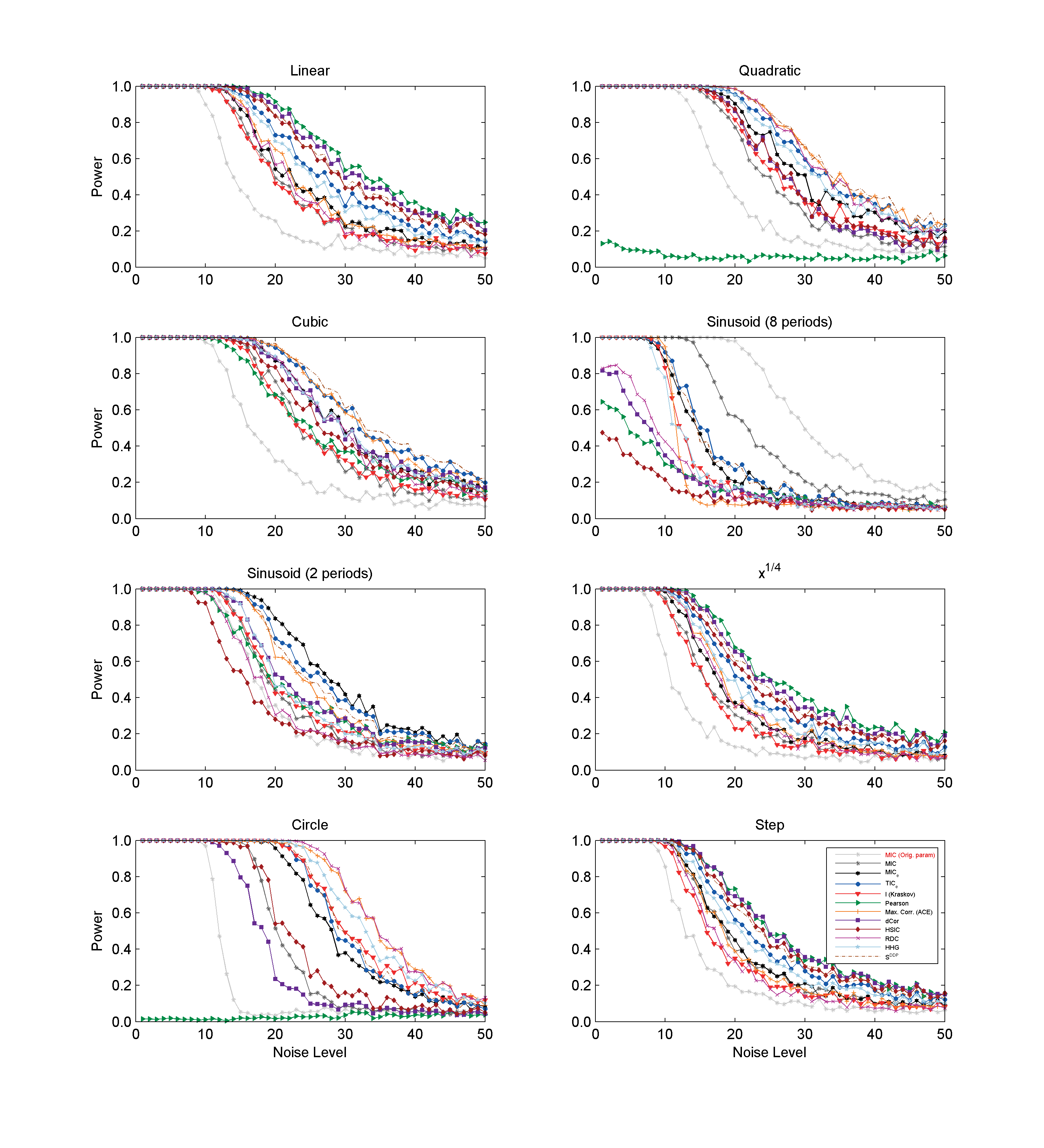}
	\caption[Power against statistical independence of tests based on measures of dependence across several relationship types]{
	    A re-creation of the power analysis performed by Simon and Tibshirani~\citep{simon2012comment}, with optimal parameter choices for each statistic.
	    Power against a null hypothesis of statistical independence for the relationships examined in~\cite{simon2012comment}, at $50$ noise levels for each relationship and $n=500$. For each statistic that has a parameter, an optimal value for the parameter was chosen as described in Figure~\ref{fig:indivRelPowerParamSweep}. (For a version with $n=100$ see supplementary materials.)
	    }
	\label{fig:indivRelPowerOptimalParam}
\end{figure}

Finally, to obtain a more fine-grained picture of the power of the methods we consider on specific relationship types, we also re-created the specific power analysis from~\citet{simon2012comment} with optimal parameter choices for each method, as above. The results are shown in Figure~\ref{fig:indivRelPowerOptimalParam}. Note that, in order to maximize our ability to discern between power curves generated by different tests within each relationship type, in this analysis we followed \citet{simon2012comment} by plotting power as a function of absolute noise level rather than the population $R^2$. This differs from the analyses above, and means that power levels are not directly comparable across relationship types.

Similarly to our other results, the optimal parameter choices used here cause the power of tests based on several of the statistics included in this analysis to be better than previously reported \citep{simon2012comment, gorfine2012comment, lopez2013randomized, kinney2014equitability, jiang2014non}. For instance, we again see here that the power of $\MIC$ is substantially improved. We additionally see that the power of $\MICestE$ and $\TICestE$ is quite good across this set of relationships. This analysis also illustrates that each measure of dependence tested indeed has its own strengths and weaknesses. For example, distance correlation and $\HSIC$ are relatively better powered to detect linear dependence than $\MICestE$ and $\TICestE$, but are relatively worse at simultaneously detecting most of the other forms of dependence tested. In contrast, $\DDP$ appears to have a similar profile to that of $\TICestE$, which again makes sense given the fact that $\DDP$, like $\TICestE$, is also a grid-based method with a mutual information-based score that aggregates by summation.

\subsection{Discussion}
In this section we analyzed the power of independence tests based on several leading measures of dependence, including $\TICestE$ and $\MICestE$, on the set of relationships chosen by Simon and Tibshirani \citep{simon2012comment}. Our analysis differs from previous ones in that we have aggregated results across relationship types, performed parameter sweeps for all the methods that have parameters, and examined a large set of methods and sample sizes.

Our main finding is that $\TICestE$, along with Heller and Gorfine's $\DDP$, provides state-of-the-art performance on average over the relationship types examined. This is significant because $\TICestE$ is trivial to compute once $\MICestE$ has been computed: $\TICestE$ is the sum of the entries of a matrix whose maximal entry is $\MICestE$.\footnote{
The parameter $\alpha$ of $\TICestE$ that leads to optimal power against independence may not equal the parameter $\alpha$ used for the computation of $\MICestE$ if, for instance, the latter is being computed with equitability as a goal. In this case, the total runtime will equal the runtime of the method with the greater value of $\alpha$, since increasing $\alpha$ just grows the portion of the equicharacteristic matrix that is computed. In most situations, we expect that the value of $\alpha$ desired for $\MICestE$ will be greater than that desired for $\TICestE$ since the former will be run with equitability in mind, and so $\TICestE$ will be a trivial side-product of the computation of $\MICestE$.}
Moreover, the power of $\TICestE$ on individual relationship types remained high across relationship types; there was no one relationship type that testing for independence using $\TICestE$ would cause us to overlook with high probability. Our results therefore point to a promising and computationally efficient strategy for exploratory data analysis: first, simultaneously compute both $\MICestE$ and $\TICestE$ on all variable pairs in a data set. Then discard pairs declared insignificant by $\TICestE$ and examine the $\MICestE$ scores of the remaining pairs. This way, the multiple-testing burden is borne by the state-of-the-art power of $\TICestE$, but the significant relationships can still be ranked equitably using $\MICestE$. We remark that using $\DDP$ together with $\MICestE$ in an analogous strategy would not be optimal for two resaons. First, such a strategy would be slower, both because $\DDP$ must be computed independently of $\MICestE$ whereas $\TICestE$ need not be, and because $\DDP$ itself is slower to compute than $\MICestE$/$\TICestE$. (See Section~\ref{sec:runtime} for more on running times.) Second, since the power against independence of $\DDP$ appears more sensitive to alternative hypothesis relationship type, it seems that filtering relationships by $\DDP$ is more likely to result in important relationships being eliminated prematurely because of their relationship type.

Our analysis also showed that the power against independence of tests based on $\MICestE$ is greater than that of tests based on its predecessor, $\MIC$, and in particular that $\MICestE$ yields tests with power close to the state of the art. However, these results require a setting of the parameter $\alpha$ of $\MICestE$ that differs from that used for optimal equitability, suggesting a trade-off between power against independence and equitability that we study in the following section. Additionally, we found that the power against independence of most of the methods tested varies considerably across different alternative hypothesis function types, whereas this sensitivity is substantially weaker for $\MICestE$ and $\TICestE$.

Finally, we observed that, at least in the bivariate setting, the performance of many of the leading methods appears quite similar, even on real data. This last observation leads us to question whether the magnitude of a method's power against independence ought to be the only measure of that method's utility. There are cases in which the answer is `yes', such as when we wish to perform an independence test between two high-dimensional variables whose result is the end-goal of our analysis. However, in data exploration scenarios in which existing measures of dependence already reliably identify thousands of relationships, it may be more important to be able to prioritize those relationships for follow-up, rather than to discover a small number of additional relationships whose strength, and therefore scientific promise, is uncertain. Solving the data exploration problem well requires us not just to maximize the number of relationships we detect, but also to think about how the statistic we choose to use will influence {\em which} relationships we find. Indeed, this issue is what inspired the original work on MIC and equitability \cite{MINE}, but we believe the questions regarding the right frameworks for understanding data exploration problems continue to pose numerous interesting challenges.

\section{The power-equitability trade-off and parameter choice for $\MICestE$}
\label{sec:power_equitability_tradeoff}
The above analyses establish that $\MICestE$ can be both highly equitable and provide high-powered tests for detecting deviations from independence. However, in each analysis the parameter $\alpha$ of $\MICestE$ was chosen to optimize the objective in question, and the parameter value that yields optimal equitability is different from the value that yields optimal power against independence. This suggests that there may be a trade-off between these two objectives that is being captured by the choice of this parameter~\citep{reshef2013equitability}.

Such a trade-off also seems plausible given the equivalence proven in \citet{reshef2015equitability} between equitability and power against a range of null hypotheses corresponding to different relationship strengths. After all, if equitablity is about simultaneously achieving high power against many null hypotheses, then ``no free lunch"-type considerations imply that to attain this objective we may have to give up some of the power we previously had against the specific null hypothesis of independence.

Here we establish that such a trade-off does indeed appear to exist within each of the parametrized methods we consider. We then discuss the implications of this trade-off for how one should choose parameters when using $\MICestE$ in practice.

\subsection{Demonstrating the power-equitability trade-off}
We examined the equitability and power against independence of $\MICestE$ for values of $\alpha$ ranging from 0.25 to 0.9, at a sample size of 500. By plotting worst-case equitability against average power for each value of $\alpha$, we sought to understand whether there is a Pareto front of equitability/power beyond which we cannot seem to advance. The existence of such a boundary would support the existence of a power-equitability trade-off. We performed a similar analysis for all of the statistics whose power against independence and equitability we assessed.

\begin{figure}
	\centering
	\includegraphics[clip=true, trim = 0.05in 0.1in 0in 0.35in, width=\textwidth]{\pathToFigures/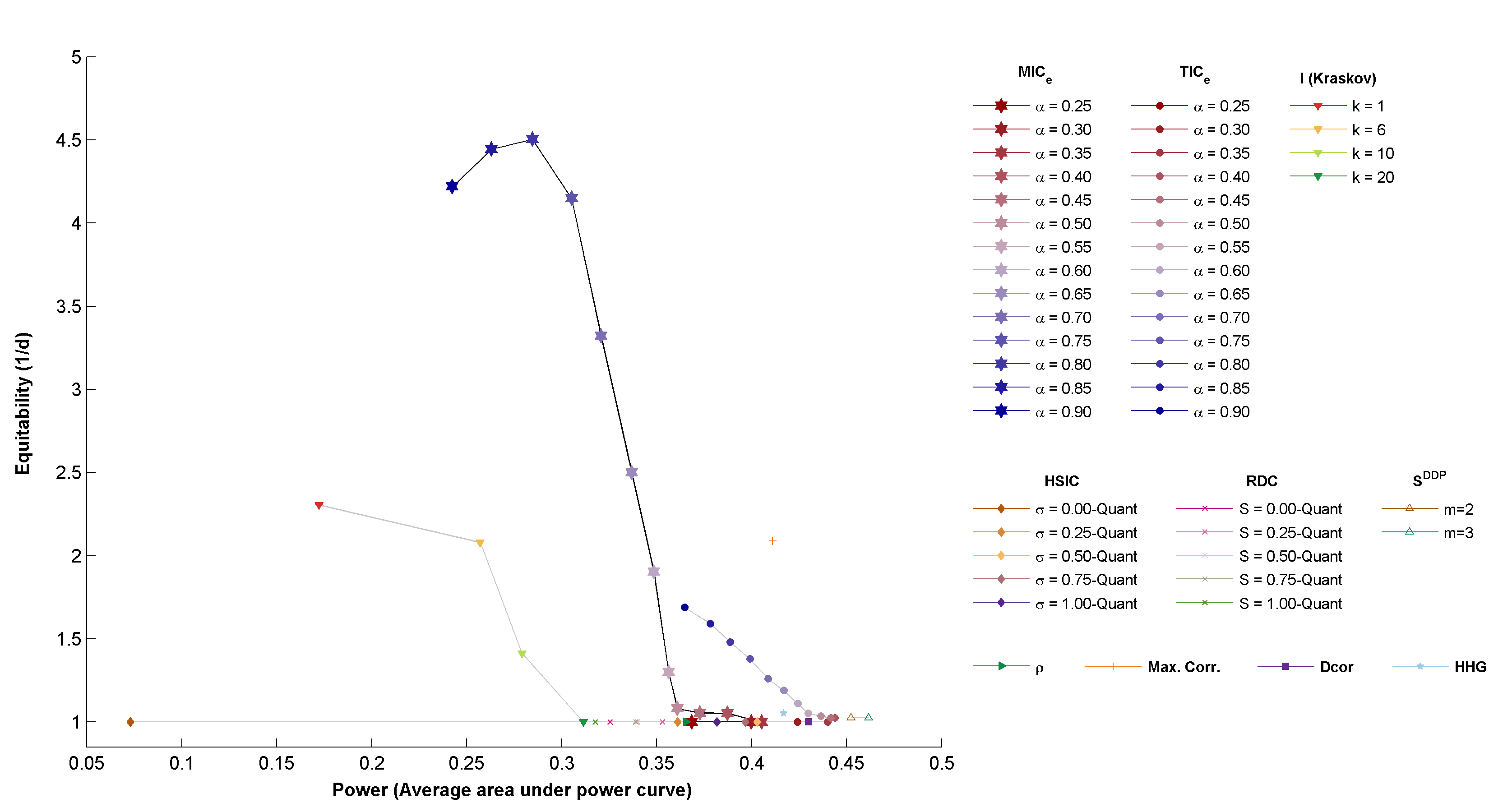}
	\caption[The power-equitability trade-off]{The trade-off between equitability and power against statistical independence across methods.
	    For each method, average power as quantified in Figure~\ref{fig:powerRanking}a is plotted as well as the worst-case equitability under the same model, with $n=500$. For every parametrized method, a point is plotted for each value of the parameter in question. The points corresponding to $\MICestE$ are emphasized. Since each coordinate is strictly preferable to all coordinates below and to the left of it, there is a Pareto ``power-equitability" front. The methods with points along this front are $\MICestE$, maximal correlation, $\TICestE$, and $\DDP$.}
	\label{fig:power_equitability_tradeoff}
\end{figure}

Figure~\ref{fig:power_equitability_tradeoff} shows that every parametrized method with a non-trivial level of equitability does indeed exhibit such a trade-off. In the case of $\MICestE$, the trade-off is captured by the parameter $\alpha$, which controls the maximal grid resolution used by the statistic. This is consistent with the bias-variance analysis in \citet{reshef2015estimating}, which showed that low values of $\alpha$ lead to better performance in the low-signal regime while larger values of $\alpha$ lead to better performance in mid-to-high-signal regimes. It is also consistent with the intuition that disallowing high-resolution grids may increase power against independence but will allow only coarse-grained distinguishability among distributions, while allowing high-resolution grids might enable distinguishing between distributions that may be more similar to each other.

Figure~\ref{fig:power_equitability_tradeoff} is also a useful summary of how the different methods we considered compare to each other along these two dimensions (for this sample size and set of relationships). Specifically, if one point is both above and to the right of another then it is strictly preferable. Thus, the figure shows a Pareto front of methods that offer optimal performance with respect to power against independence and equitability. This front includes $\MICestE$, maximal correlation, $\TICestE$, and $\DDP$.

\subsection{Choosing parameters for $\MICestE$/$\TICestE$: a practical guide}
\label{sec:choosingAlpha}
We now give some guidelines for setting parameters for $\MICestE$/$\TICestE$ more generally. The two parameters required by these statistics are the parameter $\alpha$ discussed above, which governs the maximal grid resolution $B(n)$ of the estimator according to $B(n) = n^\alpha$, and $c$, an optional parameter that controls a speed-versus-optimality trade-off in the algorithm. We discuss each of these in turn.

\subsubsection{Choosing $\alpha$}
There are two main considerations involved in choosing $\alpha$. The first, which is suggested by the analysis above, is how much we care about power against independence relative to equitability (i.e., power at distinguishing cleaner relationships from noisier relationships). The second consideration is whether we expect to see complex relationships in our data. These considerations can can be reframed in terms of hypothesis testing as follows:
\begin{enumerate}
\item \textbf{Is our null hypothesis statistical independence or presence of a weak dependence?}\\
It follows from the above analysis that when using $\MICestE$ (or, more likely, $\TICestE$) to generate tests for statistical dependence one should use a lower value of $\alpha$, while if one is interested in equitability, a larger $\alpha$ is required.
\item \textbf{What is our most complex alternative hypothesis?}\\
Since $\alpha$ places an upper bound on the resolution of grids that can be explored by the estimators, it restricts the complexity of structure that can be detected. Thus, as the relationship class of interest grows to include more complex structure relative to sample size, the value of $\alpha$ should be increased accordingly.
\end{enumerate}

\paragraph{Balancing the two considerations}
For the specific values of $\alpha$ that maximized power against independence of $\TICestE$ and equitability of $\MICestE$, respectively, in our analyses, see Appendix~\ref{app:ParamSettings}. The tables generally show that a) when optimizing for statistical power against independence in the sample-size regimes analyzed here, one should use an $\alpha$ that leads to $B(n)$ being approximately between $4$ (for less complex alternative hypotheses) and $12$ (for more complex alternative hypotheses)\footnote{
Of course, for even more complex alternative hypotheses, a larger $B(n)$ will lead to better performance, provided the sample size allows for detection of the level of complexity in question. In particular, we suspect that $B(n) > \omega(1)$ is necessary for consistency against all alternatives of the resulting independence test. Note however that this hypothesis applies only to $\MICestE$/$\TICestE$ and not to $\MIC$/TIC, because even just estimating the first entry $M(X,Y)_{2,2}$ of the population characteristic matrix yields a statistic that is consistent against all alternatives. (See, e.g., Lemma 6.7 in the supplemental online materials of \citet{MINE}.)},
and b) when optimizing for equitability, one should use an $\alpha$ approximately between $0.5$ (when $n$ is larger) and $0.75$ (when $n$ is smaller).

\paragraph{Equitability and computational efficiency}
For large $n$, the parameters suggested above for equitability are likely needlessly computationally expensive. This is because as $n$ grows, the maximal allowed grid resolution of the statistic $B(n) = n^\alpha$ will outstrip the complexity of most alternative hypotheses that we are liable to encounter in practice.

For example, at $n=5,000$, $B(n) = 70$ provides good equitability on the set of functions and noise models tested in this paper. If this level of equitability is acceptable to us, we may set $\alpha = \log_n 70$ for $n \geq 5,000$, which means that $B(n) = 70$ always. Given that the runtime of the search procedure in $\MICestE$ is $O(n^{5\alpha/2})$, which is $O(n)$ for $\alpha = 0.4$, a less extreme version of this strategy that maintains consistency and gives asymptotically linear runtime is to allow $\alpha$ to decrease for large $n$ until $\alpha = 0.4$ is reached, and then to keep it at $0.4$. In the example above, this happens around $n=40,000$. And indeed, the equitability of $\MICestE$ at this sample size with $\alpha = 0.4$ appears quite good.

For more on how to balance runtime and equitability, see Figure~\ref{fig:equitability_runtime}, which graphs equitability on our set of functional relationships against runtime as $\alpha$ and $n$ are varied, as well as Table~\ref{tab:MIC_params_runtime}, which suggests values of $\alpha$ at several sample sizes that yield 80\% of the best observed equitability for $\MICestE$ at each sample size, and the discussion in the next section, where we examine the runtime of $\MICestE$ compared to other statistics.

\subsubsection{Choosing $c$}
The parameter $c$ determines the coarseness of the discretization of the grid search in the algorithm that computes $\MICestE$, with larger values of $c$ corresponding to finer discretization \citep{reshef2015estimating}. Characterizing the effect of $c$ on the bias and variance of $\MICestE$ is an important avenue of future work. However, using $c=5$ seems to provide good performance in most settings, and in more computationally constrained settings setting even $c=1$ appears to result in only moderate performance loss \citep{reshef2015estimating}.

\section{Runtime analysis}
\label{sec:runtime}
Computational efficiency is often desirable when evaluating dependence, and here we assess the runtimes associated with the set of measures of dependence examined.

\subsection{Setting up the analysis}
Since the runtime of $\MICestE$/$\TICestE$ depends on parameter choice, results for $\MICestE$ are presented for parameter settings recommended for maximizing equitability, maximizing power against independence, and attaining reasonable equitability on a limited computational budget. The third set of parameters was computed by searching at each sample size for the parameters that resulted in the fastest runtime while still yielding 80\% of the best observed equitability at that sample size. All the parameters used for $\MICestE$/$\TICestE$ in this analysis are detailed in Table~\ref{tab:MIC_params_runtime}.

The only other method whose runtime is affected by its parameter was $\DDP$. Since $\DDP$ did not achieve non-trivial levels of equitability, we set its parameter to the value that maximized power against independence.\footnote{Since the runtime of $\DDP$ as a function of its integer-valued parameter $m$ is $O(n^{m-1})$ for $m = 2, 3, 4$, the choice of $m$ heavily affects the runtime. This is significant because the parameter setting that maximizes power against independence can be computed in different ways that lead to different values of $m$: when power is measured by average area under the power curve, $m=3$ performed the best with $m=2$ a close second; in contrast, when power is measured via the minimum $R^2$ necessary to achieve a certain level of power, $m=4$ was the best with $m=3$ and $m=2$ performing significantly worse. (See Section~\ref{subsubsec:power_param_sweep_description} for a description of these methods of quantifying power.) We therefore have chosen $m=3$ here. However, the correct choice of parameter for this statistic will likely depend on the use-case and the available computational budget. For the performance of $\DDP$ with other values of $m$, see the online supplement.} For statistics whose runtimes did not depend on parameter choice, defaults were used (see Appendix~\ref{sec:runtimeParams}).

\begin{table}[t]
\begin{center}
{\scriptsize
\begin{tabu} to \textwidth {X[c] | X[c] X[c] X[c] X[c] X[c] X[c]}
	\toprule
	Sample Size			&	$\rho$				&	Max. Corr.			&	$\RDC$					&	$\dCor$				&	$\HSIC$ 			& $\HHG$		\\
	\midrule
	50					&	0.0001				&	0.0004				&	0.0015					&	0.0010				&	0.0016			&	0.0017	\\
	100					&	0.0001				&	0.0005				&	0.0014					&	0.0014				&	0.0032			&	0.0063	\\
	500					&	0.0001				&	0.0014				&	0.0023					&	0.0504				&	0.0847			&	0.2185	\\
	1,000				&	0.0002				&	0.0025				&	0.0035					&	0.3518				&	0.4886			&	1.0956	\\
	5,000				&	0.0002				&	0.0119				&	0.0129					&	6.1402				&	6.5975			&	34.0171	\\
	10,000				&	0.0002				&	0.0239				&	0.0251					&	25.9859				&	25.7333			&	465.3222	\\
	\bottomrule
\end{tabu}}\\
\vspace{0.1in}
{\scriptsize
\begin{tabu} to \textwidth {X[c] | X[c] X[c] X[c] X[c] X[c] X[c]}
	\toprule
	Sample Size			&	$\MICestE$ [P]	&	$\MICestE$ [FE]	&	$\MICestE$ [E]		&	$\MIC$							&	$\MI$ (Kraskov)        & $\DDP (m=3)$ \\
	\midrule
	50					&	0.0004			&	0.0009			&	0.0021				&	0.0015							&	0.0096			    &   0.0010 \\
	100					&	0.0005			&	0.0012			&	0.0052				&	0.0061							&	0.0100			    &   0.0023 \\
	500					&	0.0018			&	0.0079			&	0.1630				&	0.2187							&	0.0122			    &   0.0529 \\
	1,000				&	0.0037			&	0.0172			&	0.1992				&	0.9628							&	0.0150			    &   0.2122 \\
	5,000				&	0.0195			&	0.0974			&	0.3398				&	18.7627							&	0.0427			    &   5.7464 \\
	10,000				&	0.0398			&	0.1819			&	0.6835				&	66.2238							&	0.0927			    &   23.4473 \\
	\bottomrule
\end{tabu}}
\caption[Average runtimes of algorithms for computing measures of dependence]{Average runtimes, in seconds, of algorithms for computing measures of dependence over 100 trials of uniformly distributed, independent samples at a range of sample sizes. Results for $\MICestE$, are presented for three sample-size-dependent parameter settings that optimize for maximal power against independence ([P]), 99\% of optimal equitability ([E]), and 80\% of optimal equitability (fast equitability, [FE]). For a list of the parameters used in each of these settings, see Table~\ref{tab:MIC_params_runtime}. $\TICestE$ is ommitted because its runtime is very similar to $\MICestE$ [P]. In this analysis, the Kraskov mutual information estimator was run using a pre-compiled C binary, $\MIC$ was computed approximately using the APPROX-MIC algorithm \citep{MINE} in Java, and $\MICestE$ was run in Java. The other statistics were run using their respective R functions/packages. Note that $\dCor$ was run with the standard R package, which is $O(n^2)$; as of this writing there is a faster estimator of the same population quantity that is computable in time $O(n \log n)$ \citep{huo2014fast}.}
\label{tab:runtimes}
\end{center}
\end{table}

\subsection{Results}
The results of our runtime analysis, found in Table~\ref{tab:runtimes}, show several things. First, $\MICestE$ with all three of the parameter settings given is substantially faster than the previously introduced $\MIC$ statistic from~\citet{MINE} run using default parameters. This matches the theoretical analysis in \citet{reshef2015estimating}, which shows that the complexity of the search procedure in $\MICestE$ is $O(n^{5\alpha/2})$ whereas the complexity of the search procedure in the APPROX-MIC algorithm used to compute $\MIC$ is $O(n^{4\alpha})$. Second, even when equitability is prioritized, the runtime of $\MICestE$ is comparable with or faster than that of most of the other leading measures of dependence. The two exceptions to this are RDC and maximum correlation, which are both quite fast even at very large sample sizes.

We note one interesting feature of the runtime of $\MICestE$. Since estimating $\popMIC$ involves a search procedure, runtimes for estimating it are substantially faster when data contain less noise; as such, the runtimes on statistically independent presented in Table~\ref{tab:runtimes} represent worst-case performance. When run on data drawn from a noiseless linear relationship at the same sample sizes, $\MICestE$ ran 10\%-75\% faster across the range of sample sizes tested when using settings that optimize for equitability, 5\%-50\% faster across the sample sizes tested when using settings intended to achieve equitability on a limited computational budget, and 10\%-30\% faster across the sample sizes tested when using settings that optimize for power against independence. The runtime of $\DDP$ exhibited a similar phenomenon, but the runtimes of the other methods were insensitive to the level of structure present and did not exhibit this effect.

\subsection{Discussion}
In this section we analyzed the runtimes of $\MICestE$/$\TICestE$ alongside other leading measures of dependence at sample sizes ranging from $50$ to $10,000$. Our main finding is that $\MICestE$/$\TICestE$ is faster than or comparable to most of the other methods tested, and is much faster than its predecessor $\MIC$. Specifically, with parameters chosen to yield state-of-the-art power for $\TICestE$ and approximately $80\%$ of the best achievable equitability for $\MICestE$, both statistics can be computed on a sample size of $5,000$ in $97$ milliseconds. For a data set with $n=5,000$ consisting of $1,000$ variables, this translates into a total runtime of $8.1$ minutes to compute both statistics for all variable pairs on a cluster with $100$ nodes. These numbers imply that analysis of even relatively large data sets is possible using $\MICestE$ and $\TICestE$.

We emphasize that our results represent a snapshot based on currently available implementations. Just as $\MICestE$ has provided an improvement over APPROX-MIC, and just as recent advances are providing ways for estimating distance correlation in time $O(n \log n)$ rather than $O(n^2)$, we expect that with time algorithmic improvements will allow for more efficient computation of some of the newer methods analyzed here.

\section{Conclusion}
\label{sec:discussion}
In this paper, we presented an in-depth empirical evaluation of the equitability, power against independence, and runtime of several leading measures of dependence, including two new statistics introduced in \citet{reshef2015estimating}. Our aims were to give an accessible exposition of equitability and its relationship to power against independence, provide the community with a comprehensive and rigorous side-by-side comparison of existing methods, and evaluate the new statistics against the existing state of the art.  Our main findings were as follows.

\begin{enumerate}

\item {\sl Equitability.} $\MICestE$, the estimator of the population $\MIC$ introduced in \citet{reshef2015estimating}, generally has superior and more robust equitability with respect to $R^2$ than other measures of dependence. In some specific settings (models with no noise in the independent variable and $n=5,000$), mutual information estimation achieves superior equitability in our experiments, but its equitability is otherwise highly variable and often poor, particularly at lower sample sizes. Maximal correlation achieves some degree of equitability over the models examined, but all other statistics tested have very poor equitability.

\item {\sl Power against independence}. $\TICestE$, a statistic introduced in \citet{reshef2015estimating}, shares state-of-the-art power against independence with Heller and Gorfine's $\DDP$, with both methods generally performing very well and alternately outperforming each other in different settings. $\MICestE$ also has power against independence that is competitive with the state of the art, albeit under parameter settings that differ from those that confer good equitability. Moreover, the power of independence testing using $\TICestE$ and $\MICestE$ is much less sensitive than that of the other methods examined to alternative hypothesis relationship type. The original statistic $\MIC$ has substantially higher power against independence than has been reported in previous analyses when a different parameter setting is used. Finally, distance correlation, maximal correlation, $\HHG$, $\HSIC$, and $\RDC$ also had good power against independence.

\item {\sl Power/equitability tradeoff.} The parameter $\alpha$ in the estimator $\MICestE$ corresponds to a trade-off between power against independence and equitability that is consistent with the characterization of equitability given in \citet{reshef2015equitability}. Lower values of $\alpha$ lead to higher power against a null of independence at the expense of power against null hypotheses representing weak relationship strength (i.e., equitability), while higher values of $\alpha$ lead to better equitability at the expense of power against independence.

\item {\sl Runtime.} $\MICestE$ and $\TICestE$, each of which can be trivially computed once the other has been obtained, have runtimes that allow them to be run together even on large samples in reasonable time. This runtime compares favorably with that of other complex measures of dependence such as $\DDP$, $\dCor$, $\HSIC$, and $\HHG$. The fastest measures of dependence were maximal correlation and the randomized dependence coefficient. There is a large variety of runtimes across the measures of dependence examined.
\end{enumerate}

There are several important takeaways from our results. First, they suggest that using $\MICestE$ and $\TICestE$ in tandem to filter relationships and rank them by strength is a statistically sound and computationally efficient strategy for exploratory data analysis. In particular, one can imagine a system in which first $\TICestE$ is computed for all relationships and only the significant ones are kept, and then $\MICestE$ with equitability-optimized parameters is examined only for the latter set. Since $\TICestE$ enjoys high power against independence on a wide range of alternative hypothesis relationship types, pre-filtering with $\TICestE$ in this way will not result in important relationships being overlooked due to their relationship type. Any measure of dependence deemed to have sufficient power on a broad range of alternative hypotheses can be substituted for $\TICestE$. However, since $\TICestE$ and $\MICestE$ can be computed simultaneously, and since $\TICestE$ offers state-of-the-art power against independence, using $\TICestE$ appears to be a preferable choice in such a scenario.

Second, the fact that many measures of dependence performed similarly in our analysis of power against independence, as well as in analyses of real data sets that others have performed (see, e.g., \cite{heller2014consistent}), suggests that power against independence may not be where the true challenge lies for bivariate relationships, and that we ought to demand more of the measures of dependence that we use in this setting. Equitability is one attempt to formulate a more ambitious goal, as is the concept of low detection threshold introduced in \citet{reshef2015equitability}, but there may well be other possibilities. Of course, for higher-dimensional relationships, even just power against independence is very difficult to achieve, and many of the methods evaluated here are quite useful in that setting.

Finally, the comprehensiveness of our results provides significant understanding of the comparative performance of various measures. To our knowledge, our analyses are the most exhaustive to date in that they evaluate a large swath of measures of dependence side-by-side along a number of dimensions (equitability, power against independence, and runtime); over a wide range of models, relationship types, and sample sizes; and with parameters that are optimized for each individual statistic in each analysis. Our hope is that the full set of results, which can be downloaded in bulk at \onlineSupplementLink ~will be a resource to the community that enables more consistent, direct comparisons between different measures of dependence, and facilitates a precise discussion of the trade-offs and assumptions associated with each one in various settings.

While the results presented here make a compelling case for the use of $\MICestE$ and $\TICestE$ and provide insight into the trade-offs between different measures of dependence, there are some important limitations for both the new statistics and the comparisons we performed. First, in this paper we evaluated only equitability with respect to $R^2$ on noisy functional relationships, whereas the definition we give of equitability explicitly acknowledges the possibility of using other properties of interest besides $R^2$ and standard relationships that are not noisy functional relationships. We feel that $R^2$ is an important property of interest that is intuitive and familiar to many practitioners, but equitability with respect to other properties of interest merits study as well, and the methods tested here may perform much better or worse when their equitability is evaluated with respect to other properties of interest.

Additionally, though an attempt at comprehensiveness was made, we did limit our scope to the set of noisy functional relationships in \citet{MINE} for equitability and the relationships introduced in \citet{simon2012comment} for power against independence. While we feel each of these suites of relationships provides reasonable insight into the performance of the methods in question, there are relationships that, when added to these suites, result in extremely poor performance for all the methods tested. Characterizing those relationships theoretically in both the setting of equitability and that of power against independence is important if we are to fully understand the strengths and weaknesses of each of these methods. This is an important direction for future work.

Measures of dependence are useful in a variety of settings and identifying which measures of dependence provide superior performance in the face of different objectives, assumptions, and constraints is critical. For each separate goal, we must understand both which measure of dependence is most appropriate and also which parameter settings lead to the best performance. Such an understanding provides insight into the inherent trade-offs of different methods, allowing us to navigate the landscape of measures of dependence more effectively and --- ultimately --- to better understand our data.

\section{Acknowledgments}
The authors would like to acknowledge E Airoldi, K Arnold, H Finucane, A Gelman, M Gorfine, A Gretton, T Hashimoto, R Heller, J Hern\'andez-Lobato, J Huggins, T Jaakkola, A Miller, J Mueller, A Narayanan, G Szekely, J Tenenbaum, and R Tibshirani for constructive conversations and useful feedback.

 \bibliographystyle{plainnat}		
\bibliography{\pathToCommon/References}

\begin{thebibliography}{32}
\providecommand{\natexlab}[1]{#1}
\providecommand{\url}[1]{\texttt{#1}}
\expandafter\ifx\csname urlstyle\endcsname\relax
  \providecommand{\doi}[1]{doi: #1}\else
  \providecommand{\doi}{doi: \begingroup \urlstyle{rm}\Url}\fi

\bibitem[Breiman and Friedman(1985)]{breiman1985estimating}
Leo Breiman and Jerome~H Friedman.
\newblock Estimating optimal transformations for multiple regression and
  correlation.
\newblock \emph{Journal of the American statistical Association}, 80\penalty0
  (391):\penalty0 580--598, 1985.

\bibitem[Cover and Thomas(2006)]{Cover2006}
T.~Cover and J.~Thomas.
\newblock \emph{Elements of Information Theory}.
\newblock New York: John Wiley \& Sons, Inc, 2006.

\bibitem[Csisz{\'a}r and Shields(2004)]{csiszar2004information}
Imre Csisz{\'a}r and Paul~C Shields.
\newblock Information theory and statistics: A tutorial.
\newblock \emph{Communications and Information Theory}, 1\penalty0
  (4):\penalty0 417--528, 2004.

\bibitem[Emilsson et~al.(2008)Emilsson, Thorleifsson, Zhang, Leonardson, Zink,
  Zhu, Carlson, Helgason, Walters, Gunnarsdottir, et~al.]{emilsson2008genetics}
Valur Emilsson, Gudmar Thorleifsson, Bin Zhang, Amy~S Leonardson, Florian Zink,
  Jun Zhu, Sonia Carlson, Agnar Helgason, G~Bragi Walters, Steinunn
  Gunnarsdottir, et~al.
\newblock Genetics of gene expression and its effect on disease.
\newblock \emph{Nature}, 452\penalty0 (7186):\penalty0 423--428, 2008.

\bibitem[Gorfine et~al.(2012)Gorfine, Heller, and Heller]{gorfine2012comment}
M.~Gorfine, R.~Heller, and Y.~Heller.
\newblock Comment on ``{D}etecting novel associations in large data sets''.
\newblock \emph{Unpublished (available at
  http://emotion.technion.ac.il/$\sim$gorfinm/files/science6.pdf on 11 Nov.
  2012)}, 2012.

\bibitem[Gretton et~al.(2005)Gretton, Bousquet, Smola, and
  Sch{\"o}lkopf]{gretton2005measuring}
Arthur Gretton, Olivier Bousquet, Alex Smola, and Bernhard Sch{\"o}lkopf.
\newblock Measuring statistical dependence with hilbert-schmidt norms.
\newblock In \emph{Algorithmic learning theory}, pages 63--77. Springer, 2005.

\bibitem[Gretton et~al.(2008)Gretton, Fukumizu, Teo, Song, Sch{\"o}lkopf, and
  Smola]{gretton2008kernel}
Arthur Gretton, Kenji Fukumizu, Choon~Hui Teo, Le~Song, Bernhard Sch{\"o}lkopf,
  and Alex~J Smola.
\newblock A kernel statistical test of independence.
\newblock 2008.

\bibitem[Gretton et~al.(2012)Gretton, Borgwardt, Rasch, Sch{\"o}lkopf, and
  Smola]{gretton2012kernel}
Arthur Gretton, Karsten~M Borgwardt, Malte~J Rasch, Bernhard Sch{\"o}lkopf, and
  Alexander Smola.
\newblock A kernel two-sample test.
\newblock \emph{The Journal of Machine Learning Research}, 13\penalty0
  (1):\penalty0 723--773, 2012.

\bibitem[Heller et~al.(2013)Heller, Heller, and Gorfine]{heller2013consistent}
Ruth Heller, Yair Heller, and Malka Gorfine.
\newblock A consistent multivariate test of association based on ranks of
  distances.
\newblock \emph{Biometrika}, 100\penalty0 (2):\penalty0 503--510, 2013.

\bibitem[Heller et~al.(2014)Heller, Heller, Kaufman, Brill, and
  Gorfine]{heller2014consistent}
Ruth Heller, Yair Heller, Shachar Kaufman, Barak Brill, and Malka Gorfine.
\newblock Consistent distribution-free $k$-sample and independence tests for
  univariate random variables.
\newblock \emph{arXiv preprint arXiv:1410.6758}, 2014.

\bibitem[Hoeffding(1948)]{hoeffding1948non}
Wassily Hoeffding.
\newblock A non-parametric test of independence.
\newblock \emph{The Annals of Mathematical Statistics}, pages 546--557, 1948.

\bibitem[Huo and Szekely(2014)]{huo2014fast}
Xiaoming Huo and Gabor~J Szekely.
\newblock Fast computing for distance covariance.
\newblock \emph{arXiv preprint arXiv:1410.1503}, 2014.

\bibitem[Jiang et~al.(2014)Jiang, Ye, and Liu]{jiang2014non}
Bo~Jiang, Chao Ye, and Jun~S Liu.
\newblock Non-parametric k-sample tests via dynamic slicing.
\newblock \emph{Journal of the American Statistical Association}, \penalty0
  (just-accepted):\penalty0 00--00, 2014.

\bibitem[Kinney and Atwal(2014)]{kinney2014equitability}
Justin~B. Kinney and Gurinder~S. Atwal.
\newblock Equitability, mutual information, and the maximal information
  coefficient.
\newblock \emph{Proceedings of the National Academy of Sciences}, 2014.
\newblock \doi{10.1073/pnas.1309933111}.

\bibitem[Kraskov et~al.(2004)Kraskov, Stogbauer, and Grassberger]{Kraskov}
A.~Kraskov, H.~Stogbauer, and P.~Grassberger.
\newblock Estimating mutual information.
\newblock \emph{Physical Review E}, 69, 2004.

\bibitem[Linfoot(1957)]{linfoot1957informational}
E.H. Linfoot.
\newblock An informational measure of correlation.
\newblock \emph{Information and Control}, 1\penalty0 (1):\penalty0 85--89,
  1957.

\bibitem[Lopez-Paz(2015)]{lopez2015personal}
David Lopez-Paz.
\newblock Personal communication, April 2015.

\bibitem[Lopez-Paz et~al.(2013)Lopez-Paz, Hennig, and
  Sch{\"o}lkopf]{lopez2013randomized}
David Lopez-Paz, Philipp Hennig, and Bernhard Sch{\"o}lkopf.
\newblock The randomized dependence coefficient.
\newblock In \emph{Advances in Neural Information Processing Systems}, pages
  1--9, 2013.

\bibitem[Moon et~al.(1995)Moon, Rajagopalan, and Lall]{moon1995estimation}
Young-Il Moon, Balaji Rajagopalan, and Upmanu Lall.
\newblock Estimation of mutual information using kernel density estimators.
\newblock \emph{Physical Review E}, 52\penalty0 (3):\penalty0 2318--2321, 1995.

\bibitem[Murrell et~al.(2014)Murrell, Murrell, and Murrell]{Murrell2014comment}
Ben Murrell, Daniel Murrell, and Hugh Murrell.
\newblock R2-equitability is satisfiable.
\newblock \emph{Proceedings of the National Academy of Sciences}, 2014.
\newblock \doi{10.1073/pnas.1403623111}.
\newblock URL
  \url{http://www.pnas.org/content/early/2014/04/29/1403623111.short}.

\bibitem[Paninski(2003)]{paninski2003estimation}
Liam Paninski.
\newblock Estimation of entropy and mutual information.
\newblock \emph{Neural computation}, 15\penalty0 (6):\penalty0 1191--1253,
  2003.

\bibitem[R{\'e}nyi(1959)]{renyi1959measures}
Alfred R{\'e}nyi.
\newblock On measures of dependence.
\newblock \emph{Acta mathematica hungarica}, 10\penalty0 (3):\penalty0
  441--451, 1959.

\bibitem[Reshef et~al.(2013)Reshef, Reshef, Mitzenmacher, and
  Sabeti]{reshef2013equitability}
David Reshef, Yakir Reshef, Michael Mitzenmacher, and Pardis Sabeti.
\newblock Equitability analysis of the maximal information coefficient, with
  comparisons.
\newblock \emph{arXiv preprint arXiv:1301.6314}, 2013.

\bibitem[Reshef et~al.(2011)Reshef, Reshef, Finucane, Grossman, McVean,
  Turnbaugh, Lander, Mitzenmacher, and Sabeti]{MINE}
David~N Reshef, Yakir~A Reshef, Hilary~K Finucane, Sharon~R Grossman, Gilean
  McVean, Peter~J Turnbaugh, Eric~S Lander, Michael Mitzenmacher, and Pardis~C
  Sabeti.
\newblock Detecting novel associations in large data sets.
\newblock \emph{Science}, 334\penalty0 (6062):\penalty0 1518--1524, 2011.

\bibitem[Reshef et~al.(2014)Reshef, Reshef, Mitzenmacher, and
  Sabeti]{reshef2014comment}
David~N. Reshef, Yakir~A. Reshef, Michael Mitzenmacher, and Pardis~C. Sabeti.
\newblock Cleaning up the record on the maximal information coefficient and
  equitability.
\newblock \emph{Proceedings of the National Academy of Sciences}, 2014.
\newblock \doi{10.1073/pnas.1408920111}.
\newblock URL
  \url{http://www.pnas.org/content/early/2014/08/07/1408920111.short}.

\bibitem[Reshef et~al.(2015{\natexlab{a}})Reshef, Reshef, Finucane, Sabeti, and
  Mitzenmacher]{reshef2015estimating}
Yakir~A Reshef, David~N Reshef, Hilary~K Finucane, Pardis~C Sabeti, and Michael
  Mitzenmacher.
\newblock Measuring dependence powerfully and equitably.
\newblock \emph{arXiv preprint arXiv:1505.02213}, 2015{\natexlab{a}}.

\bibitem[Reshef et~al.(2015{\natexlab{b}})Reshef, Reshef, Sabeti, and
  Mitzenmacher]{reshef2015equitability}
Yakir~A Reshef, David~N Reshef, Pardis~C Sabeti, and Michael Mitzenmacher.
\newblock Equitability, interval estimation, and statistical power.
\newblock \emph{arXiv preprint arXiv:1505.02212}, 2015{\natexlab{b}}.

\bibitem[Sejdinovic et~al.(2013)Sejdinovic, Sriperumbudur, Gretton, Fukumizu,
  et~al.]{sejdinovic2013equivalence}
Dino Sejdinovic, Bharath Sriperumbudur, Arthur Gretton, Kenji Fukumizu, et~al.
\newblock Equivalence of distance-based and rkhs-based statistics in hypothesis
  testing.
\newblock \emph{The Annals of Statistics}, 41\penalty0 (5):\penalty0
  2263--2291, 2013.

\bibitem[Simon and Tibshirani(2012)]{simon2012comment}
N.~Simon and R.~Tibshirani.
\newblock Comment on ``{D}etecting novel associations in large data sets''.
\newblock \emph{Unpublished (available at
  http://www-stat.stanford.edu/$\sim$tibs/reshef/comment.pdf on 11 Nov. 2012)},
  2012.

\bibitem[Speed(2011)]{speed2011correlation}
T.~Speed.
\newblock A correlation for the 21st century.
\newblock \emph{Science}, 334\penalty0 (6062):\penalty0 1502--1503, 2011.

\bibitem[Sz{\'e}kely et~al.(2007)Sz{\'e}kely, Rizzo, Bakirov,
  et~al.]{szekely2007measuring}
G{\'a}bor~J Sz{\'e}kely, Maria~L Rizzo, Nail~K Bakirov, et~al.
\newblock Measuring and testing dependence by correlation of distances.
\newblock \emph{The Annals of Statistics}, 35\penalty0 (6):\penalty0
  2769--2794, 2007.

\bibitem[Szekely and Rizzo(2009)]{szekely2009brownian}
G.J. Szekely and M.L. Rizzo.
\newblock Brownian distance covariance.
\newblock \emph{The Annals of Applied Statistics}, 3\penalty0 (4):\penalty0
  1236--1265, 2009.

\end{thebibliography}

\newpage
\appendix
\counterwithin{table}{section}
\counterwithin{figure}{section}
\section{Data generation}
\subsection{Definitions of functions used}
\label{app:functionsUsed}

Tables~\ref{table:fctSuiteEquitability} and~\ref{table:fctSuitePower} contain the definitions of the functions used to assess the equitability and statistical power against independence, respectively, of measures of dependence throughout this paper.  The functions used for all analyses of power against independence (Table~\ref{table:fctSuitePower}) are taken from~\cite{simon2012comment}.

\begin{table}[H]
\centering
{\scriptsize
\begin{tabular}{clll}
\toprule
\textbf{\#} &    \textbf{Function Name} 			& \textbf{Definition}											& \\
\midrule
1 & Cosine, High Freq				& $y = \cos(14\pi x)$											& $x \in [0,1]$ \\
\hline
2 & Cosine, Non-Fourier Freq [Low]		& $y = \cos(7\pi x)$											& $x \in [0,1]$ \\
\hline
3 & Cosine, Varying Freq [Medium]		& $y = \sin(5\pi x (1+x)) $										& $x \in [0,1]$ \\
\hline
4 & Cubic						& $y = 4x^3 + x^2 - 4x$										& $x \in [-1.3,1.1]$ \\
\hline
5 & Cubic, Y-stretched				& $y = 41(4x^3 + x^2 - 4x)$									& $x \in [-1.3,1.1]$ \\
\hline
6 & Exponential [$10^x$]				& $y = 10^x$												& $x \in [0,10]$ \\
\hline
7 & Exponential [$2^x$]				& $y = 2^x$												& $x \in [0,10]$ \\
\hline
8 & L-shaped						& $y = \begin{cases}
									x/99		& \text{if } x \leq \frac{99}{100} \\
									1		& \text{if } x > \frac{99}{100}
								\end{cases} $											& $x \in [0,1]$ \\
\hline
9 & Line							& $y = x$													& $x \in [0,1]$ \\
\hline
10 & Linear+Periodic, High Freq		& $y = \frac{1}{10}\sin(10.6(2x-1)) + \frac{11}{10}(2x-1)$				& $x \in [0,1]$ \\
\hline
11 & Linear+Periodic, High Freq 2		& $y = \frac{1}{5}\sin(10.6(2x-1)) + \frac{11}{10}(2x-1)$				& $x \in [0,1]$ \\
\hline
12 & Linear+Periodic, Low Freq		& $y = \frac{1}{5}\sin(4(2x-1)) + \frac{11}{10}(2x-1)$ 					& $x \in [0,1]$ \\
\hline
13 & Linear+Periodic, Medium Freq		& $y = \sin(10\pi x) + x$										& $x \in [0,1]$ \\
\hline
14 & Lopsided L-shaped				& $y = \begin{cases}
									200x 						& \text{if } x < \frac{1}{200} \\
									-198x + \frac{199}{100}		& \text{if } \frac{1}{200} \leq x < \frac{1}{100} \\
									-\frac{x}{99} + \frac{1}{99}	& \text{if } x \geq \frac{1}{100}
								\end{cases} $											& $x \in [0,1]$ \\
\hline
15 & Parabola						& $y = 4x^2$												& $x \in [-\frac{1}{2},\frac{1}{2}]$\\
\hline
16 & Sigmoid						& $y = \begin{cases}
									0						& \text{if } x \leq \frac{49}{100} \\
									50(x-\frac{1}{2}) + \frac{1}{2}	& \text{if } \frac{49}{100} \leq x \leq \frac{51}{100} \\
									1						& \text{if } x > \frac{51}{100}
								\end{cases} $											& $x \in [0,1]$ \\
\hline
17 & Sine, High Freq					& $y = \sin(16\pi x)$											& $x \in [0,1]$\\
\hline
18 & Sine, Low Freq					& $y = \sin(8\pi x)$											& $x \in [0,1]$\\
\hline
19 & Sine, Non-Fourier Freq [Low]		& $y = \sin(9\pi x)$											& $x \in [0,1]$ \\
\hline
20 & Sine, Varying Freq [Medium]		& $y = \sin(6\pi x (1+x)) $										& $x \in [0,1]$ \\
\hline
21 & Spike						& $y = \begin{cases}
									20 						& \text{if } x < \frac{1}{20} \\
									-18x + \frac{19}{10}		& \text{if } \frac{1}{20} \leq x < \frac{1}{10} \\
									-\frac{x}{9} + \frac{1}{9}	& \text{if } x \geq \frac{1}{10}
								\end{cases} $											& $x \in [0,1]$ \\
\bottomrule
\end{tabular}}
\caption[Definitions of the functions used to analyze equitability.]{Definitions of the functions used to analyze equitability. Under noise/sampling models containing noise in the independent variable or independent-variable marginal distributions other than $E_{f(X)}$ or $U_{f(X)}$, functions 6, 8, 14, 16, and 21 were excluded due to poor performance across all methods tested. This is presumably due to the fact that a) horizontally perturbing points in a very steep portion of a function drastically changes the distribution in question, and b) sampling uniformly along the x-axis drastically under-samples a large part of the graph of a function if that graph contains very steep portions.}
\label{table:fctSuiteEquitability}
\end{table}

\begin{table}[H]
\centering
{\scriptsize
\begin{tabular}{lll}
\toprule
\textbf{Function Name} 			& \textbf{Definition}											& \\
\midrule
Line							& $y = x$													& $x \in [0,1]$ \\
\hline
Quadratic						& $y = 4x^2$												& $x \in [-\frac{1}{2},\frac{1}{2}]$\\
\hline
Cubic						& $y = 128(x-\frac{1}{3})^3-48(x-\frac{1}{3})^3-12(x-\frac{1}{3})$			& $x \in [0, 1]$ \\
\hline
Sinusoid (8 periods)				& $y = \sin(16\pi x)$											& $x \in [0,1]$\\
\hline
Sinusoid (2 periods)				& $y = \sin(4\pi x)$											& $x \in [0,1]$\\
\hline
$x^{1/4}$						& $y = x^{1/4}$												& $x \in [0,1]$ \\
\hline
Circle						& $y = \pm\sqrt{1-(2x-1)^2}$					& $x \in [0,1]$ \\
\hline
Step						& $y = \begin{cases}
									0						& \text{if } x \leq \frac{1}{2} \\
									1						& \text{if } x > \frac{1}{2}
								\end{cases} $											& $x \in [0,1]$ \\
\bottomrule
\end{tabular}}
\caption[Definitions of the functions used to analyze statistical power against independence.]{Definitions of the functions from~\cite{simon2012comment} used to analyze statistical power against independence.}
\label{table:fctSuitePower}
\end{table}

\subsection{Generating a sample from a distribution with a specified $R^2$}
\label{app:dataGen}
Given a noisy functional relationship of the form $(X + \ep, f(X) + \ep')$, the $R^2$ of the relationship is the correlation between $f(X + \ep)$ and $f(X) + \ep'$. Many of the equitability and power analyses performed in this work require the ability to set $\ep$ and $\ep'$ such that the resulting distribution has a given population $R^2$.

In the case that $\ep = 0$ and the variance of $\ep'$ is known, the $R^2$ of the distribution has a closed form expression given in \citet{MINE}. If we specialize that expression to the case we consider in this paper, wherein $\ep' \sim \mathcal{N}(0, \sigma^2)$, and then solve for $\sigma$, we obtain the following expression.
\[
\sigma(R^2) = \sqrt{ \mbox{var}(f(X)) \left(\frac{1}{R^2} - 1\right)}
\]

In cases that include noise in the independent variable, we set $\ep$ and ($\ep'$ if the noise model requires) by binary search, using the sample $R^2$ of a very large sample as an estimate of the population $R^2$.

\section{Additional equitability results}
\label{app:additionalEquitabilityResults}

\begin{table}[H]
	\centering
	\includegraphics[clip=true, trim = 0in 0.75in 1.375in 0in, width=\textwidth]{\pathToFigures/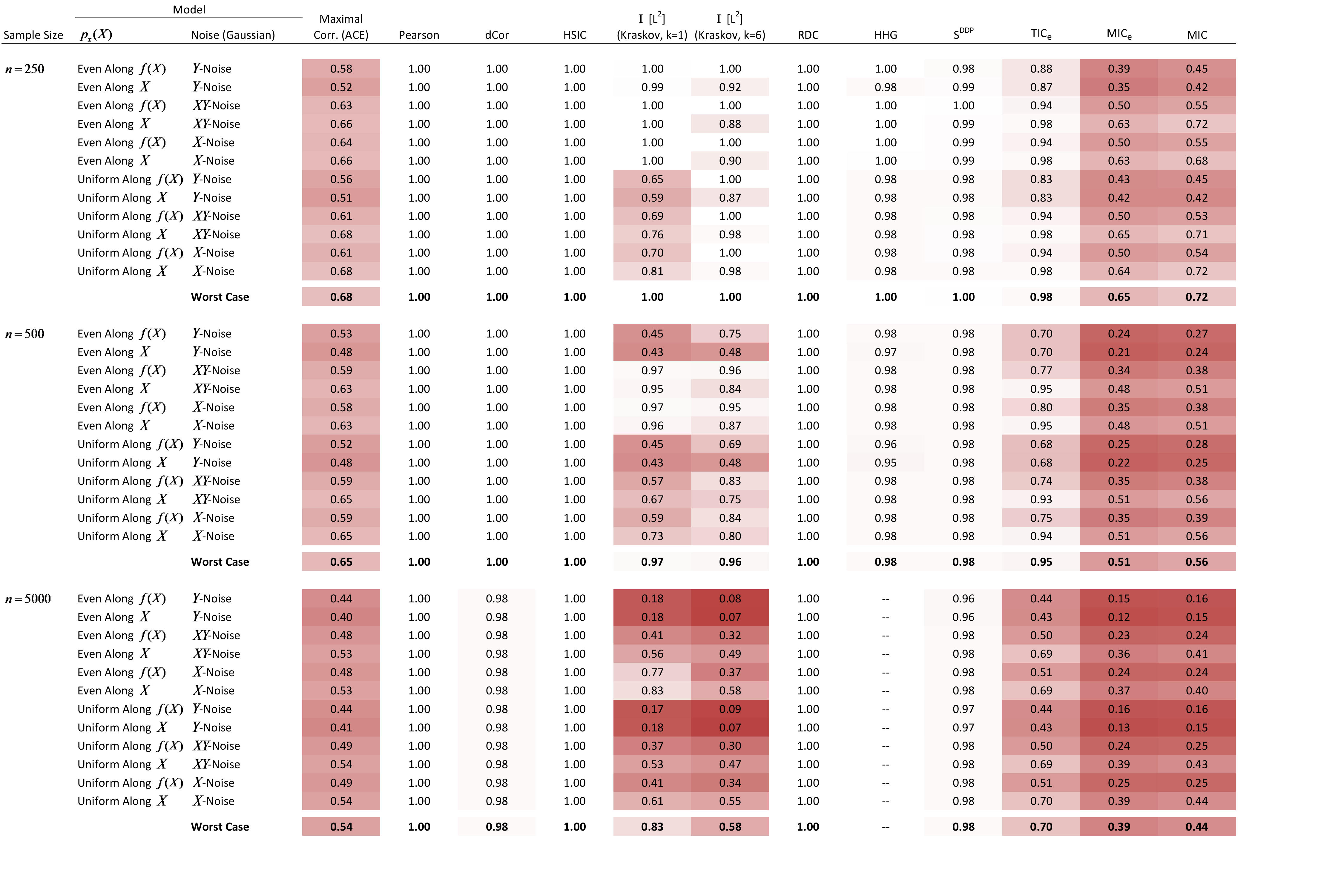}
	\caption[Summary of worst-case equitability of measures of dependence for a variety of noise/sampling models and sample sizes]{
	A summary of the worst-case equitability of measures of dependence for a variety of noise models, independent-variable marginal distributions, and sample sizes. \textit{[Smaller values correspond to better equitability.]}
	Each number is a worst-case interpretable interval length for a given statistic in a given setting. Therefore, smaller numbers indicate shorter interpretable intervals and more equitable behavior. Table cells are colored proportionally (red = interval of length 0; white = interval of length 1). The equitability of $\MICestE$ is relatively robust to factors like noise models, independent variable marginal distributions, and sample size. Figures analogous to Figures~\ref{fig:equitabilityAnalysis_evenCurve_XYNoise} and~\ref{fig:equitabilityAnalysis_evenCurve_YNoise} for all the settings presented in this table are included in the online supplementary materials. For statistics whose performance was dependent on parameter settings, we present for each sample size the best results across parameter values tested. Results are not presented for $\HHG$ for $n=5,000$ as it was prohibitively computationally expensive to analyze at this sample size.}
	\label{table:equitability_worstCase}
\end{table}

\begin{table}[H]
	\centering
	\includegraphics[clip=true, trim = 0in 0.75in 1.375in 0in, width=\textwidth]{\pathToFigures/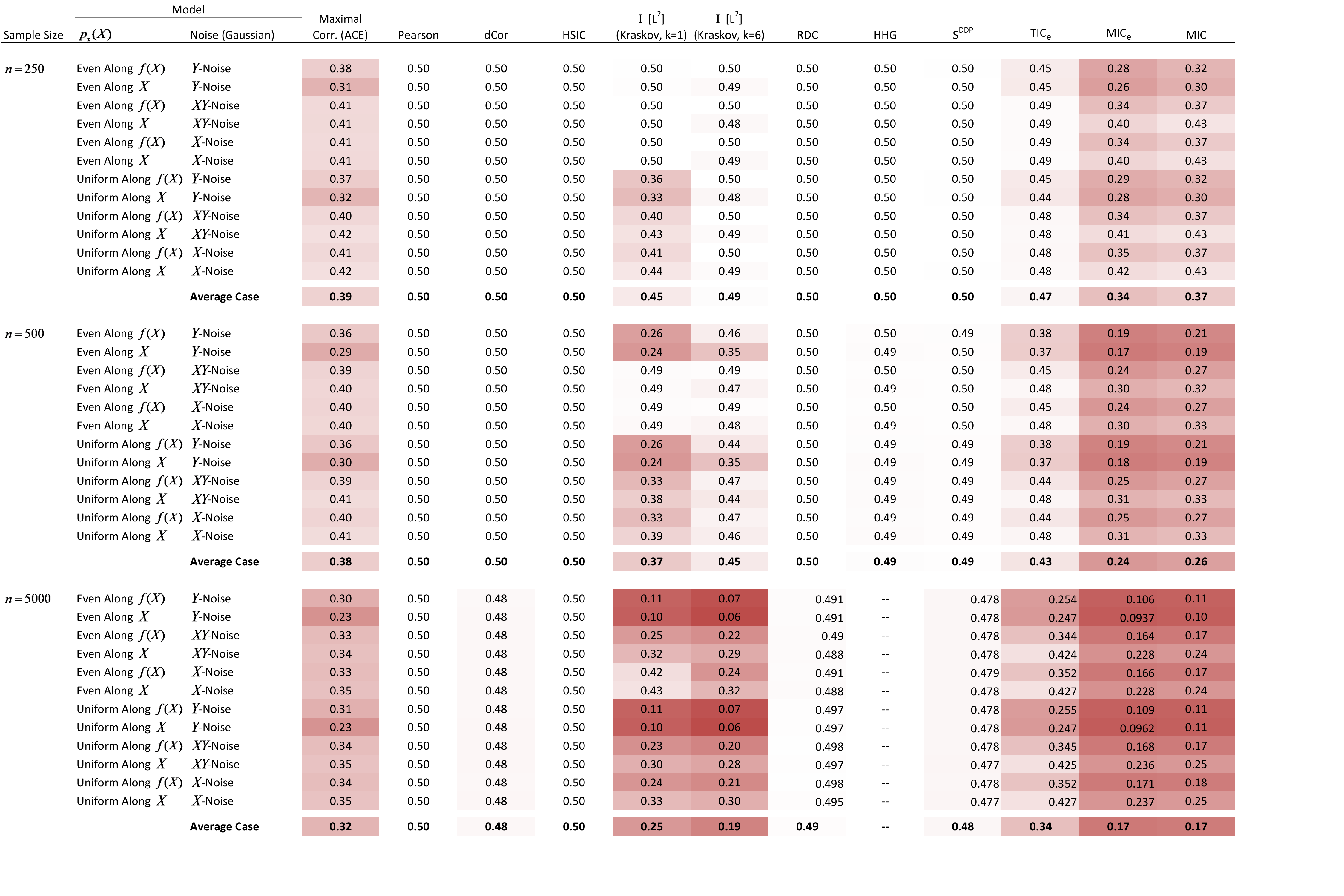}
	\caption[Summary of average-case equitability of measures of dependence for a variety of noise/sampling models and sample sizes]{
	A summary of the average-case equitability of measures of dependence for a variety of noise models, independent-variable marginal distributions, and sample sizes. \textit{[Smaller values correspond to better equitability.]}
	Each number is an average interpretable interval length for a given statistic in a given setting. Therefore, smaller numbers indicate shorter interpretable intervals on average and more equitable behavior. Table cells are colored proportionally (red = interval of length 0; white = interval of length 1). The equitability of $\MICestE$ is relatively robust to factors like noise models, independent variable marginal distributions, and sample size.  Figures analogous to Figures~\ref{fig:equitabilityAnalysis_evenCurve_XYNoise} and~\ref{fig:equitabilityAnalysis_evenCurve_YNoise} for all the settings presented in this table are included in the online supplementary materials. For statistics whose performance was dependent on parameter settings, we present for each sample size the best results across parameter values tested. Results are not presented for $\HHG$ for $n=5,000$ as it was prohibitively computationally expensive to analyze at this sample size.}
	\label{table:equitability_aveCase}
\end{table}

\begin{figure}[H]
	\centering
	\begin{minipage}[b][\textheight][t]{0.62\linewidth}
		\includegraphics[clip=true, trim = 0.15in 1in 0.7in 0.6in, width=\textwidth]{\pathToFigures/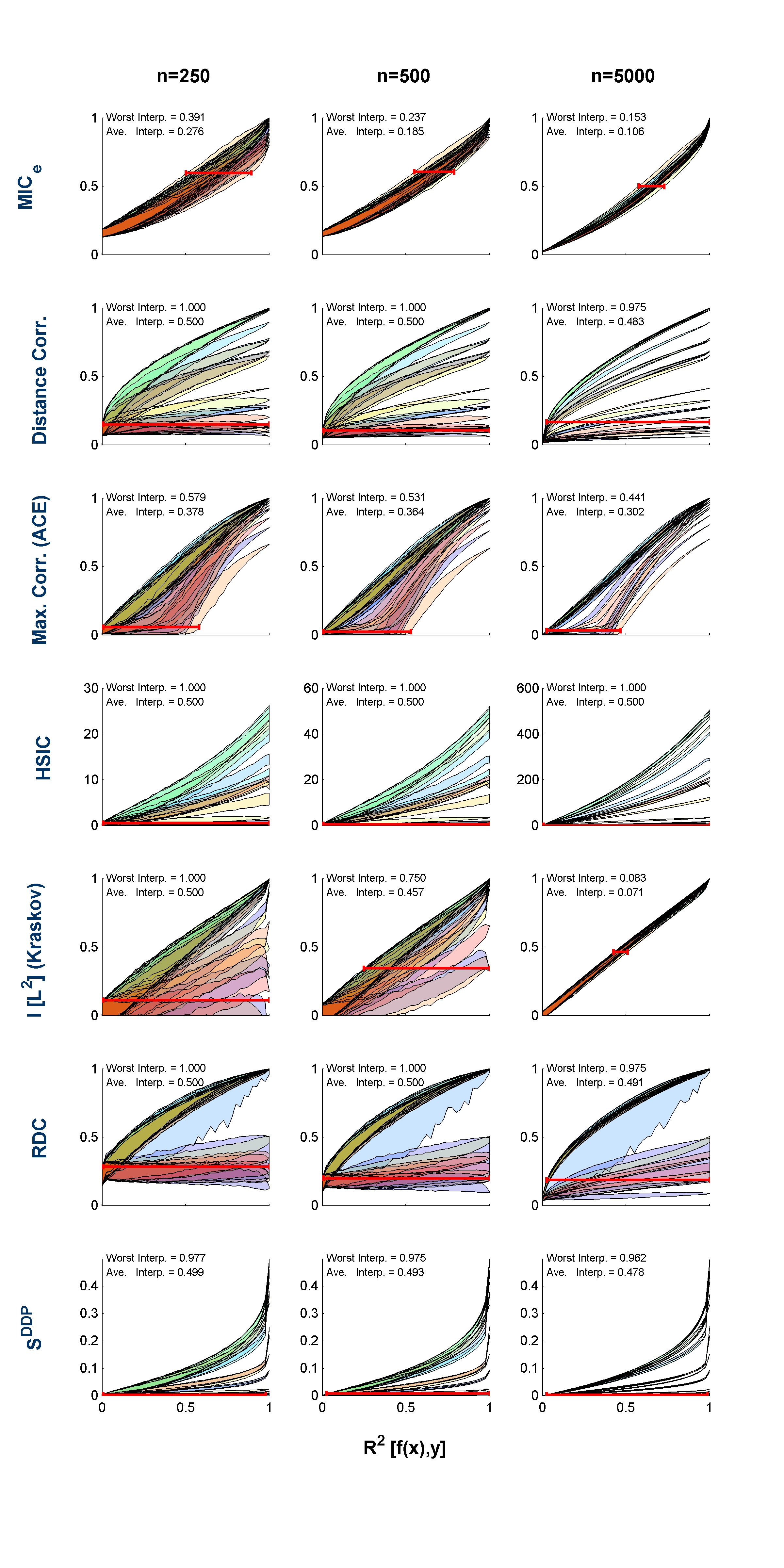}
	\end{minipage}
	\hfill
	\begin{minipage}[b][\textheight][t]{0.34\linewidth}
		\includegraphics[clip=true, trim = 4.7in 0.25in 2.25in 0.35in, width=0.9\textwidth]{\pathToFigures/PearsonEquitabilityPlotWithLegend.png}
		\caption[The equitability of measures of dependence on noisy functional relationships, with noiseless independent variable]{The equitability of measures of dependence on a set $\Q$ of noisy functional relationships with alternative noise model and marginal distribution. \textit{[Narrower is more equitable.]}
		The relationships take the form $(X, f(X)+\ep')$ where $\ep'$ is normally distributed with varying amplitude, and relationship strength is quantified by $\Phi = R^2$. The plots were constructed as described in Figure~\ref{fig:whatIsEquitability}. In contrast to its poor equitability under the noise model used in Figure~\ref{fig:equitabilityAnalysis_evenCurve_XYNoise}, the Kraskov mutual information estimator, represented using the squared Linfoot correlation, is quite equitable under this noise model at large sample sizes. At the low and mid-range sample sizes, $\MICestE$ remains more equitable.
		For every parametrized statistic whose parameter meaningfully affects equitability, results are presented at each sample size using parameter settings that maximize equitability across all twelve of the noise/marginal distributions tested at that sample size. See Tables~\ref{table:equitability_worstCase} and~\ref{table:equitability_aveCase} for a summary of the equitability of these measures of dependence under those additional models, as well as the supplemental materials for the corresponding figures.}
	\label{fig:equitabilityAnalysis_evenCurve_YNoise}
	\end{minipage}
\end{figure}

\begin{figure}[H]
	\centering
	\begin{minipage}[b][\textheight][t]{0.62\linewidth}
		\includegraphics[clip=true, trim = 0.9in 0.825in 2.65in 0.65in, height=\textheight]{\pathToFigures/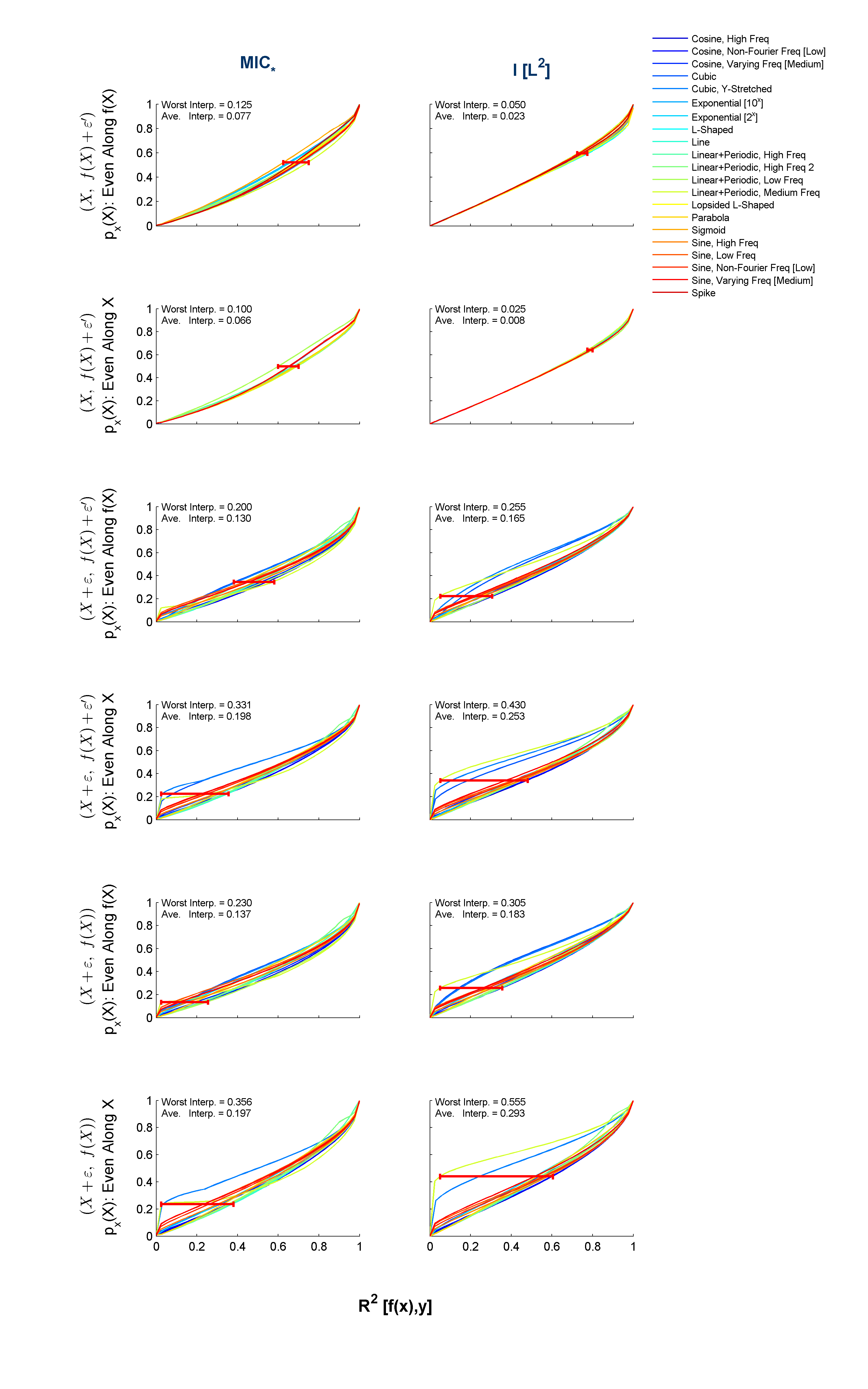}
	\end{minipage}
	\hfill
	\begin{minipage}[b][\textheight][t]{0.34\linewidth}
		\includegraphics[clip=true, trim = 7.45in 12.5in 0.45in 0.35in, width=0.9\linewidth]{\pathToFigures/Infinite_Data_Limit_Equitability_of_MIC_vs_MI__EmpiricalR2s_.png}
		\caption[The equitability of $\popMIC$ and mutual information in the infinite data limit]{
		    The equitability of $\popMIC$ and mutual information in the infinite data limit. \textit{[Narrower is more equitable.]}
    		Six combinations of noise models and independent variable marginal distributions were analyzed. The values of $\popMIC$ were computed using the newly introduced algorithm from \citet{reshef2015estimating}. In each plot, the worst-case interpretable interval is indicated by a red line, and both the worst- and average-case equitability are listed. Mutual information values are represented in terms of the squared Linfoot correlation. In the large-sample limit, mutual information is more equitable than $\popMIC$ in settings where there is noise only in the dependent variable, while $\popMIC$ has superior equitability otherwise.}
	\label{fig:equitabilityAnalysis_InfDataLimit}
	\end{minipage}
\end{figure}

\begin{figure}[H]
	\centering
	\begin{minipage}[t!]{0.66\linewidth}
	\includegraphics[clip=true, trim = 0.125in 1.0in 0.125in 0.8in, width=0.95\linewidth]{\pathToFigures/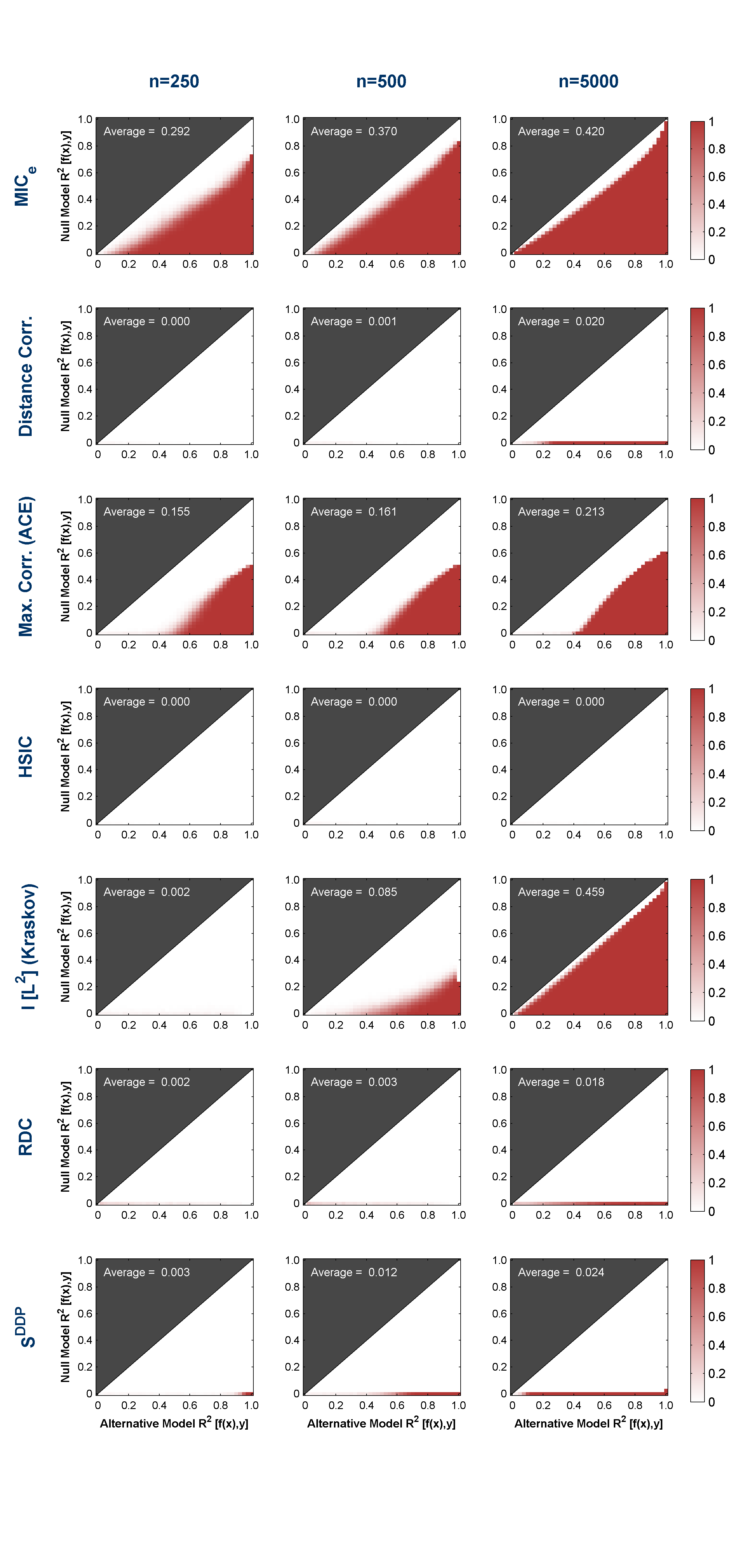}
	\end{minipage}
	\hfill
	\begin{minipage}[t!][0.9\textheight][t]{0.30\linewidth}
	\caption[The equitability of measures of dependence on noisy functional relationships with noise in the dependent variable only, visualized in terms of power.]{
    	The equitability of measures of dependence on noisy functional relationships with noise in the dependent variable only, visualized in terms of power. \textit{[Redder is more equitable.]}
    	The set $\Q$ of noisy functional relationships analyzed is the same as in Figure~\ref{fig:equitabilityAnalysis_evenCurve_YNoise}, and relationship strength is again quantified by $\Phi = R^2$.  Plots were generated as in Figure~\ref{fig:equitabilityAndPower_evenCurve_XYNoise}.
    	In contrast to its performance under the noise model used in Figure~\ref{fig:equitabilityAnalysis_evenCurve_XYNoise}, the Kraskov mutual information estimator yields powerful tests under this noise model at large sample sizes. At the low and mid-range sample sizes, tests based on $\MICestE$ remain more powerful.
    	For every parametrized statistic whose parameter meaningfully affects equitability, results are presented at each sample size using parameter settings that maximize equitability across all twelve of the noise/marginal distributions tested at that sample size. See Tables~\ref{table:equitability_worstCase} and~\ref{table:equitability_aveCase} for a summary of the equitability of these measures of dependence under those additional models, as well as the supplemental materials for the corresponding figures.}
	\label{fig:equitabilityAndPower_evenCurve_YNoise}
	\end{minipage}
\end{figure}

\section{Parameter sweeps for power against independence}
\label{app:additionalPowerResults}

\begin{figure}[H]
	\centering
	\includegraphics[clip=true, trim = 0.85in 0.45in 0.825in 0.4in, width=0.9\textwidth]{\pathToFigures/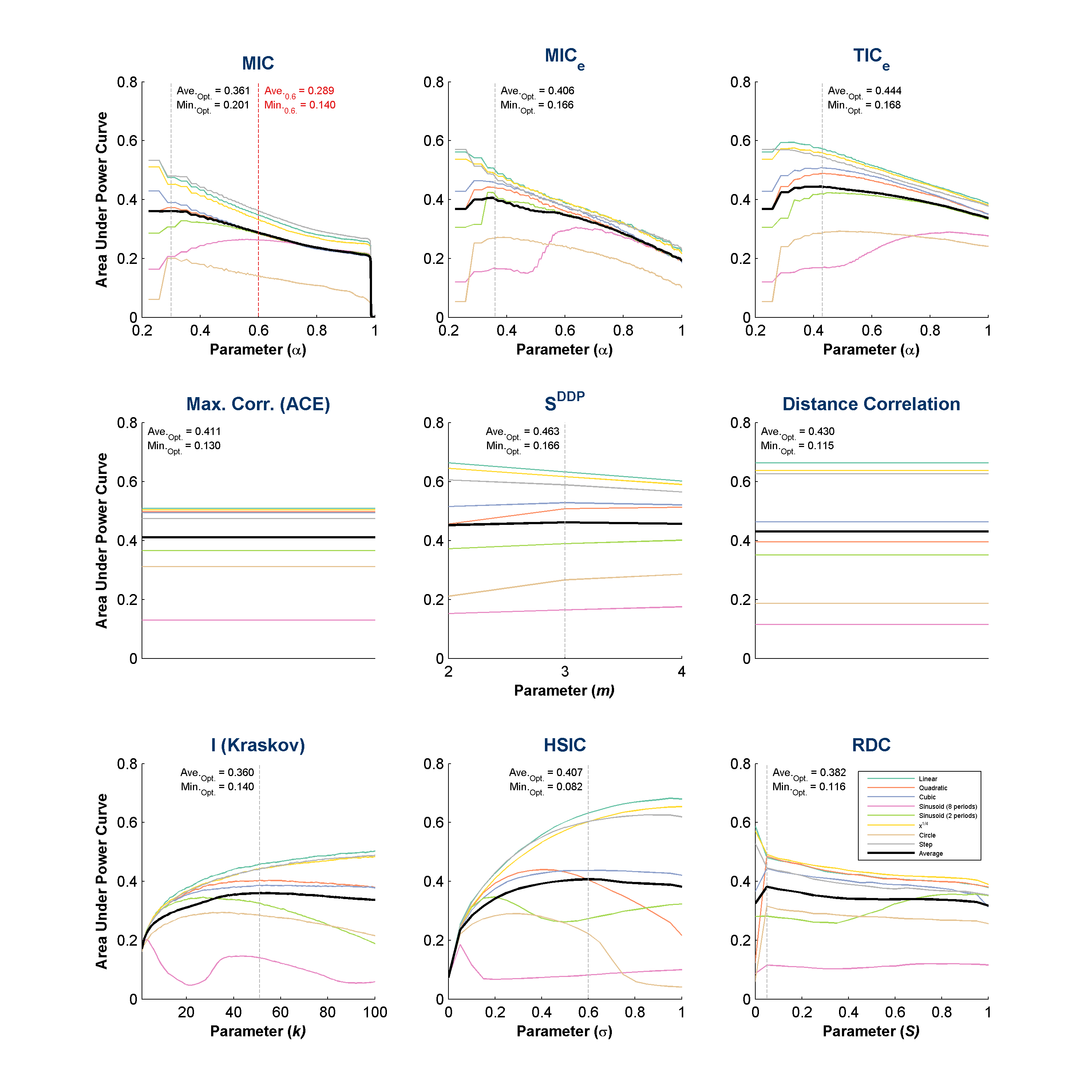}
	\caption[Power against independence as a function of the parameter of each measure of dependence]{
	    Power against independence as a function of the parameter of each measure of dependence. \textit{[Higher is more powerful.]}
	    For each measure of dependence, we computed power curves over a range of parameters using the relationships from \citet{simon2012comment}. In order to aggregate the power of a given test across relationship types, all power curves were computed as functions of the $R^2$ of the noisy relationship comprising the alternative hypothesis, and the area under each power curve was computed.  Here, we show for each statistic the area under the power curve for each relationship type as a function of that statistic's parameter.  The black line represents the average area under the power curves across all relationship types, and the vertical dotted line represents the optimal parameter setting.  Both the average and worst-case performance across relationship types are listed for the optimal parameter setting of each statistic.  For the $\MIC$ statistic from~\cite{MINE}, the red line represents the default parameter setting, which was used by Simon and Tibshirani. This parameter setting turns out to be poor for testing for independence; it is better suited for achieving equitability. For testing for independence, lower values of the parameter are better suited, though these incur a cost in terms of equitability. (See Figure~\ref{fig:power_equitability_tradeoff}.)}
	\label{fig:indivRelPowerParamSweep}
\end{figure}

\begin{figure}[H]
	\centering
	\includegraphics[clip=true, trim = 0.7in 0.45in 0.825in 0.4in, width=0.9\textwidth]{\pathToFigures/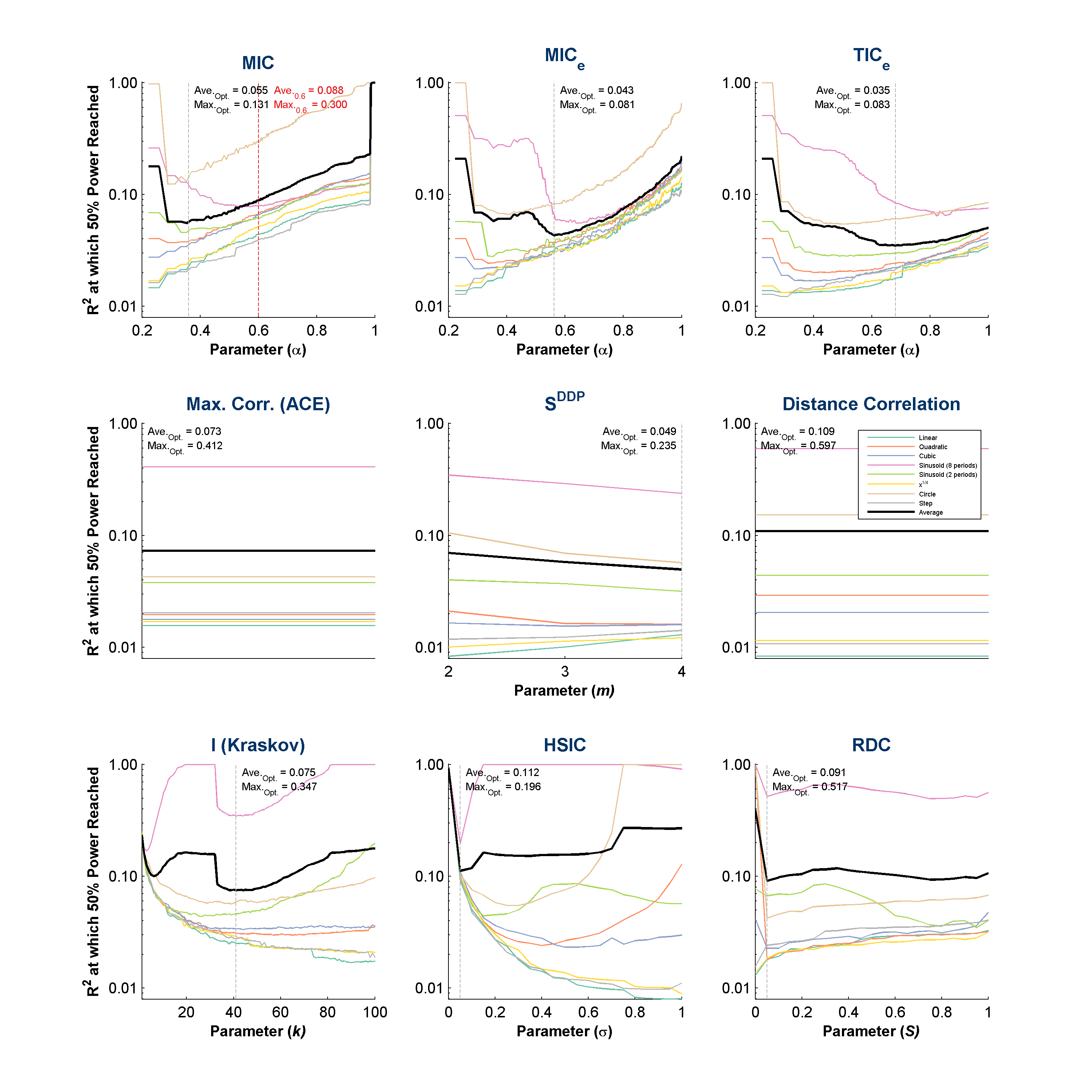}
	\caption[An alternative analysis of power against independence as a function of the parameter of each measure of dependence.]{
	    Power against independence as a function of the parameter of each measure of dependence, with overall power quantified differently than in Figure~\ref{fig:indivRelPowerParamSweep}. \textit{[Lower is more powerful.]}
	    As in Figure~\ref{fig:indivRelPowerParamSweep}, we compute power curves for a range of parameters of each measure of dependence using the relationships from \citet{simon2012comment}.  Here, in order to aggregate the power of a given test across relationship types, the power curve of each test was computed as a function of the $R^2$ of the noisy relationship being tested, and the $R^2$ at which $50\%$ power is achieved for each relationship type was determined.  This number is graphed for each relationship type and statistic as a function of that statistic's parameter.  The black line represents the average $R^2$ at which $50\%$ power is achieved across all relationships tested, and the vertical dotted line represents the optimal parameter setting.  Both the average and the worst-case performance across relationship types are listed for the optimal parameter setting of each statistic. For the $\MIC$ statistic from~\cite{MINE}, the red line represents the default parameter setting, which was used by Simon and Tibshirani. This parameter setting turns out to be poor for testing for independence; it is better suited for achieving equitability. For testing for independence, lower values of the parameter are better suited, though these incur a cost in terms of equitability. (See Figure~\ref{fig:power_equitability_tradeoff}.)}
	\label{fig:indivRelPowerParamSweepMaxR2}
\end{figure}

\section{The Equitability-runtime trade-off}
\begin{figure}[H]
	\centering
	\includegraphics[clip=true, trim = 1.55in 0in 0.05in 0in, width=\textwidth]{\pathToFigures/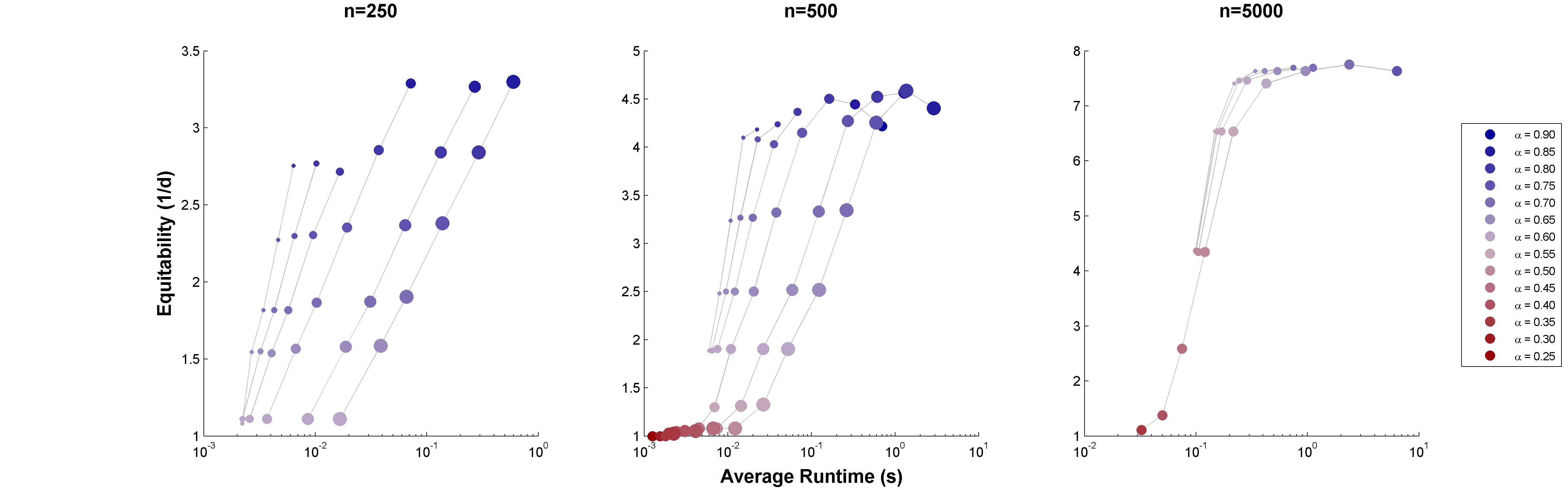}
	\caption[The relationship between equitability and runtime of $\MICestE$ at $n=250,500,5000$.]{
	    The relationship between equitability and runtime of $\MICestE$.
	    Sample sizes are $n=250$ (left), $500$ (middle), and $5000$ (right). Each plot shows, as $\alpha$ varies, the worst-case equitability of $\MICestE$ with the given value of $\alpha$ on the model used in Figure~\ref{fig:power_equitability_tradeoff} graphed against the runtime of $\MICestE$ with the same value of $\alpha$. The multiple series in every plot correspond to different values of $c$, with marker size indicating the size of $c$. The values of $c$ used are $1$, $2$, $3$, $5$, $10$, and $15$. ($c=10$ and $c=15$ are ommitted from the analysis for $n=5,000$.) As $\alpha$ increases, we generally see a rise in equitability but also in runtime.
	    }
	\label{fig:equitability_runtime}
\end{figure}

\section{Parameter values used in analyses}
\label{app:ParamSettings}
Parameter sweeps were performed for all methods in evaluating their equitability and statistical power against independence.

\subsection*{Parameter values used in equitability analyses}
\label{sec:equitabilityParams}
For each method, results are presented for the parameter values tested that maximized worst-case equitability across all models $\Q$ examined, at each sample size (see Table~\ref{tab:MIC_params_equitability}). Results for all parameter values tested, including for some methods not included in the figures here due to space constraints, can be found in the online supplement at \onlineSupplementLink.

In the case of $\RDC$ and $\HSIC$ the parameter values tested did not have a strong effect on equitability, so we present performance for the default / rule of thumb parameter values.  That is, the random sampling parameters, $(S_x, S_y)$, of $\RDC$ and the RBF kernel bandwidth parameters, $(\sigma_x, \sigma_y)$, used for $\HSIC$ were set independently for each of the two samples being tested to the Euclidean distance empirical median (values of $\left\{0-, 25-, 50-, 75-, 100-\right\}$\%-ile pairwise distances were also tested for these parameters). For $\RDC$, the number of random features was set to $k=10$. For the Kraskov mutual information estimator, $k=1$, $k=6$, $k=10$, and $k=20$ were tested. In the case of $\DDP$, values of $m > 3$ were prohibitively computationally expensive to run for this analysis. For $\MICestE$, at $n=250$, $500$, and $5,000$, the ranges of $\alpha$ tested were $\left\{0.60, 0.65, ..., 0.80, 0.85\right\}$, $\left\{0.25, 0.30, ..., 0.85, 0.90\right\}$, and $\left\{0.35, 0.40, ..., 0.70, 0.75\right\}$, respectively.

\begin{table}[h]
{\small
\centering
\begin{tabular}{cccccccccc}
\toprule
\multirow{2}{*}{\textbf{Sample size}}   & \multicolumn{2}{c}{\textbf{$\MICestE$}}   & \multicolumn{2}{c}{\textbf{$\TICestE$}}   & \textbf{$\DDP$}   & \textbf{I (Kraskov)}  & \multicolumn{2}{c}{\textbf{$\RDC$}}   & \textbf{$\HSIC$} \\
                                        & $\alpha$      & $c$                       & $\alpha$      & $c$                       & $m$               &   $k$                 &   $S_x, S_y$          &   $k$         &   $\sigma_x, \sigma_y$    \\ \midrule
                   
$250$               &   0.75        &   15          &   0.80        &   3           &   2           &   6       &  Median pair. dist.   &   10      &   Median pair. dist.      \\
$500$               &   0.80        &   5           &   0.80        &   3           &   2           &   6       &  Median pair. dist.   &   10      &   Median pair. dist.      \\   
$5,000$             &   0.65        &   3           &   0.70        &   3           &   2           &   6       &  Median pair. dist.   &   10      &   Median pair. dist.      \\ \bottomrule
\end{tabular}
\caption[Parameters used in the equitability analyses.]{Parameters used in the equitability analyses.}
\label{tab:MIC_params_equitability}
}
\end{table}

\subsection*{Parameter values used in statistical power analyses}
\label{sec:powerParams}
Tables~\ref{tab:MIC_params_power_AUC} and~\ref{tab:MIC_params_power_PowerThreshold} summarize the optimal parameters identified for tests for independence based on the methods examined, using area under the power curves and a $50$\% power threshold, respectively, as the optimization criterion. The parameters in Table~\ref{tab:MIC_params_power_AUC} were used to generate the power curves in Figure~\ref{fig:indivRelPowerOptimalParam}. The parameter ranges tested for each statistic can be observed from Figures~\ref{fig:indivRelPowerParamSweep} and ~\ref{fig:indivRelPowerParamSweepMaxR2}.

\begin{table}[h]
{\footnotesize
\centering
\begin{tabular}{cccccccccccc}
\toprule
\multirow{2}{*}{\textbf{Sample size}}   & \multicolumn{2}{c}{\textbf{$\MICestE$}}   & \multicolumn{2}{c}{\textbf{$\TICestE$}}   & \multicolumn{2}{c}{\textbf{$\MIC$}}   & \textbf{$\DDP$}   & \textbf{I (Kraskov)}  & \multicolumn{2}{c}{\textbf{$\RDC$}}   & \textbf{$\HSIC$} \\
                                        & $\alpha$      & $c$                       & $\alpha$      & $c$                       & $\alpha$      & $c$                       & $m$               &   $k$                 &   $S_x, S_y$          &   $k$         &   $\sigma_x, \sigma_y$    \\ \midrule
                   
$100$               &   0.48        &   5           &   0.50        &   5           &   0.40        &   5           &   3           &   13       &  5\%-ile pair. dist.   &   10      &   45\%-ile pair. dist.      \\
$500$               &   0.35        &   5           &   0.38        &   5           &   0.30        &   5           &   3           &   50       &  5\%-ile pair. dist.   &   10      &   60\%-ile pair. dist.      \\   \bottomrule
\end{tabular}
\caption[Best parameters for testing for independence, identified by maximizing the average area under the power curves generated by a given test for the set of relationships examined.]{Best parameters for testing for independence, identified by maximizing the average area under the power curves generated by a given test for the set of relationships examined.}
\label{tab:MIC_params_power_AUC}
}
\end{table}

\begin{table}[h]
{\footnotesize
\centering
\begin{tabular}{cccccccccccc}
\toprule
\multirow{2}{*}{\textbf{Sample size}}   & \multicolumn{2}{c}{\textbf{$\MICestE$}}   & \multicolumn{2}{c}{\textbf{$\TICestE$}}   & \multicolumn{2}{c}{\textbf{$\MIC$}}   & \textbf{$\DDP$}   & \textbf{I (Kraskov)}  & \multicolumn{2}{c}{\textbf{$\RDC$}}   & \textbf{$\HSIC$} \\
                                        & $\alpha$      & $c$                       & $\alpha$      & $c$                       & $\alpha$      & $c$                       & $m$               &   $k$                 &   $S_x, S_y$          &   $k$         &   $\sigma_x, \sigma_y$    \\ \midrule
                   
$100$               &   0.74        &   5           &   0.96        &   5           &   0.48        &   5           &   5           &   12       &  5\%-ile pair. dist.   &   10      &   30\%-ile pair. dist.      \\
$500$               &   0.56        &   5           &   0.68        &   5           &   0.36        &   5           &   4           &   41       &  5\%-ile pair. dist.   &   10      &   5\%-ile pair. dist.      \\   \bottomrule
\end{tabular}
\caption[Best parameters for testing for independence, identified by minimizing the $R^2$ across the set of relationships examined for which the average power of a given test remained above $50$\%.]{Best parameters for testing for independence, identified by minimizing the average across relationship types of the minimal $R^2$ for which the power of a given test remained above $50$\%.}
\label{tab:MIC_params_power_PowerThreshold}
}
\end{table}

\subsection*{Parameter values used in runtime analyses}
\label{sec:runtimeParams}
For methods whose runtime did not strongly depend on parameter settings, default parameter values were used.  That is, the Kraskov mutual information estimator was run using $k=6$, and the random sampling parameters, $(S_x, S_y)$, of $\RDC$ and the RBF kernel bandwidth parameters, $(\sigma_x, \sigma_y)$, used for $\HSIC$ were set independently for each of the two samples being tested to the Euclidean distance empirical median. In the case of $\RDC$, the number of random features was set to $k=10$, as in the runtime analysis in~\cite{lopez2013randomized}. The parameters used for $\MICestE$ are presented in Table~\ref{tab:MIC_params_runtime}.

\begin{table}[h]
\centering
\begin{tabular}{ccccccc}
\toprule
\multirow{2}{*}{\textbf{Sample size}} & \multicolumn{2}{c}{\textbf{Power}} & \multicolumn{2}{c}{\textbf{Fast equitability}} & \multicolumn{2}{c}{\textbf{Equitability}} \\
                    & $\alpha$      & $c$           & $\alpha$      & $c$           & $\alpha$      & $c$       \\ \midrule
                   
$50$                &   0.54        &   5           &   0.75        &   3           &   0.85        &   5       \\
$100$               &   0.48        &   5           &   0.70        &   2           &   0.80        &   5       \\
$500$               &   0.36        &   5           &   0.65        &   1           &   0.80        &   5       \\
$1,000$             &   0.32        &   5           &   0.60        &   1           &   0.75        &   4       \\
$5,000$             &   0.26        &   5           &   0.50        &   1           &   0.65        &   1       \\
$10,000$            &   0.24        &   5           &   0.45        &   1           &   0.60        &   1       \\ \bottomrule
\end{tabular}
\caption[Parameters used in the runtime analysis of $\MICestE$.]{Parameters used in the runtime analysis of $\MICestE$ presented in Table~\ref{tab:runtimes}.}
\label{tab:MIC_params_runtime}
\end{table}
For $\MICestE$, the three sample-size-dependent parameter settings optimize for maximal power against independence, 80\% of optimal equitability (fast equitability), and 99\% of optimal equitability. For sample sizes for which results were not available, parameter values were estimated via interpolation/extrapolation using a power curve.  As pointed out in Section~\ref{sec:choosingAlpha}, these parameter settings depend on the set of relationships being examined, and, for example, for relationship suites with less complex relationships than the ones examined in the analyses here, lower values of $\alpha$ would perform well and be more computationally efficient.

\end{document}